%% file: main.tex
\documentclass[aps,rmp,reprint,amsmath,amssymb,graphicx,longbibliography]{revtex4-1}

\usepackage{helvet}
\usepackage{bm}
\usepackage{epsfig}
\usepackage{amsfonts,amsmath,amssymb,amsthm,amscd}
\usepackage{color}
\usepackage{dcolumn}
\usepackage{array}
\usepackage{float}
\usepackage[]{units}

\usepackage[normalem]{ulem}

\usepackage{graphicx}
\usepackage{epstopdf}
\usepackage{array}

\usepackage{slashed}
\usepackage{gensymb}
\usepackage{newtxtext,newtxmath}

\usepackage[colorlinks=true,linkcolor=blue,citecolor=blue]{hyperref}
\usepackage{cleveref}
\usepackage{our}
\usepackage{mathrsfs}
\usepackage{makecell,longtable}
\newfont{\cirth}{cirth scaled\magstep2}
\usepackage{ifthen}
\newboolean{referencesPerChapter}
\setboolean{referencesPerChapter}{true} 
\def\bib{\ifthenelse{\boolean{referencesPerChapter}}
        {\bibliography{references_fixed}}
        {}
    }

\graphicspath{{figures/}{../figures/}}
\usepackage{subfiles}

\usepackage{totcount}
\newtotcounter{citnum}
\def\oldbibitem{} \let\oldbibitem=\bibitem
\def\bibitem{\stepcounter{citnum}\oldbibitem}

\usepackage{color}

\begin{document}
\setboolean{referencesPerChapter}{false}

\title{Charged particle motion and radiation in strong electromagnetic fields}

\author{A. Gonoskov}
\author{T. G. Blackburn}
\author{M. Marklund}
\affiliation{Department of Physics, University of Gothenburg, SE-41296 Gothenburg, Sweden}
\author{S. S. Bulanov}
\affiliation{Lawrence Berkeley National Laboratory, Berkeley, California 94720, USA}

\date{\today{}}

\begin{abstract}
The dynamics of charged particles in electromagnetic fields is an essential component of understanding the most extreme environments in our Universe. In electromagnetic fields of sufficient magnitude, radiation emission dominates the particle motion and effects of quantum electrodynamics (QED) in strong fields are crucial, which triggers electron-positron pair cascades and counterintuitive particle-trapping phenomena. As a result of recent progress in laser technology, high-power lasers provide a platform to create and probe such fields in the laboratory. With new large-scale laser facilities on the horizon and the prospect of investigating these hitherto unexplored regimes, we review the basic physical processes of radiation reaction and QED in strong fields, how they are treated theoretically and in simulation, the new collective dynamics they unlock, recent experimental progress and plans, as well as possible applications for high-flux particle and radiation sources.

\end{abstract}

\maketitle
\tableofcontents

\subfile{introduction}

\subfile{theory}

\subfile{numerics}

\subfile{dynamics}

\subfile{experiments}

\subfile{conclusions}

\begin{acknowledgments}
The authors acknowledge fruitful discussions with the members of the ``International Quantum Plasma Initiative'' (IQPI) collaboration, T. Heinzl, A. Ilderton, and D. Seipt, as well as with A. Arefiev, S. V. Bulanov, E. Esarey, C. H. Keitel, J. K. Koga, J. Magnusson, S. P. D. Mangles, A. Di Piazza, C. P. Ridgers, C. B. Schroeder, L. O. Silva, and A. G. R. Thomas.
S. S. B. acknowledges the support from the U.S. DOE Office of Science Offices of HEP and FES (through LaserNetUS), under Contract No. DE-AC02–05CH11231. A. G. acknowledges the support of the Swedish Research Council (Grant No. 2017-05148). M. M. acknowledges the support of the Swedish Research Council (Grant No. 2020-06768).
\end{acknowledgments}

\subfile{definitions}

\bibliographystyle{apsrmp4-1}
\bibliography{references_fixed}

\end{document}

%% file: introduction.tex
\section{Introduction}
\label{sec:Introduction}

Over the last hundred years elementary particle physics has made tremendous progress in the exploration of the high-energy regime, culminating in the Standard Model and its verification at grand-scale experimental facilities, such as LEP, LEP II, SLAC, Tevatron, and LHC.
Now, a new pathway in fundamental physics research has opened, using the electromagnetic fields of focused laser pulses to explore the high-intensity regime.
{
Here the high density and simultaneous interaction of the participating particles provide a fruitful basis for the creation of extreme environments, 
which can only be described theoretically using nonperturbative methods in quantum electrodynamics (QED).
While powerful, this theory is insufficient outside highly idealised interaction configurations or when collective dynamics dominate.
Understanding how single-particle QED processes couple to large-scale plasma phenomena and vice versa is a significant unsolved problem.
Experimental investigation of this little-explored regime is therefore of fundamental interest, as well as being of practical importance.
This includes the effects of extreme environments on the next generation of particle accelerators and the formation of dense electron-positron plasmas, which occur in astrophysical phenomena, and could be exploited to create extraordinary particle and radiation sources.
}
In light of this, the purpose of this review is to assess the current state of the art in our theoretical understanding of the strong-field regime and upcoming experimental programs, aimed at the exploration of such opportunities.

    \begin{figure}[tbp]
    \centering
    \includegraphics[width=\linewidth]{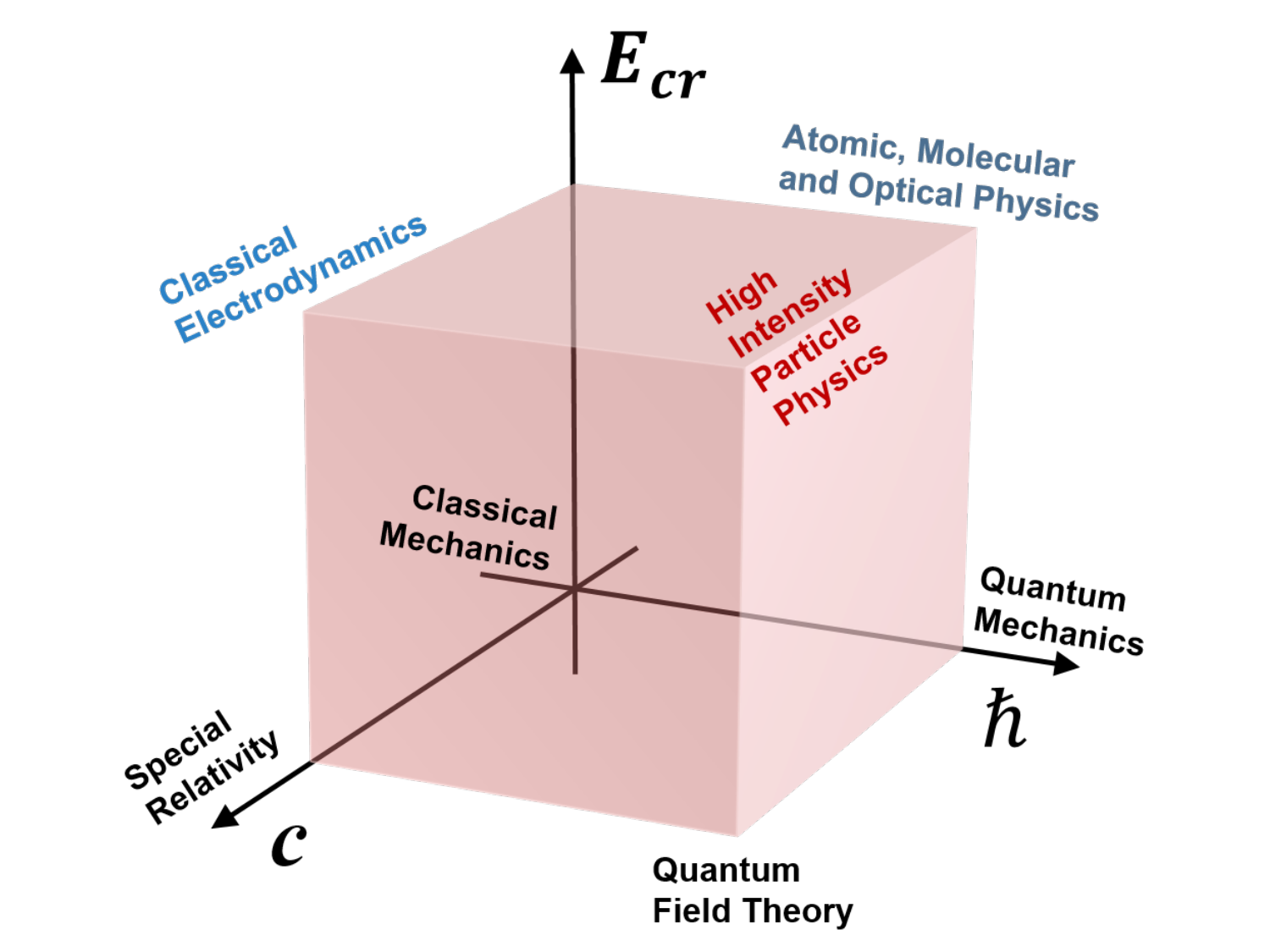}
    \caption{%
        The cube of theories: its three axes correspond to relativistic ($c$), quantum ($\hbar$), and high-intensity effects ($\Ecrit$) and its vertices to the theories: $(0,0,0)$ is classical mechanics, $(c,0,0)$ is Special Relativity, $(0,\hbar,0)$ is Quantum Mechanics, $(c,\hbar,0)$ is Quantum Field Theory, $(c,0,E_{cr})$ is the classical Electrodynamics, $(0,\hbar,E_{cr})$ is atomic, molecular and optical physics, and $(c,\hbar,E_{cr})$ is the High Intensity Particle Physics.
        Here $\Ecrit = m^2 c^3 / (e \hbar)$ is the critical field of QED, where $m$ is the electron mass and $e$ is the elementary charge.
        Adapted from \citet{bulanov.pra.2013}.%
    }
    \label{cube}
    \end{figure} 

Increasingly strong electromagnetic (EM) fields significantly modify particle dynamics, giving rise to phenomena that are not encountered either in classical or perturbative quantum theories of these interactions.
To illustrate this, consider the ``cube of theories'' shown in \cref{cube}, proposed in \citet{bulanov.pra.2013} by analogy with that shown in \citet{okun.spu.1991}.
This cube has three orthogonal axes, defined by the constants $c$ (the speed of light in vacuum), $\hbar$ (the reduced Planck's constant) and $\Ecrit$ (the critical field of quantum electrodynamics, which does work equal to the electron rest energy over a single Compton length).
These three axes give the strength of relativistic, quantum, and strong-field effects, respectively.
Each vertex of the cube corresponds to a physical theory: $(0,0,0)$ is non-relativistic mechanics; $(c,0,0)$ is special relativity; and $(0,\hbar,0)$ is quantum mechanics.
Quantum field theory, which combines quantum and relativistic effects, is found at $(c,\hbar,0)$.
Classical electrodynamics corresponds to the vertex $(c,0,E_{cr})$ and atomic, molecular and optical physics to $(0,\hbar,E_{cr})$.
The vertex that interests us is located at $(c,\hbar,E_{cr})$, which encompasses strong field quantum electrodynamics (SFQED).

The `strong-field' aspect of quantum electrodynamics significantly alters the physics of the interactions, as we will discuss in this review.
With the increase of field strength, the interaction between particles and electromagnetic fields enters the so-called `radiation-dominated' regime, and then, as the particle energies increase, nonlinear quantum effects come into play.
These phenomena fall under the umbrella of `high intensity particle physics', a new, emerging branch of physics that occupies a significant region in the Standard Model parameter space and is largely unexplored.
It is crucial for understanding extreme astrophysical environments, including neutron stars, magnetars, black holes, and will be important for the design of future lepton and $\gamma$-$\gamma$ colliders.

The reason that interest in this field has grown signficantly in recent years is the progress in the development of large-scale, high-intensity laser facilities~\cite{danson.hplse.2019}, in the `petawatt revolution' following the invention of chirped pulse amplification (CPA)~\cite{strickland.oc.1985}.
Indeed, the intensities achieved over the past decade have now reached $10^{23}~\Wcmsqd$~\cite{yoon.optica.2021}.
Until recently, experimental studies were limited to single example from the 1990s, namely E144 at SLAC, where a 46.5 GeV electron beam interacted with a $10^{18}$ W/cm$^2$ laser pulse, producing high energy photons, which in turn transformed into electron-positron pairs in the electromagnetic field of the laser pulse~\cite{bula.prl.1996,burke.prl.1997}.
The situation changed with two experiments on radiation reaction at the Rutherford Appleton Laboratory, which reported radiative energy loss, and quantum corrections thereto, of an electron beam in the interaction with a counterpropagating, high-intensity laser pulse~\cite{cole.prx.2018,poder.prx.2018}.

Today, experiments on radiation emission and pair creation in the strong-field regime form part of the planned experimental programs at almost every major petawatt or multipetawatt laser facility, including the Extreme Light Infrastructure (ELI)~\cite{weber.mre.2017,gales.rpp.2018}, Apollon~\cite{papadopoulos.hpl.2016}, the Station of Extreme Light~\cite{cartlidge.science.2018}, the Center for Relativistic Laser Science (CoReLS)~\cite{yoon.optica.2021}, J-KAREN-P~\cite{kiriyama.hedp.2020}, the Omega Laser Facility~\cite{bromage.hplse.2019}, and the Zetawatt-Equivalent Ultrashort Pulse Laser System (ZEUS)~\cite{nees.cleo.2020}, as well as conventional accelerator facilities~\cite{abramowicz.arxiv.2019,meuren.exhilp.2019}.

\subsection{The strong electromagnetic field}
\label{sec:KeyParameters}

When we characterize an electromagnetic field as `strong' or `weak', we implicitly compare its magnitude to a scale of some kind. The theory of quantum electrodynamics defines such a scale through the electron mass $m$, which yields a characteristic energy $m c^2$ and length $\lc = \hbar / (m c)$, and the elementary charge $e$:
    \begin{equation}
    \Ecrit = \frac{m^2 c^3}{e \hbar} = 1.323 \times 10^{18}~\text{V}\text{m}^{-1}.
    \label{eq:CriticalField}
    \end{equation}
We refer to this quantity as the \emph{critical field of quantum
electrodynamics}. Such a field does work equal to the electron rest energy
over a (reduced) Compton length $\lc=3.86\times 10^{-11}$ cm; the gyroradius $r_c$
of an electron in a magnetic field of equivalent strength,
$\Bcrit = \Ecrit/c = 4.41 \times 10^9~\mathrm{T}$, would be
$r_c = \lc$.
$\Ecrit$ is often named the \emph{Schwinger field}, in honour of \citet{schwinger.pr.1951}, although it appears earlier in the work of {\citet{sauter.zp.1931} and \citet{heisenberg.zp.1936}.}
It is constructed from fundamental constants and is
therefore ubiquitous in the study of quantum effects in
strong electromagnetic fields~\citep{dipiazza.rmp.2012}.

It is possible to define an equivalent of \cref{eq:CriticalField}
within the framework of classical electrodynamics, by exchanging
the Compton length $\lc = \hbar / (m c)$ for the classical
electron radius $r_e = e^2 / (4 \pi \epsilon_0 m c^2)=2.82 \times 10^{-13}$ cm:
    \begin{equation}
    \Ecrit^\text{clas} =
        \frac{4 \pi \epsilon_0 m^2 c^4}{e^3} =
        \frac{\Ecrit}{\alpha} =
        1.813 \times 10^{20}~\text{V}\text{m}^{-1}.
    \label{eq:ClassicalCriticalField}
    \end{equation}
This \emph{classical critical field} is larger than the
QED critical field by a factor of $1/\alpha$, where $\alpha
\simeq 1/137$ is the fine-structure constant, which hints
that quantum effects lead the hierarchy of corrections
to the usual Lorentz force equation (see \cref{sec:TheProblemOfRR}).

Both classical and quantum electrodynamics are relativistic
theories, so the comparison of a field strength to $\Ecrit$
(or $\Ecrit^\text{clas}$) must be done in a way that is invariant
under Lorentz transformations.
The quantities available to us are the particle four-momentum
$p^\mu = \gamma m (c, \vec{v})$
and the field tensor $F_{\mu\nu}$.
Here $\gamma = (1 - \vec{v}^2/c^2)^{-1/2}$ is the
Lorentz factor associated with three-velocity $\vec{v}$.
The equivalent quantity for a photon with wavevector $\vec{k}$ and energy $\hbar \omega$ is the four-vector $\hbar k^\mu = (\hbar \omega/c) (1, \vec{n})$, where $\vec{n}$ is its (unit) direction of propagation.

The symmetry properties of $F_{\mu\nu}$ restrict the number of
non-zero invariant combinations of $p^\mu$ and $F_{\mu\nu}$ to four,
if the field is constant and homogeneous~\cite{ritus.jslr.1985,baier1998}.
Two are purely field-dependent and two are particle-dependent as well.
The former two are the Poincare invariants \cite{schwinger.pr.1951,dunne.tp.2005}:
    \begin{align}
    \InvariantF &=
        -\frac{F_{\mu\nu} F^{\mu\nu}}{2 \Ecrit^2} =
        \frac{\vec{E}^2 - c^2 \vec{B}^2}{\Ecrit^2}
    \label{eq:InvariantF},
    \\
    \InvariantG &=
        -\frac{F_{\mu\nu} \tilde{F}^{\mu\nu}}{4 \Ecrit^2} =
        \frac{c \vec{B}\cdot\vec{E}}{\Ecrit^2}.
    \label{eq:InvariantG}
    \end{align}
Here we use the dual field tensor $\tilde{F}_{\mu\nu} =
\tfrac{1}{2} \epsilon_{\mu\nu\alpha\beta} F^{\alpha\beta}$,
where $\epsilon_{\mu\nu\alpha\beta}$ is the antisymmetric
tensor, and implicitly sum over repeated indices.
Replacing $F_{\mu\nu}$ with $\tilde{F}_{\mu\nu}$ in
\cref{eq:Chi,eq:ChiGamma} would yield the same invariants.
Additional combinations are possible if the field is not
homogeneous (in the four-dimensional sense), involving terms
of the form $\partial_\mu F^{\alpha\beta}$.
There is no contribution from terms such as $\partial_\mu F^{\mu\nu}$
and $\partial_\mu \tilde{F}^{\mu\nu}$, as these are zero
by virtue of Maxwell's equations.

The physical meaning of these invariants is as follows.
A field which satisfies $\InvariantF > 0$, $\InvariantG = 0$
is \emph{electric-type} and is unstable against spontaneous
decay to electron-positron pairs~\cite{schwinger.pr.1951}.
This occurs for any field magnitude in principle, but
it is exponentially suppressed unless $\InvariantF \gtrsim 1$.
(This is equivalent to $\vec{E}^2 \gtrsim \Ecrit^2$ if $\vec{E}^2 \gg c^2 \vec{B}^2$.)
Fields of \emph{magnetic-type}, with $\InvariantF < 0$,
$\InvariantG = 0$, are stable against vacuum pair creation,
but do induce birefringence effects~\citep{erber.rmp.1966,marklund.rmp.2006}.
{A special role is played by \emph{crossed fields}, in which the electric and magnetic fields are mutually perpendicular and have equal magnitude (up to a factor of $c$, $\left|\vec{E}\right| = c\left|\vec{B}\right|$),
because of the possibility to approximate an arbitrary electromagnetic field as a crossed field, under certain conditions~\citep{nikishov.jetp.1964,ritus.jslr.1985}.}
(We will return to this point in \cref{sec:LCFA}.)
Crossed fields are not susceptible to vacuum pair creation,
regardless of their amplitude~\cite{heisenberg.zp.1936,schwinger.pr.1951}.

The two invariants which mix particle and field properties are the \emph{quantum nonlinearity parameters}:\footnote{The literature includes alternative notation for $\chi_e$, such as $\Upsilon$~\cite{erber.rmp.1966,chen.prd.1992,harding.rpp.2006} and $\eta$~\cite{bell.prl.2008}.
Note also that it is possible for $\chi_\gamma$ to be defined by normalizing the photon energy $\hbar \omega$ to $2mc^2$, rather than $mc^2$ as we have done here, particularly in the astrophysical literature~\cite{erber.rmp.1966,harding.rpp.2006}.
The rationale for doing so is that electron-positron pair creation by a photon creates two particles of mass $m$, and so $2 m c^2$ is a natural energy scale.}
    \begin{align}
    \begin{split}
    \chi_e &=
        \frac{\sqrt{-(F_{\mu\nu} p^\nu)^2}}{m c \Ecrit}
    \\
    &=
        \gamma \frac{\sqrt{(\vec{E} + \vec{v} \times \vec{B})^2 - (\vec{E}\cdot\vec{v}/c)^2}}{\Ecrit}
    \label{eq:Chi}
    \end{split}
    \\
    \begin{split}
    \chi_\gamma &=
        \frac{\hbar \sqrt{-(F_{\mu\nu} k^\nu)^2}}{m c \Ecrit}
    \\
    &=
        \frac{\hbar \omega}{m c^2}
        \frac{\sqrt{(\vec{E} + \left(c^2 \vec{k}/\omega\right) \times \vec{B})^2 - \left(\vec{E}\cdot\left(c\vec{k}/\omega\right)\right)^2}}{\Ecrit}
    \label{eq:ChiGamma}
    \end{split}
    \end{align}
Hereafter, wherever the distinction is not necessary, we use the term particle in relation to both photons and electrons (positrons), denoting the corresponding quantities by the subscript $\gamma$ and $e$ respectively.
{We omit the subscript in situations where it is clear whether we refer to a photon, or to an electron or positron.}

For the electron, $\chi_e$ compares the field strength observed in its instantaneous rest frame with $\Ecrit$~\cite{ritus.jslr.1985},
{or the typical energy of the radiation spectrum ($\simeq \hbar\omega_c$, the synchrotron critical frequency: see \cref{eq:ClassicalEmissionRate}) with its energy $\gamma m c^2$~\cite{SokolovTernov}}.
It therefore parametrizes the importance of quantum corrections to the particle dynamics and radiation emission.

A photon does not have a rest frame, so a physical interpretation of $\chi_\gamma$ is obtained by considering the following argument~\cite{bassompierre.plb.1995}.
In order for a photon with momentum $\hbar\omega \vec{n} / c$, embedded in an electromagnetic field, to create an electron-positron pair, there must be a transfer of energy from the field of $\Delta \Energy \simeq m^2 c^4 / (\hbar \omega)$.
This energy transfer must take place over $\Delta t \simeq m c / (e E_\text{eff})$, the time taken for the electron and positron to gain a transverse momentum of $m c$ if the field has effective magnitude $E_\text{eff} = \gamma \abs{\vec{E}_\perp + c \vec{n} \times \vec{B}}$ (we have assumed $\hbar\omega \gg m c^2$ and $\gamma \gg 1$).
The uncertainty principle requires that $\Delta \Energy \Delta t < \hbar$, or equivalently, $\chi_\gamma > 1$.
This condition then defines the regime where electron-positron pair creation becomes probable.

In a {null} field, where $\InvariantF = \InvariantG = 0$, it is $\chi_e$ and $\chi_\gamma$ which control the importance of quantum effects.
{A crossed field is a null field where the electric and magnetic fields have the same magnitude (up to a factor of $c$) and are perpendicular to each other.}
A particularly important case of crossed field is the plane electromagnetic wave, wherein the electric and magnetic fields oscillate with angular frequency $\omega_0$ and the phase fronts propagate along {a} null wavevector $\kappa^\mu$.
Here the dynamics are characterized by an additional
invariant parameter~\cite{nikishov.jetp.1964}:\footnote{
Use of $a_0$ is conventional in the laser and
plasma-physics communities, but it can be denoted
as $\xi$ in the SFQED literature~\cite{dipiazza.rmp.2012}.}
    \begin{equation}
    a_0 =
        \frac{1}{mc^2}
        \frac{e \sqrt{-(F_{\mu\nu} p^\nu)^2}}{\kappa_\mu p^\mu} =
        \frac{e E_0}{m c \omega_0},
    \label{eq:a0}
    \end{equation}
where $E_0 = \abs{\vec{E}} = c \abs{\vec{B}}$.
$a_0$ is called the \emph{normalized amplitude} of the wave, the field amplitude in relativistic units (i.e. in units of $mc\omega_0/e$), the
\emph{strength parameter}, or the \emph{normalized vector
potential}.
The reason for the last of these is that $a_0$ is often defined
as $a_0 = e \sqrt{-A_\mu A^\mu} / (m c)$, using the wave's
four-potential $A^\mu$; this gives an equivalent result under
the conventional choice of Lorenz gauge $\partial_\mu A^\mu = 0$
[see discussion of gauge invariance in \citet{reiss.pra.1979,heinzl.oc.2009}].
As the transverse momentum acquired by an electron in a
wave with potential $A^\mu$ is $p^\mu_\perp = e A^\mu$,
{$a_0 = \abs{p_\perp} / (m c)$} measures the importance of relativistic
effects, i.e. the classical nonlinearity of the dynamics.

The parameter $a_0$ also functions as a measure of the multiphoton, or nonlinear, nature of the electromagnetic interaction.
Consider the motion of an electron, driven by a circularly polarized EM wave, in its average rest frame:
in this frame, the electron performs circular motion with radius $r'$, angular frequency $\omega' = 1/r'$ and constant Lorentz factor $\gamma' = \sqrt{1 + a_0^2}$, emitting radiation with a characteristic frequency of $\omega_c' = \gamma'^3 / r'$.
The effective harmonic order, which indicates how many photons are absorbed, $n = \omega_c' / \omega' \simeq a_0^3$.
From the quantum perspective, \cref{eq:a0} compares
the work done by the electromagnetic wave over a Compton length,
$e E_0 \lc$, to the energy of a photon of that wave,
$\hbar\omega_0$; if $a_0 \gg 1$, the interaction is unavoidably
nonlinear in the number of participating photons.

An alternative interpretation relates $a_0$ to the \emph{Keldysh parameter} $\gamma_K = \omega_0 \sqrt{2 m V_0} / (e E_0)$, which characterizes the ionization of an atomic electron with binding energy $V_0$ by an electromagnetic wave with amplitude $E_0$ and frequency $\omega_0$~\cite{keldysh.jetp.1965,popov.pu.2004}. If $\gamma_K \ll 1$, the {tunneling} regime, the ionization occurs on a time scale much shorter than the period of the wave and the electron tunnels through the potential barrier; if $\gamma_K \gg 1$, the multiphoton regime, simultaneous absorption of several photons promotes the electron to the continuum. The laser strength parameter $a_0$ is equivalent to $1/\gamma_K$ for $V_0 \simeq m c^2$; thus we identify $a_0 \gg 1$ as the {tunneling} regime for SFQED processes~\cite{dipiazza.rmp.2012}, such as electron-positron pair creation~\cite{reiss.jmp.1962}. Finally, we note that a `critical' normalized vector potential $a_0$ may be defined by setting $E_0 = \Ecrit$ in \cref{eq:a0}: 
    \begin{equation}
    \acrit = m c^2 / (\hbar \omega_0).   
    \end{equation}
This is equivalent to $\acrit \approx 4.1 \times 10^5 \lambda^{-1}~[\micron]$.

\subsection{Relevant environments}
\label{sec:Environments}

We now turn to the question of what fields satisfy the
criteria to be `strong', as defined in \cref{sec:KeyParameters},
and where they can be found.
Here we briefly discuss the {five} major cases:
the magnetic fields surrounding compact astrophysical objects,
such as pulsars, magnetars and black holes;
the boosted, collective Coulomb fields of ultrarelativistic,
dense lepton bunches, as found at the final focus of
conventional particle accelerators;
the coherently summed, nuclear electric fields observed
by ultrarelativistic leptons travelling through a crystal
along an axis of symmetry;
and the electromagnetic fields produced by focusing of
high-power lasers.

A commonality between these {five} cases is the
exploitation of the relativistic factor $\gamma$ in
\cref{eq:Chi}; even though an electromagnetic field may
be `weak' in the lab frame ($\abs{\vec{E}}/\Ecrit \ll 1$),
it can still induce nonlinear quantum effects if probed
by an ultrarelativistic particle, with $\gamma$ sufficiently
large that $\chi \simeq \gamma \abs{\vec{E}} / \Ecrit \not\ll 1$.
In the majority of scenarios that follow, the origin of
this ultrarelativistic particle is external to the
field, i.e. electrons are accelerated by some
mechanism to $\gamma \gg 1$ \emph{before} encountering
the electromagnetic field.

A distinct class of interactions is possible with
high-intensity lasers, which can, at large $a_0$,
accelerate initially stationary electrons to
sufficiently high energy to make radiation losses
and quantum effects important.
In this case the laser plays the role of both accelerator
and target~\cite{bell.prl.2008}.
This fact is crucial to understanding the richness of
the dynamics induced in this scenario, which we discuss
in \cref{sec:Dynamics}.

It is unknown whether fields that satisfy $\InvariantF \simeq 1$,
rather than $\chi \simeq 1$, can be created 
{using lasers}
~\cite{fedotov.prl.2010,bulanov.prl.2010s}, as they should trigger electron-positron cascades that
efficiently absorb and reflect the electromagnetic energy (see \cref{sec:Cascades}).

\subsubsection{Highly magnetized astrophysical environments}
\label{sec:AstrophysicalEnvironments}

Magnetic fields of extraordinary strength can be found around compact objects, including neutron stars and black holes. A pulsar, for example, is a highly magnetized, rotating neutron star, formed after the supernova of a massive star and the collapse of its core. Characteristic values of the magnetic field strength at the surface $B_s$ (as inferred from the spin-down rate) and rotation period are $10^{7}$--$10^{9}$~T and $1.5$~ms--8~s, respectively~\cite{harding.rpp.2006}.
{The rotation drives a wind of relativistic particles and powers the emission of electromagnetic radiation along the magnetic axis of the pulsar;}
as energy is lost, the pulsar spins down and the rotation period increases. 
In the `standard model' of a pulsar, the electric field that is induced by the rotation of its magnetic field is shorted out by electron-positron plasma~\cite{goldreich.apj.1969,sturrock.apj.1971}, except in small regions where particle acceleration takes place.
The creation of this pair plasma is a nonsteady process that is now studied with large-scale simulations of the pulsar magnetosphere~\cite{chen.apj.2020,schoeffler.aj.2019}.

The rotation of the pulsar magnetic field creates a voltage drop across the open field lines large enough to drive the acceleration of electrons and positrons to energies of at least 10~TeV~\cite{harding.rpp.2006}.
The combination of high particle energy and strong magnetic field in the region above the polar cap causes cascades of photon emission and electron-positron pair creation~\cite{daugherty.apj.1982}, with extremely high multiplicity~\cite{timokhin.apj.2015}.
[Photons with energies below the pair creation threshold can instead drive a \emph{photon splitting} cascade~\cite{adler.ap.1971,baring.apj.2001}.]
Coherent emission from pulsars is suspected to be linked to the dynamics of the pair cascade~\cite{sturrock.apj.1971,philippov.prl.2020,cruz.apj.2020}.

Even larger magnetic fields, $B_s \sim 10^{10}$ to $10^{11}$~T~\cite{olausen.apjs.2014}, are believed to exist near magnetars~\cite{duncan.apj.1992}, giving rise to exotic effects such as vacuum birefringence, observation of which was reported by \citet{mignani.mnras.2017} (see also \citet{capparelli.epjc.2017}).
Gamma ray emission~\cite{hirotani.apj.2016} and pair creation~\cite{crinquand.prl.2020} also occur around black holes.
Some models of the supermassive black hole at the centre of M87 predict that millimeter-wave radio emission is dominated by electron-positron pairs produced by, e.g. photon-photon collisions, that are accelerated by unscreened electric fields~\cite{akiyama.apj.2019}.

\subsubsection{Beam-beam interactions}
\label{sec:BeamBeam}

Lepton colliders use high-energy particle beams to probe fundamental physics at the smallest spatial scales~\cite{schael.pr.2013}.
These are conventionally electron-positron, but there are projects to built muon-antimuon, and electron-electron colliders {using both conventional and plasma technologies \cite{roadmap.doe.2016,alegro.arxiv.2019,alegro.arxiv.2020,roadmap.arxiv.2022}}. As the cross sections for the sought after processes are usually small, 
{high particle flux is needed and therefor particle beam has to be focused}
to small radii at the interaction point. 
The combination of high Lorentz factor and charge density leads to the creation of strong, collective electromagnetic fields, which can deflect the particles of the oncoming beam~\cite{yokoya.1992}.
This deflection, which is focusing (defocusing) for electron-positron (electron-electron) collisions, is known as \emph{disruption}~\cite{hollebeek.nim.1981,chen.prd.1988} and it is associated with the emission of radiation, called \emph{beamstrahlung}~\cite{blankenbecler.prd.1987,noble.nima.1987}.

The power and spectral properties of the beamstrahlung are controlled by the electron quantum parameter $\chi_e$ (often referred to as the `beamstrahlung parameter' and denoted by $\Upsilon$ in this context).
Its average value is given by~\cite{chen.prd.1992}
    \begin{equation}
    \avg{\chi_e} \simeq
        \frac{0.11 \Energy_0 [10~\text{GeV}] \, Q [\text{nC}]}
        {\sigma_z [\micron] (\sigma_x [\micron] + \sigma_y [\micron])}
    \label{eq:BeamstrahlungChi}
    \end{equation}
where $\Energy_0$ is the energy of the beams, $Q$ is the charge per bunch, and  the $\sigma_i$ are the root-mean-square radii of the charge distribution in the directions parallel ($z$) and perpendicular ($x,y$) to the collision axis.
The normalized amplitude $a_0$ can be obtained by treating the beam field as a half-cycle electromagnetic wave with wavelength $\lambda \simeq 4 \sigma_z$, i.e. $a_0 \simeq 4.5 Q [\text{nC}] / \sigma_\perp [\micron]$, for a round beam with $\sigma_\perp = \sigma_x = \sigma_y$~\cite{delgaudio.prab.2019}.

Beamstrahlung is viewed as undesirable because it reduces the centre-of-mass energy of the collision and because both the photons themselves, and the secondary electron-positron pairs they can create~\cite{chen.prl.1989}, are additional sources of background.
In future linear colliders, where the beam energy will exceed $100$~GeV, it can be minimized by the use of long, flat particle beams~\cite{ilc.2013,clic.2016}.
On the other hand, maximizing $\chi_e$ by the use of round beams could lead to a high-luminosity photon-photon collider~\cite{blankenbecler.prl.1988}, a high-brightness source of photons and secondary electron-positron pairs~\cite{delgaudio.prab.2019}, or facilitate the study of the fully nonperturbative regime of SFQED~\cite{yakimenko.prl.2019} (see \cref{sec:RadiativeCorrections}) {and might be more advantageous from the point of view of acceleration in the plasma based colliders \cite{roadmap.doe.2016,alegro.arxiv.2019,alegro.arxiv.2020}.}
Codes that simulate beam-beam interactions, accounting for the relevant strong-field processes of photon emission and pair creation (see \cref{sec:LCFA}), include \textsc{cain}~\cite{chen.nima.1995}, a successor to \textsc{abel}~\cite{yokoya.nima.1986}, and \textsc{guinea-pig++}~\cite{rimbault.2007}.

\subsubsection{Nuclear electric fields}

The Coulomb field of a heavy atomic nucleus can be strong
in the sense outlined in \cref{sec:KeyParameters} if the
number of protons $Z \gg 1$: for example, the
field strength at the radius of the $1s$ orbital
is given by $\InvariantF^2 \simeq (\alpha Z)^6$.
Indeed, there is a critical
value of $Z = Z_\text{cr}$ above which the lowest energy
bound state falls into the Dirac sea, i.e. the
negative-energy continuum, and spontaneous electron-positron
pair creation becomes probable~\cite{reinhardt.rpp.1977}.
This threshold is estimated to be $Z_\text{cr} =
1/\alpha \simeq 137$ by solution of the Dirac equation
for an electron in the Coulomb potential of a point-like
nucleus~\cite{darwin.prsa.1928,gordon.zp.1928};
it is increased to $Z_\text{cr} \simeq 173$ by finite-size
effects.
Such nuclei exist only transiently, at the timescale
of $10^{-23}$~s~\cite{ruffini.pr.2010}.
Nevertheless, $Z \gtrsim 80$ has proven sufficient for
observation of Delbr{\"u}ck scattering~\cite{milstein.pr.1994,schumacher.rpc.1999}
and photon splitting~\cite{akhmadaliev.prl.2002},
nonlinear QED processes that are triggered in
the polarized vacuum surrounding the nucleus.
Such processes could also be explored
using polarized $\gamma$ rays~\cite{koga.prl.2017}
or an intense optical laser~\cite{dipiazza.prl.2008,dipiazza.pra.2008}.

Even stronger electromagnetic fields are generated in
the collision of heavy nuclei that are moving with relativistic
velocity~\cite{gershtein.lnc.1973,baur.pr.2007,tuchin.ahep.2013} (see also \cite{popov.pan.2001} and references therein).
In a peripheral collision between two nuclei with atomic
number $Z$ and Lorentz factor $\gamma$, a characteristic electromagnetic field
strength of $E \simeq c B \simeq Z e \gamma / (\varepsilon_0 b^2)$ is
sustained over a duration of $\Delta t \simeq b / (\gamma c)$
in the lab frame
(peripheral indicates that $b$, the impact parameter, is greater
than the sum of the radii of the two nuclei).
The field invariants $\abs{\InvariantF}$ and $\chi_{e,\gamma}$
are much greater than unity for typical heavy-ion
collider parameters; 
however, the short duration means that $a_0 \simeq e E \Delta t / (m c^2)
\simeq Z \alpha < 1$, for a typical impact parameter
of a Compton length~\cite{baur.pr.2007}, and thus the interaction
occurs in the perturbative, rather than {tunneling}, regime (see \cref{sec:KeyParameters}).
The boosted Coulomb field of a nucleus can be treated
as a flux of \emph{quasireal} photons~\cite{baur.pr.2002},
which has allowed photon-photon scattering to be
observed for the first time~\cite{denterria.prl.2013,aaboud.nphys.2017}.
These photons can also create lepton pairs~\cite{bottcher.prd.1989,sengul.epjc.2016}.
Electromagnetic processes of this kind must be separated
from quantum chromodynamical (QCD) phenomena, such as
quark-gluon plasma formation and chiral magnetic effects~\cite{huang.rpp.2016}.

\subsubsection{Aligned crystals}
\label{sec:AlignedCrystals}

An experimental scenario in which strong electromagnetic fields
of nuclear origin extend over macroscopic volumes is afforded
by aligned crystals~\cite{baier1998} (see also the review by
\citet{uggerhoj.rmp.2005} and references therein).
Ultrarelativistic positrons travelling through
a crystal can be `channelled' along an axis of symmetry~\cite{lindhard.pl.1964,bogh.pl.1964},
where the screened electric fields of the nuclei sum coherently
to a macroscopic, continuous field of magnitude
$E \simeq 10^{13}~\text{V}\text{m}^{-1}$, which is typically
a few millimetres in size.
This corresponds to a quantum parameter $\chi_e \simeq \Energy_0 [100~\text{GeV}]$,
where $\Energy_0$ is the energy of the positron.
The equivalent strength parameter may be estimated
as $a_0 \simeq p_\perp / (mc)$, where $p_\perp \simeq \Energy_0 \theta / c$
and $\theta \ll 1$ is the angle between the trajectory and the
crystal axis.
Planar channelling occurs when $\theta < \theta_c =
\sqrt{2 U_0 / \Energy_0}$ (for positively charged particles).
Setting $\theta = \theta_c$ and taking $U_0 \simeq 50~\text{eV}$
as a typical value of the crystal potential~\cite{uggerhoj.rmp.2005},
we obtain $a_0 \lesssim 6 \Energy_0^{1/2} [100~\text{GeV}]$.
Lepton beams with the necessary 100-GeV energies, available at the
Super Proton Synchrotron at CERN, have been used to measure
photon emission~\cite{kirsebom.prl.2001,andersen.prd.2012} 
and electron-positron pair creation~\cite{belkacem.prl.1987} 
in the strong-field regime.
The effect of radiation reaction, the modification to the charge dynamics
caused by the emission of multiple hard photons,
has been measured recently in the radiation emitted
by channelled positrons~\cite{wistisen.ncomm.2018,wistisen.prr.2019}.
{In all these cases, the processes of interest must be distinguished from a background of single-nucleus effects, including bremsstrahlung~\cite{uggerhoj.rmp.2005}.}

\subsubsection{High-intensity lasers}
\label{sec:HighIntensityLasers}

Laser systems create strong fields by focusing ultrashort, energetic pulses of electromagnetic radiation, exploiting the high degree of spatial coherence to reach near diffraction-limited focal spot sizes.
The electric field strength at focus $E_0$ is related to the peak intensity $I_0$ by $E_0 = \sqrt{2 I_0 / (\epsilon_0 c)}$ (assuming linear polarization); the strength parameter follows as
    \begin{equation}
    a_0 \simeq
        0.85 I_0^{1/2} [10^{18}~\Wcmsqd] \lambda [\micron]
    \end{equation}
where $\lambda$ is the wavelength. In terms of the power $\Power$ per unit area $A$, $a_0 \simeq 270\, \Power^{1/2} [\text{PW}] \, \lambda [\micron] / A^{1/2} [\micron^2]$.

\begin{figure*}[tbp]
	\centering
	\includegraphics[width=1.0\textwidth]{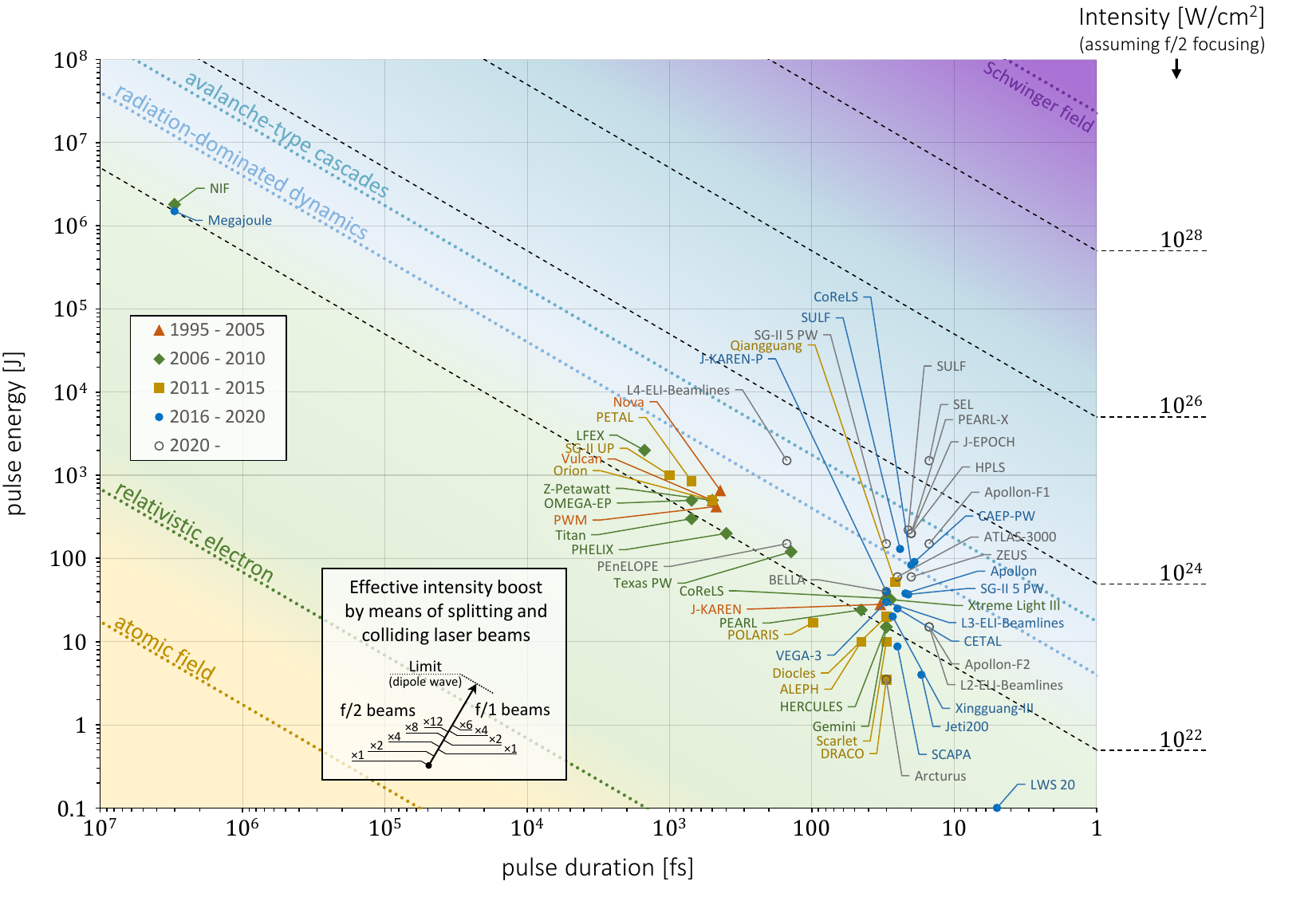}
    \caption{Laser facilities on the map of pulse energy and pulse duration. Data from \citet{danson.hplse.2019,nas.2018} and individual facilities {(see \cref{tbl:LaserData})}. 
    {For purposes of illustration, we show the equivalent peak intensity levels, assuming f/2 focusing of a linearly polarized pulse with Gaussian spatial and temporal profiles and a nominal wavelength of $\lambda = 1$~$\mu$m.}
    The physics in access is denoted by colours and characteristic intensity scales {(see Sec.~\ref{sec:PhysicalRegimes})}: $3.5 \times 10^{16}$~W/cm$^2$ (atomic field, corresponds to the field strength of 1 in atomic units (see \cite{krausz.rmp.2009})), $1.37 \times 10^{18}$~W/cm$^2$ (relativistic electron, corresponds to $a_0 = 1$), $8 \times 10^{22}$~W/cm$^2$ (radiation-dominated dynamics, corresponding to $L_{d} = \lambda$, assuming $\gamma = a_0$, see \cref{eq:InertiaLength}), $3.5 \times 10^{23}$~W/cm$^2$ (avalanche-type cascades, corresponding to $L_p = \lambda$, assuming $\gamma = a_0$, see \cref{eq:PairCreationLength}) and $4.65 \times 10^{29}$~W/cm$^2$ (corresponding to the Schwinger field strength). The insert shows the effective intensity boost that one would gain by means of splitting the laser power among the specified number of beams and colliding them so that the electric field is summed up coherently (see Appendix of \citet{gonoskov.prl.2014}). The boost is shown via the corresponding relative shift of location on the map for the cases of using f/2 (left side) and f/1 (right side) focusing, and the dotted line shows the ultimate boost given by the dipole wave {(see~\cref{sec:DipoleWaves})}.}
    \label{fig:LaserHR}
\end{figure*}

The peak powers necessary to reach $a_0 \gg 1$ are obtained by amplifying femtosecond-duration optical pulses to energies of tens or hundreds of joules.
This was made possible by the invention of `chirped pulse amplification' (CPA)~\cite{strickland.oc.1985} and the subsequent petawatt revolution has seen development of high-power laser facilities across the globe~\cite{danson.hplse.2019}.
Intensities as high as $10^{22-23}~\Wcmsqd$ have already been reported~\cite{yanovsky.oe.2008,sung.ol.2017,kiriyama.ol.2018,yoon.optica.2021}, while $10^{23-24}~\Wcmsqd$ is expected in upcoming laser facilities, such as Apollon~\cite{papadopoulos.hpl.2016}, the pillars of the Extreme Light Infrastructure~\cite{weber.mre.2017,gales.rpp.2018}, and the Shanghai Coherent Light Facility~\cite{shen.ppcf.2018}, to name but a few.
Current petawatt-scale laser facilities, and those soon to be commissioned, are shown in \cref{fig:LaserHR} in terms of their pulse energy, duration and date of commissioning.

Moreover, high-power laser systems are, or are planned to be, colocated with other experimental platforms, including conventional electron accelerators~\cite{yakimenko.prab.2019,abramowicz.arxiv.2019,meuren.arxiv.2020} and X-ray free electron lasers~\cite{shen.ppcf.2018}. It was the former that enabled the first demonstration of SFQED effects, in a landmark experiment that measured nonlinearities in Compton scattering and electron-positron pair creation at a laser intensity of $10^{18}~\Wcmsqd$~\cite{bula.prl.1996,burke.prl.1997,bamber.prd.1999}. However, recent developments in laser-driven wakefield acceleration~\cite{tajima.prl.1979} (see \cite{esarey.rmp.2009} for details) have made it possible to produce multi-GeV electron beams \cite{kim.prl.2013,wang.ncomm.2013,leemans.prl.2014,gonsalves.prl.2019} and therefore to perform `all-optical' laser-electron collision experiments~\cite{bulanov.nima.2011}. These have been realised in the dual-pulse laser systems Diocles~\cite{chen.prl.2013,yan.nphot.2017} and Gemini~\cite{sarri.prl.2014,cole.prx.2018,poder.prx.2018}. Many of the upcoming systems shown in \cref{fig:LaserHR} are \emph{multibeam} facilities, designed to deliver several laser pulses simultaneously to the same target chamber, which is, in most cases, a requirement for the facility to be able to conduct SFQED experiments~\cite{zhang.pop.2020}.

This facilitates a range of possible experimental geometries, which we discuss in \cref{sec:Geometries}.
Broadly, there are three characteristic geometries in which strong-field effects are {manifest}~\cite{bulanov.nima.2011,blackburn.rmpp.2020}: laser--electron-beam, laser-plasma, and laser-laser.
In the first, the electrons are accelerated to high energy before the collision and therefore the laser pulse acts only as the `target'; the chararacteristic quantum parameter scales as
    \begin{equation}
    \chi_e \simeq 0.18\, \Energy_0 [\text{GeV}] \, I_0^{1/2} [10^{21}~\Wcmsqd]
    \end{equation}
where $\Energy_0$ is the initial energy of the electron beam and $I_0$ the laser intensity.
{It is also possible to use the electron beam to generate high-energy photons, which then collide with the high-intensity laser: see \cref{sec:PhotonPhoton}.}
In the latter two geometries, the laser is simultaneously accelerator and target~\cite{bell.prl.2008}.
The requirement that the laser fields be strong enough to accelerate the electrons, from rest, to suitably high energies means that there is a different scaling of quantum parameter with laser intensity.
Assuming that the Lorentz factor $\gamma \simeq a_0$, which is a suitable estimate in both the laser-solid~\cite{wilks.prl.1992} and laser-laser cases~\cite{zeldovich.spu.1975}, we obtain
    \begin{equation}
    \chi_e \simeq 0.09\, I_0 [10^{23}~\Wcmsqd] \, \lambda [\micron]
    \end{equation}
for an interaction driven primarily by a laser, rather than an electron beam.
See detailed discussion in \cref{sec:Dynamics}.

\subsection{Physical regimes}
\label{sec:PhysicalRegimes}

    \begin{figure*}[tbp]
	\centering
	\includegraphics[width=0.8\textwidth]{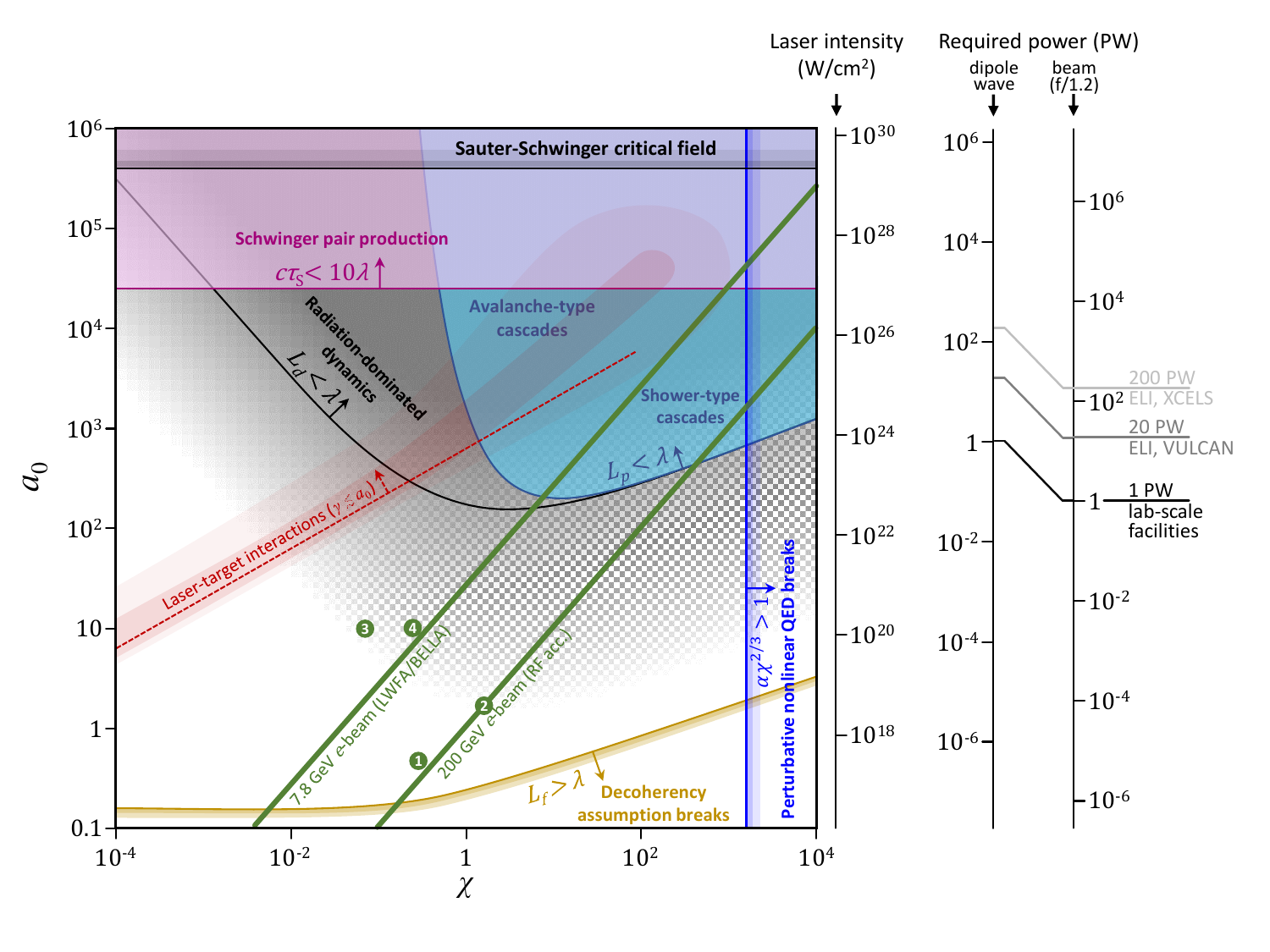}
	\caption{
	    Interaction regimes in the space of normalized field amplitude $a_0$
	    and quantum nonlinearity parameter $\chi$.
	    We identify the importance of different processes by comparing
	    key space scales with a characteristic laser wavelength
	    $\lambda = 1~\micron$:
	    the decreasing \emph{depletion length} $L_d$ [\cref{eq:InertiaLength}]
	    denotes the transition from weak to strong radiation losses
	    ($L_d > 10^3 \lambda$ in white, $L_d < \lambda$ in dark gray);
	    a \emph{quantization length} $L_q$ [\cref{eq:QuantizationLength}]
	    comparable to $\lambda$ indicates that discrete nature of radiation
	    emission is important ($L_q > \lambda$, chequered);
	    as the \emph{pair-creation length} $L_p$ [\cref{eq:PairCreationLength}]
	    becomes smaller than $\lambda$, electron-positron pair creation
	    by photons in strong fields becomes prolific ($L_p < \lambda$ in light blue);
	    the field amplitude necessary for the onset of vacuum pair creation
	    by the Schwinger mechanism is shown in violet,
	    assuming a non-zero $\InvariantF = a_0 [\hbar \omega_0 / (m c^2)]$.
	    Key limits on theoretical treatments are:
	    the field cannot be assumed to be constant if the \emph{formation length}
	    $L_f < \lambda$ [\cref{eq:FormationLength}] (yellow);
	    perturbation theory in SFQED is expected to break down
	    when $\alpha \chi^{2/3} \gtrsim 1$ (blue);
	    a general electromagnetic field cannot be approximated as crossed
	    if $\chi^2 < \InvariantF, \InvariantG$.
	    We also outline the prospects for laser experiments, in
	    interactions with a stationary plasma targets ($\gamma \lesssim a_0$, red line)
	    or ultrarelativistic electron beams (green lines).
	    Green points indicate experiments that have probed the strong-field
	    regime (in publication order): (1)~\citet{bula.prl.1996,burke.prl.1997},
	    (2)~\citet{wistisen.ncomm.2018},
	    (3)~\citet{cole.prx.2018}, (4)~\citet{poder.prx.2018}.
	    }
	\label{fig:Map}
    \end{figure*}

Having introduced what `strong' electromagnetic fields are,
and where they may be found,
we now consider the various phenomena that they induce,
and how these phenomena permit classification of different
physical regimes.
We focus in particular on high-intensity lasers,
as progress in their development has opened up new experimental
opportunities to reach regimes where radiation reaction and QED
effects play a significant role in particle dynamics
(see \cref{sec:Experiments}).
The parameter space shown in \cref{fig:Map} is that of the classical
and quantum nonlinearity parameters $a_0$ [\cref{eq:a0}] and
$\chi$ [\cref{eq:Chi,eq:ChiGamma}], which provide
Lorentz-invariant proxies for the field amplitude
and particle energy.
The importance of a given process is quantified by comparing
its characteristic length scale with the characteristic
length scale of the external field, which is the laser
wavelength $\lambda = 2 \pi c / \omega_0$ (for frequency $\omega_0$).
As most high-power lasers operate at near-infrared or
optical wavelengths (see \cref{fig:LaserHR}),
we set this to a nominal value of $\lambda = 1~\micron$.
{Scalings for the lengths introduced in this section are given in \cref{tbl:Scalings}.}

The nature of radiative energy losses can be characterized by two
length scales: one classical, the other quantum.
The \emph{depletion length} $L_d$ is the distance over
which an electron, accelerated by a strong electromagnetic field,
radiates away energy equal to its own kinetic energy:
    \begin{equation}
    L_d =
        \frac{\gamma m c^3}{\Power_\text{rad}}.
    \label{eq:InertiaLength}
    \end{equation}
Here $\Power_\text{rad}$ is the radiated power {(see \cref{p_rad})} for an electron accelerated by an electromagnetic field with normalized amplitude
$a_0$, i.e. {$\chi = \chi_e$}$= \gamma a_0 \hbar\omega_0 / (mc^2)$.
For the sake of simplicity, we neglect the self-consistent
reduction of $\Power_\text{rad}$ due to energy loss.
$L_d < \lambda$ marks the transition to the \emph{radiation-dominated
regime}~\cite{bulanov.ppr.2004,koga.pop.2005,hadad.prd.2010}.
This is equivalently expressed as $R_c = \alpha a_0 \chi > 1$
if $\chi \ll 1$, where $R_c$ is the classical radiation reaction
parameter~\cite{dipiazza.lmp.2008,harvey.prd.2011}.

The \emph{quantization length} gives the typical distance
between the emissions of individual photons, which may be
defined as the ratio of the mean energy of the photon
spectrum $\avg{\hbar \omega}$ to the radiated power:
    \begin{equation}
    L_q =
        \frac{\avg{\hbar \omega} c}{\Power_\text{rad}}.
    \label{eq:QuantizationLength}
    \end{equation}
If $L_q > \lambda$, the discrete nature of radiation emission
becomes dynamically important, leading to the phenomena of
stochastic broadening,
straggling and quenching (see \cref{sec:Nondeterminism}).
This is equivalently expressed as $R_q = \alpha a_0 < 1$
if $\chi \ll 1$~\cite{dipiazza.prl.2010}.
If $L_d \ll \lambda$ and $L_q \gtrsim \lambda$, recoils due to photon
emission are infrequent, but individually strong~\cite{gonoskov.pre.2015}.

The shading in \cref{fig:Map} characterizes radiation losses
according to these two length scales $L_d$ and $L_q$:
the tone shows their strength, from weak ($L_d > 10^3 \lambda$, white)
to strong ($L_d < \lambda$, dark gray), while the texture
indicates whether quantization effects are negligible ($L_q = 0$, uniform)
or important ($L_q > \lambda$, sharply chequered).

Finally, we can define the \emph{formation length} $L_f$ as the
typical distance over which a single photon is emitted.
Physically we may understand it as the distance travelled
before the electron is deflected sufficiently to separate
from its radiation cone~\cite{kirk.ppcf.2009}.
We have
    \begin{equation}
    L_f = \frac{c \avg{\theta^2}^{1/2}}{\omega_\text{eff}}
    \label{eq:FormationLength}
    \end{equation}
where $c/\omega_\text{eff} \simeq \lc \gamma^2 / \chi$
is the instantaneous radius of curvature of the electron trajectory
and $\avg{\theta^2}$ is the mean-square divergence of
the emitted radiation~\cite{baier1998} (see section \ref{sec:FormationLength} for detailed discussion).
{This definition is chosen for illustrative purposes: we discuss other definitions that are available in the literature, from a classical~\cite{Jackson1999} or quantum perspective~\cite{ritus.jslr.1985,baier1998}, in \cref{sec:FormationLength}}.
The condition that $L_f \ll \lambda$ is required for interference
effects to be negligible, i.e. that the emission of many photons
may be treated as sequential and incoherent~\cite{dipiazza.prl.2010}.
This is discussed in detail in \cref{sec:LCFA}.

The dominant physical process for photons propagating through a
strong electromagnetic field is electron-positron pair creation,
via the multiphoton Breit-Wheeler process,
{as it appears at first order in an expansion in $\alpha$}.
This is the generalization of the two-photon pair creation
mechanism~\cite{breit.pr.1934} to the highly multiphoton,
strong-field regime~\cite{reiss.jmp.1962,nikishov.jetp.1964}
(see \cref{sec:ComptonBreitWheeler}).
Its importance is characterized by the \emph{pair-creation length}
    \begin{equation}
    L_p = c P^{-1}_{b},
    \label{eq:PairCreationLength}
    \end{equation}
where $P_b$ is the pair-creation probability rate
{(see~\cref{gamma probability}) for a photon propagating in an electromagnetic field with normalized amplitude $a_0$. The photon energy $\hbar\omega$ is defined so that $\chi_\gamma = \chi = a_0 (\hbar \omega) (\hbar \omega_0) / (m c^2)^2$.}
If $L_p \gg \lambda$, the photon is likely to escape the region
of strong field.

The Schwinger mechanism for pair creation, in contrast to
the multiphoton Breit-Wheeler process, is driven by the field itself.
However, as stated in \cref{sec:KeyParameters}, a crossed field
such as a plane electromagnetic wave cannot produce pairs
via this mechanism, regardless of its amplitude $a_0$.
Non-crossed fields, created by the superposition of multiple,
non-copropagating, plane waves, can.
{In \cref{fig:Map} the uppermost horizontal black line, with shading on one side,} denotes the $a_0$ that is
equivalent to the critical field strength,
assuming that we have $N$ counterpropagating laser beams,
each with normalized amplitude $a_0/N$ and frequency $\omega_0$
equivalent to $1~\micron$,
i.e. $\InvariantF = a_0 [\hbar \omega_0 / (m c^2)] = 1$ from \cref{eq:InvariantF}.
It is worth noting the Schwinger pair-creation rate yields, on average, one pair within a spacetime volume of $10 \lambda^4 / c$, which is equivalent to multiple 10 cycle laser pulses focused into a $\lambda^3$ volume, for a significantly lower field strength $a_0 \simeq 2.5 \times 10^4$ \cite{narozhny.jetpl.2004,bulanov.jetp.2006};
this is shown as the magenta line in \cref{fig:Map}.

Also outlined in \cref{fig:Map} is the region where $\alpha \chi^{2/3} > 1$, where the perturbative treatment of nonlinear QED is predicted to break down~\cite{ritus.jslr.1985,narozhny.prd.1980}.
A theory that is applicable in this regime is yet to be developed
(see \cref{sec:OpenQuestions}).

The green and red lines show the expected capability of laser-based
experiments to reach large values of $\chi$, either by colliding
a beam of ultrarelativistic electrons (energy given next to the line) with a laser pulse of peak amplitude $a_0$ (green lines)
or by direct irradiation of a plasma target (red lines),
which accelerates electrons to energies of order
$\gamma \sim a_0$~\cite{wilks.prl.1992}.
The numbered green labels (1), (3) and (4) correspond to experiments
performed in the colliding beams geometry.
Label (2) indicates the equivalent $\chi$ and $a_0$ for a recent
crystals-based experiment (see \cref{sec:AlignedCrystals}).

The scales on the right hand side of \cref{fig:Map} indicate
the total laser power necessary to reach a given $a_0$,
either by focusing a single laser beam,
or by creating a dipole wave with multiple colliding beams \cite{bulanov.prl.2010m,gonoskov.prl.2013}.
The latter is the theoretical limit on the $a_0$
achievable at a given power~\cite{bassett.oa.1986,gonoskov.pra.2012}.
The scalings of the quantities discussed in this section are collected in \cref{tbl:Scalings} for convenience. Full expressions for the rates may be found in \cref{sec:ComptonBreitWheeler}.

    \begin{table}
    \caption{
        Scalings of the probability rates and their associated
        length scales with $a_0$, $\chi$ and $\gamma$.
        The latter refers either to a Lorentz factor for an electron,
        or to $\hbar \omega / (m c^2)$ for a photon, as appropriate.}
    \label{tbl:Scalings}
    \begin{ruledtabular}
    \begin{tabular}{lcc}
        Quantity
                & $\chi \ll 1$
                        & $\chi \gg 1$ \\
        \hline
        photon emission rate [$c/\lc$]
                & $1.44 \alpha \gamma^{-1} \chi$
                        & $1.46 \alpha \gamma^{-1} \chi^{2/3}$ \\
        radiation power [$mc^3/\lc$]
                & $2 \alpha \chi^2 / 3$
                        & $0.37 \alpha \chi^{2/3}$ \\
        mean photon energy [$mc^2$]
                & $0.43 \gamma \chi$
                        & $0.25 \gamma$ \\
        r.m.s. divergence angle
                & $1.1 \gamma^{-1}$
                        & $1.3 \gamma^{-1} \chi^{1/3}$ \\
        \hline
        pair creation rate [$c/\lc$]
                & $0.23 \alpha \gamma^{-1} \chi e^{-8/(3\chi)}$
                        & $0.38 \alpha \gamma^{-1} \chi^{2/3}$ \\
        \hline
        depletion length $L_d$ [$\lambda$]
                & $33 a_0^{-1} \chi^{-1}$
                        & $55 a_0^{-1} \chi^{1/3}$ \\
        mean free path $L_q$ [$\lambda$]
                & $48 a_0^{-1}$
                        & $15 a_0^{-1} \chi^{1/3}$ \\
        formation length $L_f$ [$\lambda$]
                & $0.18 a_0^{-1}$
                        & $0.21 a_0^{-1} \chi^{1/3}$ \\
        pair creation length $L_p$ [$\lambda$]
                & $95 a_0^{-1} e^{8/(3\chi)}$
                        & $57 a_0^{-1} \chi^{1/3}$
    \end{tabular}
    \end{ruledtabular}
    \end{table}

    \begin{figure}
    \includegraphics[width=0.8\linewidth]{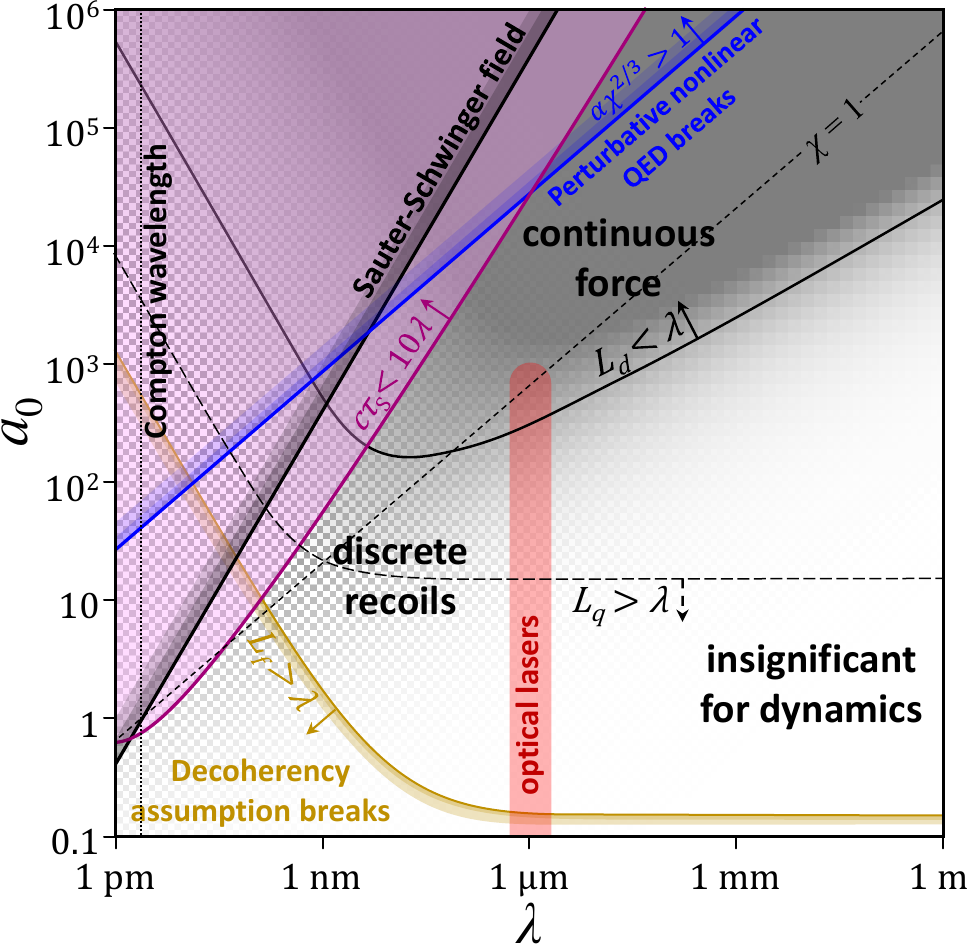}
    \caption{
        Interaction regimes as a function of the amplitude $a_0$ and the wavelength $\lambda$ of the electromagnetic wave driving an electron, using the same key as \cref{fig:Map}.
        The importance of RR effects is parametrized by $L_d$, \cref{eq:InertiaLength}, and $L_q$, \cref{eq:QuantizationLength}, which are both wavelength-dependent.
    }
    \label{fig:a0_wavelength}
    \end{figure}

\Cref{fig:Map} is plotted assuming a laser wavelength of 1~$\mu$m, which is most relevant for present-day laser technology.
{Allowing for the experimental challenges in doing so, changing the wavelength would open up new avenues for exploration of strong-field effects.}
It was noticed in \cite{esirkepov.pla.2015,bulanov.jpp.2017,zhang.pop.2020,jeong.oe.2020} that the wavelength plays an important role in determining what interaction regimes are accessible with given radiation power.
These are shown in \cref{fig:a0_wavelength} for varying wavelength $\lambda$ and amplitude $a_0$.
Here we assume that the electron energy is determined by the laser amplitude, $\gamma = a_0$, as would be approximately the case in a laser-laser or laser-plasma interaction.
While the wavelength is not Lorentz invariant, its value in the laboratory frame is nevertheless useful, as the laboratory frame coincides with the average rest frame in the case of a standing-wave interaction.
Remarkably, optical or near-infrared lasers provide opportunities to explore RR in both its quasicontinuous and discrete forms, whereas for larger (smaller) wavelengths, RR has a more continuous (discrete) form.

Ultrashort-wavelength lasers have the advantage of reaching the quantum regime, $\chi = 1$, more easily because of reduced radiation losses: note that the region $L_d < \lambda$ is reached at higher $a_0$ the more the wavelength is reduced.
While the technology able to produce very short wavelength lasers with required power is not yet developed, there are proposals to use relativistic flying mirrors~\cite{bulanov.ufn.2013} and high harmonic generation~\cite{bulanov.pop.1994, gordienko.prl.2005} to create high frequency EM pulses with ultrahigh intensity (see \cref{sec:FutureFacilities}).
On the other hand, the overlap between the radiation-dominated and quantum dominated regimes are reached almost simultaneously if $\lambda \simeq 1~\mu$m, at an $a_0$ achievable with 10 PW of laser power.
Thus, current optical or near-infrared lasers are well-placed to access a regime where relativistic plasma dynamics is strongly affected by both radiation and quantum effects.

\subsection{Overview and scope of this review}
\label{sec:Overview}

No review can be absolutely comprehensive, as the boundaries between fields of study are not themselves absolute.
Here we limit ourselves to a region of parameter space where the most significant interaction is between particles and EM fields. We do not consider the direct interaction of EM fields with themselves, which encompasses vacuum modification or decay, and point the reader to the recent reviews by \citet{mourou.rmp.2006, marklund.rmp.2006, dipiazza.rmp.2012, ehlotzky.rpp.2009, king.hplse.2016}. We focus on SFQED effects that can be experimentally studied at existing, or next-generation high-power laser facilities.
As such, the scope of this review article is the interaction of electrons, positrons, and photons with strong electromagnetic fields, emergent phenomena that arise from this interaction, possible routes towards experimental investigation, and applications thereof.

In order to facilitate this program, we organize the review article in the following way. We begin with an overview of fundamental physics concepts of SFQED. In \cref{sec:Theory}, {the essential elements of the} classical and quantum descriptions of charged particle interactions with strong fields are given. First, the development of classical radiation reaction is reviewed, {with particular emphasis on the Lorentz-Abraham-Dirac and Landau-Lifshitz equations}, and the limits of classical theory are given. Second, we discuss the multiphoton Compton and Breit-Wheeler processes, which are the basic building blocks of SFQED, the approximations customarily used to study these effects, radiative corrections, higher order processes, and phenomenological aspects.
 
In \cref{sec:Numerics} we discuss the state-of-the-art numerical methods used to simulate these interactions, including practical aspects of how the computational demand of multiscale simulations can be controlled.
 
{Charged particle dynamics, i.e. emergent phenomena that are consequences of the fundamental processes described in \cref{sec:Theory}, is covered in \cref{sec:Dynamics}.}
First, the typical interaction setups are discussed, which correspond to different types of charged particle dynamics in strong fields. We show how the increase of field strength and particle energy leads to non-trivial evolution of particle dynamics, from well-characterized classical trajectories to trapping in different configurations of EM fields, and from chaotic motion to strange attractors and limit cycles.
Here we encounter \emph{cascades}, one of the most fascinating processes of SFQED, where a few initial particles are transformed into many, high-energy photons and electron-positron pairs via iterated multiphoton Compton and Breit-Wheeler processes.
These cascades can be so intense that they can change the properties of the strong field itself.
 
Though such particle dynamics requires laser intensities that are not yet available, existing facilities have been able to conduct ground-breaking experiments on the topic. We review these in \cref{sec:Experiments}, beginning with those that were carried out at conventional accelerator facilities, where the laser was only used as a target for high-energy electron beams, to all-optical setups, where electrons were accelerated by laser-plasma interactions and then collided with laser pulses.
We also discuss possible future directions, including the generation and diagnosis of ultrastrong fields.

The study of particle dynamics in strong EM fields opens up possibilities for generation of various particle and radiation sources, which are discussed in section \ref{sec:Applications}.
 
In the course of this review, we repeatedly stress that there are many unanswered questions and yet-to-be studied phenomena in SFQED. We summarize those in \cref{sec:OpenQuestions}, along with a description of possible future experimental programs.
We conclude in \cref{sec:Conclusions}.
At the end of the paper we list commonly used symbols with their definitions.

\bib

%% file: theory.tex
\section{Theory of strong electromagnetic fields}
\label{sec:Theory}

\subsection{Classical radiation reaction}
\label{sec:ClassicalRR}

\subsubsection{The problem of radiation reaction}
\label{sec:TheProblemOfRR}

We begin with a brief historical perspective on the problem of accounting for the recoil from the electromagnetic radiation emitted by an accelerated charge, a puzzle that has engaged the community for more than a century. More in-depth discussions can be found in \citet{caldirola.ncim.1979, klepikov.spu.1985, weinberg.1995, Rohrlich2007}.
In a picture of electrodynamics where matter is treated as a continuous fluid, recoil due to radiation emission is self-consistently embedded in the Vlasov-Maxwell system~\cite{vlasov.1938}, which implies the existence of a continuity equation for both energy and momentum.
However, the discovery of the electron by Thompson in 1897 raised conceptual difficulties, because the electromagnetic field energy of a point charge diverges~\cite{pryce.prsa.1938}.

An early attempt to resolve these difficulties was made by \citet{lorentz.1892,lorentz.1904}, who modelled the electron as a sphere of small, but finite, radius $r$, such that the entirety of the electron mass could be attributed to its electromagnetic energy $e^2/\left(6 \pi \epsilon_0 c^2 r\right)$.
Under acceleration, an imbalance between the electromagnetic forces exerted between different parts of the sphere leads to an additional force $\propto \frac{\rmd^3 \vec{x}}{\rmd t^3}$, where $\vec{x}$ is the position of the charge centre.
The magnitude of this force is consistent with the momentum loss due to radiation emission, as described by the non-relativistic Larmor formula~\cite{larmor.1897}.
A relativistic force, using the relativistic formulation of the radiated power~\cite{heaviside.nature.1902} was presented in \citet{abraham.ap.1902}.

Left unresolved, however, was the question of what held the charge together.
A perfectly rigid sphere would not be consistent with relativity, whereas deformation of the sphere would indicate internal degrees of freedom of unknown origin.
This motivated Dirac to revisit the problem.
He derived a relativistic equation of motion by enforcing energy-momentum conservation of the electromagnetic flux through a small surface surrounding a charge of yet unknown structure~\cite{dirac.prsa.1938}.
Notably, no higher order terms (in the radius of the charge $r$) need to be omitted.
The equation of motion, referred to as the Lorentz-Abraham-Dirac (LAD) equation, is given by~\cite{dirac.prsa.1938}:
	\begin{align}
	\frac{\rmd p^\mu}{\rmd \tau}
	    &= -\frac{e F^{\mu\nu} p_\nu}{m c} + g^\mu,
	\\
	g^\mu_\text{LAD} &= \tau_\text{rad} 
		\!\left[
			\frac{\rmd^2 p^\mu}{\rmd \tau^2}
			+ \frac{p^\mu}{(mc)^2} \frac{\rmd p^\nu}{\rmd \tau} \frac{\rmd p_\nu}{\rmd \tau}
		\right],
	\label{eq:LAD}
	\end{align}
where $\tau_\text{rad} = 2 \alpha \lambdabar / (3 c) = 2 r_e / (3 c)$ and $\tau$ is proper time.
{The second term, $g^\mu$, is the additional four-force that arises from radiation reaction.}
An interesting consequence is that the observed mass of the electron, i.e. the scaling factor $m$ between velocity $\frac{\rmd x^\mu}{\rmd \tau}$ and momentum $p^\mu = m \frac{\rmd x^\mu}{\rmd \tau}$, is the difference between a divergent quantity, electromagnetic in origin, and a `bare' mass; this predates the renormalization later applied in quantum field theory.
The LAD model is not completely self-consistent, which is manifested by pathological solutions that describe self-accelerating (runaway) charges: see \citet{spohn.epl.2000, burton.cp.2014}.
These solutions can be eliminated by imposing a boundary condition of zero acceleration in the distant future~\cite{dirac.prsa.1938}.
However, the pathological behaviour can be restarted and then terminated by hard collisions~\cite{eliezer.mpcps.1943}.
Another approach is to reduce the order of equation, which we discuss in \cref{sec:FromLADtoLL}.
Other classical theories of radiation reaction have been proposed by \cite{eliezer.prsa.1948, mo.prd.1971, bonnor.prsa.1974, herrera.prd.1977, caldirola.ncim.1979, ford.pla.1991, sokolov.jetp.2009} (see the review by \cite{burton.cp.2014}).

Distinguishing between these models is perhaps best accomplished by appealing to a more fundamental theory of electrodynamics.
The development of quantum theory led to a paradigm shift in how the `localization' of the charge of an electron could be understood.
Furthermore, in a scattering matrix approach, conservation of momentum applies directly to the in and out states, which can contain arbitrary numbers of additional particles, including photons.
The quantum formalism established a broader framework that now accompanies discussions of radiation reaction and its implications in strong-field environments~\cite{ruffini.pr.2010,dipiazza.rmp.2012}.
We discuss the limits of the classical theory in \cref{sec:ClassicalLimits} and radiation reaction, as the classical limit of a QED result, in \cref{sec:QuantumRR}.

\subsubsection{From Lorentz-Abraham-Dirac to Landau-Lifshitz}
\label{sec:FromLADtoLL}

Pathological behaviour in solutions of the Lorentz-Abraham-Dirac equation, \cref{eq:LAD}, is driven by the second derivative of momentum.
It should be noted that these pathological solutions arise because of the specific form of the second derivative, not its mere presence~\cite{burton.cp.2014}.
The `critical surface', which is the manifold of \emph{physical} solutions to the LAD equation, is repulsive: small deviations from it grow and the solutions run away to infinity~\cite{spohn.epl.2000}.
These runaways can be prevented by requiring that the acceleration vanish in the asymptotic future~\cite{dirac.prsa.1938}, which provides the necessary third boundary condition after the initial position $x^\mu(0)$ and momentum $p^\mu(0)$; however, there follows the question of the solution's uniqueness~\cite{spohn.epl.2000}.
{Moreover, eliminating the runaways in this fashion leads to the problem of `pre-acceleration', where the charge moves in advance of the applied force~\cite{klepikov.spu.1985,Rohrlich2007}.}

Landau and Lifshitz proposed to eliminate the second derivative by a reduction of order process, expanding \cref{eq:LAD} in the small parameter $\tau_\text{rad}$~\cite{landau.lifshitz}.
This requires that radiation reaction is weak, as compared to the Lorentz force, in the instantaneous rest frame of the charge~\cite{eliezer.prsa.1948}.
At lowest order, this has the effect of replacing $\frac{\rmd p_\mu}{\rmd\tau} \to e F_{\mu\nu} p^\nu / (mc)$ in \cref{eq:LAD}, yielding the radiation-reaction force in the Landau-Lifshitz {(LL)} prescription:
	\begin{multline}
	g^\mu_\text{LL} =
	    \frac{\tau_\text{rad}}{(\lc/c)^2}
		\!\left[
			-\frac{(\partial_\sigma F^{\mu\nu}) p_\nu p^\sigma}{m c \Ecrit / \lc}
			+ \frac{F^{\mu\nu} F_{\nu\sigma} p^\sigma}{\Ecrit^2}
			- \chi_e^2 p^\mu
		\right].
	\label{eq:LandauLifshitz}
	\end{multline}
Here we anticipate \cref{sec:ClassicalLimits} by expressing the result in terms of quantum parameters such as $\Ecrit$ and $\chi_e$; nevertheless, all factors of $\hbar$ cancel as required.
We see that \cref{eq:LandauLifshitz} is first order in the momentum and therefore that only two boundary conditions, the initial position $x^\mu(0)$ and momentum $p^\mu(0)$, are required to solve it.
{\citet{spohn.epl.2000} argues that solutions to the Landau-Lifshitz equation lie on the same manifold as physical solutions of the LAD equation.}
Moreover, for given $x^\mu(0)$ and $p^\mu(0)$, there is exactly one solution on the critical surface, which also satisfies the asymptotic boundary condition $\lim_{\tau \to \infty} \frac{\rmd p^\mu}{\rmd \tau} \to 0$~\cite{spohn.epl.2000}.
{Investigations of higher order approximations to the LAD equation, in the non-relativistic~\cite{zhang.ptep.2013} and relativistic cases~\cite{ekman.prd.2021}, show that the effect of RR is to drive the particle towards the regime where LL is valid.}
More details concerning the relation between the LAD and Landau-Lifshitz equations may be found in the reviews by \citet{dipiazza.rmp.2012} (section VI) and \citet{burton.cp.2014}.

Implementations of classical radiation reaction in plasma simulation codes have focused almost exclusively on the Landau-Lifshitz equation, which we discuss in \cref{sec:Numerics}.
It is possible to solve Landau-Lifshitz equation analytically for the trajectory of a charge in a plane EM wave~\cite{heintzmann.zpa.1972,dipiazza.lmp.2008,hadad.prd.2010,harvey.prd.2011}, as well as to obtain analytical results for aspects of its radiation spectrum~\cite{dipiazza.plb.2018}.
The lightfront momentum $p^- \equiv \kappa.p/\kappa^0$ (here $\kappa^\mu$ is the wavevector) is not conserved under radiation reaction~\cite{harvey.prd.2011}.
(This is true in general, but \citet{harvey.prd.2011} work only to first order in $\tau_\text{rad}$, reflecting the fact the Landau-Lifshitz equation is an expansion in that parameter.)
Thus a charge can gain or lose energy in a plane EM wave, in violation of the Lawson-Woodward theorem~\cite{lawson.ieee.1979}.
Note that continuing the expansion of \cref{eq:LAD} to all orders in $\tau_\text{rad}$, and resuming the resulting divergent series, recovers the problematic solutions of the LAD equation~\cite{zhang.ptep.2013}.

Differences between the predictions of the LAD and Landau-Lifshitz equations are small if the external electromagnetic field's wavelength $\lambda$ and magnitude $E_0$ satisfy $\lambda \gg r_e$ and $E_0 \ll \Ecrit^\text{clas} = \Ecrit / \alpha$ in the particle's rest frame~\cite{landau.lifshitz}.
Note that the restriction on wavelength is a restriction on the timescale of variation of the field.
If the Landau-Lifshitz equation is used in a situation where there are abrupt jumps in the electromagnetic field strength, errors in energy and momentum conservation follow~\cite{baylis.pla.2002}: see \citet{shi.ap.2019} for discussion of signatures in a proposed experiment.

    \begin{figure}
    
    \includegraphics[width=\linewidth]{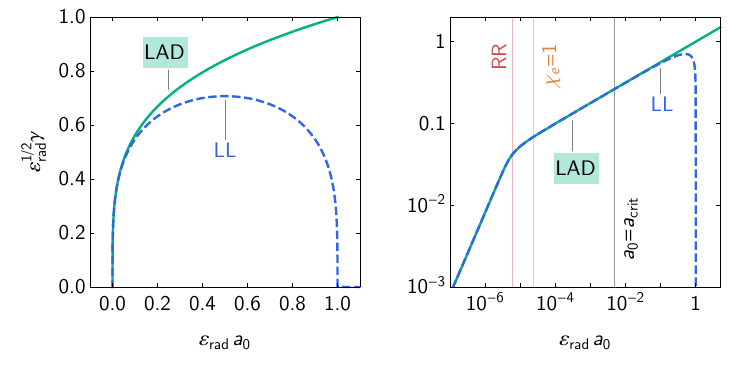}
    \caption{
        The Lorentz factor $\gamma$ of an electron in a rotating electric field of normalized amplitude $a_0$, as predicted by the LAD and Landau-Lifshitz equations, \cref{eq:LAD,eq:LandauLifshitz} respectively: linear (left) and log-scaled (right).
        Vertical lines indicate, from left to right: (red) onset of classical radiation reaction effects, \cref{eq:CRRThreshold}; (orange) a quantum parameter of unity, \cref{eq:QRRThreshold}; (black) the QED-critical field strength, \cref{eq:CriticalField}.
        Here $\erad = 1.47 \times 10^{-8}$, equivalent to a wavelength of $0.8~\micron$.
        Adapted from \citet{bulanov.pre.2011}.}
    \label{fig:LADvsLL}
    \end{figure}

Furthermore, the Landau-Lifshitz and Ford-O'Connell equations~\cite{ford.pla.1991} have been compared for the collision of an ultrarelativistic electron and a pulsed, plane EM wave in \citet{kravets.pre.2013}, where it is shown that energy losses due to radiation reaction prevent the electron entering the regime where the two equations predict significantly different behaviour.

Radiative energy losses are why the threshold for quantum effects is reached long before the breakdown of the Landau-Lifshitz equation.
Consider the steady-state energy of an electron, orbiting in a rotating electric field that has amplitude $a_0$ and angular frequency $\omega_0$ (a model for fields at the magnetic node of a circularly polarized standing wave).
Differences between the LAD and Landau-Lifshitz equations become visible at $\erad a_0 \simeq 0.1$,
where $\erad = (2\alpha/3) [\hbar \omega_0 / (mc^2)]$:
see \cref{fig:LADvsLL} which is adapted from \citet{bulanov.pre.2011}.
However, the quantum parameter $\chi_e$ is already unity when $\erad a_0 = (2\alpha/3)^2 \ll 0.1$ and a QED-critical field strength is reached at $\erad a_0 = 2\alpha/3 < 0.1$.
Indeed, considering the interaction of an electron with a plane EM wave (normalized amplitude $a_0$) reveals that the terms neglected in the Landau-Lifshitz equation are of order $\alpha \chi_e$ or $\alpha \chi_e / a_0$, which are sub-leading with respect to quantum corrections~\cite{dipiazza.plb.2018}.
This makes it essential to consider radiation reaction from a quantum perspective.
We conclude that the LAD and LL equations are indistinguishable in the parameter space where a classical description of RR is valid.

\subsubsection{Limits of the classical theory}
\label{sec:ClassicalLimits}

The failure of the classical theory of radiation reaction is evident in the fact that the characteristic timescale, $\tau_\text{rad}$, appearing in the LAD equation is smaller than the Compton time $\lc/c$ by a factor of the fine-structure constant $\alpha$~\cite{burton.cp.2014}.
Thus insight from quantum electrodynamics plays an essential role in investigation of the classical theory~\cite{dipiazza.rmp.2012}.
We discuss the QED approach to particle dynamics and radiation in the following section.
First we distinguish two different, possible, quantum effects from the classical perspective.

For an ultrarelativistic particle (Lorentz factor $\gamma \gg 1$) in a magnetic field (strength $B < \Bcrit$), there are two types of quantum effects~\cite{baier.jetp.1968}.
The first is associated with the noncommutation of the particle's dynamical variables and is of order $\hbar \omega_0 / (\gamma m c^2) \simeq B / (\gamma^2 \Bcrit) \ll 1$, where $\omega_0$ is the cyclotron frequency:
thus the particle motion becomes more `classical' as its energy $\gamma$ increases.
The second is associated with recoil and is of order $\hbar \omega_c / (\gamma m c^2) \simeq \gamma B / \Bcrit \gtrsim 1$, where {$\omega_c \propto \gamma^3 \omega_0$} is the typical frequency of the emitted radiation~\cite{schwinger.pr.1949}.
Thus in many scenarios of interest, one need only take into account the commutators of the particle's dynamical variables and the field of the emitted photon~\cite{baier.jetp.1968}.
We may therefore discuss quantum effects principally in terms of their effect on the spectrum of radiation emitted by an accelerating charge.

Consider the electric and magnetic fields arising from a charge in arbitrary motion, its trajectory defined by a position $\vec{r}(t)$ and velocity $\vec{\beta}(t)$, which were derived by \citet{lienard.ee.1898,wiechert.an.1900}.
Fourier analysis yields the energy radiated per unit photon energy $\hbar\omega$ in the far-field~\cite{Jackson1999}:
	\begin{equation}
	\frac{\rmd \Energy}{\rmd \hbar\omega} =
		\frac{\alpha \omega^2}{4 \pi^2}
		\int \rmd \Omega
		\left|
		    \int
		    \rmd t
		    \,
			\vec{n} \times (\vec{n} \times \vec{\beta})
			e^{i \omega (t - \vec{n}\cdot\vec{r}/c)}
		\right|^2,
	\label{eq:SpectralIntensity}
	\end{equation}
where $\vec{n}$ is the observation direction and the integral is taken over all time $t$ and solid angle $\Omega$.
The energy radiated, as predicted by \cref{eq:SpectralIntensity}, is consistent with the energy loss predicted by \cref{eq:LandauLifshitz}~\cite{thomas.prx.2012,schlegel.njp.2012,martins.ppcf.2016}.
Simplification is possible if the interference between emission from different parts of the trajectory is negligible~\cite{esarey.pre.1993}.
Then the energy spectrum may be expressed as the integral of the instantaneous spectral emission rate $\frac{\rmd W}{\rmd \hbar\omega}$ over the particle trajectory, assuming that, at high $\gamma$, the radiation is emitted predominantly in direction parallel to the electron's instantaneous velocity~\cite{SokolovTernov,baier1998}:
    \begin{align}
    \frac{\rmd \Energy}{\rmd \hbar\omega} &\simeq
        \int\! \hbar\omega \frac{\rmd W}{\rmd \hbar\omega} \,\rmd t
    \label{eq:IncoherentSpectrum}
    \\
    \frac{\rmd W}{\rmd \hbar\omega} &=
        \frac{\alpha}{\sqrt{3}\pi\hbar\gamma^2}
        \int_\xi^\infty \!K_{5/3}(y) \,\rmd y,
	&
	\xi &= \frac{2 \hbar\omega}{3 \chi_e \gamma m c^2} = {\frac{\omega}{\omega_c}.}
	\label{eq:ClassicalEmissionRate}
    \end{align}
where $K_\nu$ is a modified Bessel function of order $\nu$.
The $\xi$ parameter is the ratio of $\omega$ to a critical frequency $\omega_c$, which we have written in terms of $\chi_e$ to emphasize that it depends on the charge's instantaneous proper acceleration (see \cref{sec:KeyParameters}).
With $\chi_e = \gamma B / \Bcrit$, for example, we obtain $\omega_c = 3 \gamma^3 \omega_0 / 2$, where the cyclotron frequency $\omega_0 = e B / (\gamma m)$.
The radiation spectrum from an ensemble of electrons can be obtained by summing \cref{eq:IncoherentSpectrum} over the ensemble~\cite{reville.apj.2010,wallin.pop.2015}.

	\begin{figure}
	\centering
	\includegraphics[width=\linewidth]{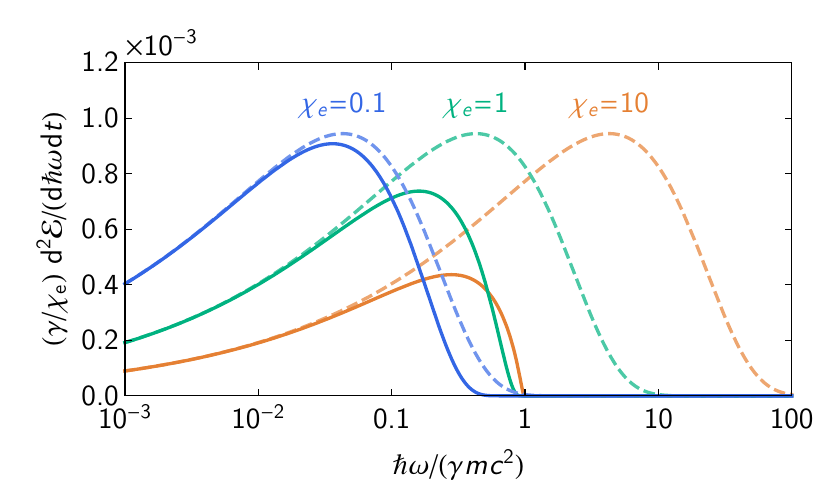}
	\caption{
		Radiation at unphysically large frequencies is predicted by classical theory if $\chi_e \gtrsim 1$:
		classical power spectra (dashed) and quantum-corrected power spectra (solid) at $\chi_e = 0.1$ (blue), $1$~(green) and $10$~(orange), using \cref{eq:ClassicalEmissionRate} and \cref{eq:RecoilCorrection} respectively.}
	\label{fig:QuantumCorrections}
	\end{figure}

\Cref{eq:ClassicalEmissionRate} predicts that half the power radiated is radiated to frequencies that satisfy $\xi > 1$ (the spectrum peaks at $\xi \simeq 0.29$).
However, this means that, for $\chi_e \gtrsim 1$, the majority of the radiation is carried by photons with more energy, individually, than the electron itself: see \cref{fig:QuantumCorrections}.
Preventing this makes quantum corrections, i.e. introducing $\hbar$, essential.
First-order quantum effects are manifest by substituting $\omega \to \omega [1 + \hbar\omega/(\gamma m c^2)]$ in the definition of $\xi$~\cite{schwinger.pnas.1954}, which reduces the total radiated power by the factor $1 - 55\sqrt{3}\chi_e/16$~\cite{schwinger.pnas.1954,sokolov.dan.1953}.
This is a pure recoil effect: spin effects first appear at second order in $\hbar$.
We discuss those in detail in \cref{sec:Spin}.
To guarantee that the spectrum is cut off at $\hbar\omega = \gamma m c^2$, it suffices to make the heuristic substitution $\xi \to \xi_q$~\cite{lindhard.pra.1991}
{
    \begin{align}
    \xi_q &= \frac{2 f_c}{3 \chi_e (1 - f_c)},
    &
    f_c &= \frac{\hbar\omega}{\gamma m c^2},
    \label{eq:RecoilCorrection}
    \end{align}
where $f_c$ is the fraction of the electron energy transferred to the photon.}
\Cref{eq:ClassicalEmissionRate} with \cref{eq:RecoilCorrection} is exactly the radiation spectrum of a spin-zero electron in QED~\cite{baier.jetp.1968}.
The classical result and Schwinger's correction are reproduced at zeroth and first order in $\hbar$.

\Cref{fig:QuantumCorrections} shows that introducing this cutoff significantly reduces the radiated power for $\chi_e \gtrsim 1$; momentum conservation requires that the magnitude of the radiation reaction force is reduced in tandem.
As such, if $\chi_e$ is not too large, the Landau-Lifshitz equation can be corrected phenomenologically by scaling $g^\mu_\text{LL}$ by the same factor by which the radiation power is reduced (see \cref{sec:QuantumRR}).
However, if the emission of even a single photon can significantly affect the electron dynamics, it is clear that radiation reaction in the quantum regime is fundamentally of a different character to its classical counterpart.
Moreover, as the photons emitted by electrons with $\chi_e \gtrsim 1$ can satisfy $\chi_\gamma \gtrsim 1$, consideration must be given to the possibility of electron-positron pair creation~\cite{reiss.jmp.1962,nikishov.jetp.1964}, a process without a classical analogue.
Thus we must turn to quantum electrodynamics, an inherently many-particle theory.

\subsection{Strong-field quantum electrodynamics}
\label{sec:StrongFieldQED}

\subsubsection{Overview}
\label{sec:SFQEDintro}

\begin{figure}
    \includegraphics[width=0.9\linewidth]{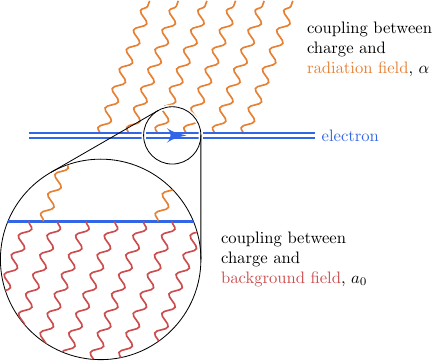}
    \caption{
        Nonperturbativity in strong-field quantum electrodynamics arises in two ways: from the coupling between the charge and the background (external) field, shown in red, and the coupling between the charge and the radiation (quantized) field, shown in orange.
        For $a_0 > 1$, the former interaction must be taken into account exactly, i.e. to all orders in $a_0$, as indicated by the double fermion lines in a diagrammatic representation.
        For fields with sufficient magnitude or duration, higher order contributions to the coupling with the radiation field can dominate lower order terms.}
    \label{fig:Nonperturbativity}
    \end{figure}

`Strong-field' quantum electrodynamics has the distinct property that the electromagnetic field, with which charged particles interact, is split into two: a fixed, classical background (an external field) and a fluctuating, quantized component (see \cref{fig:Nonperturbativity}). The interaction with the former is treated \emph{exactly} in the Furry picture of QED~\cite{furry.pr.1951}, which permits the interaction to be arbitrarily strong; the coupling to the quantized, `radiation' field is treated perturbatively. For an electromagnetic wave, where the coupling to the background field is controlled by the parameter $a_0$, this means that theoretical calculations are not restricted to the perturbative regime $a_0 \ll 1$. This is essential because the characteristic number of photons $n$ absorbed when an electron  emits a single, high-energy photon, $n \sim a_0^3$, becomes very large if $a_0 > 1$ (see \cref{sec:KeyParameters}). SFQED is therefore nonperturbative in the sense that results are correct to all orders in the expansion parameter $a_0$. The exact treatment of the background field means that the interaction occurs with the entire electromagnetic field, rather than with `individual' photons thereof. As such, predictions are sensitive to the spatiotemporal structure of the field, e.g. pulse shape and duration: this was first studied within SFQED for photon emission by \citet{boca.pra.2009,mackenroth.prl.2010,seipt.pra.2011} and for pair creation by~\citet{heinzl.plb.2010}.

\begin{figure}
    \includegraphics[width=0.7\linewidth]{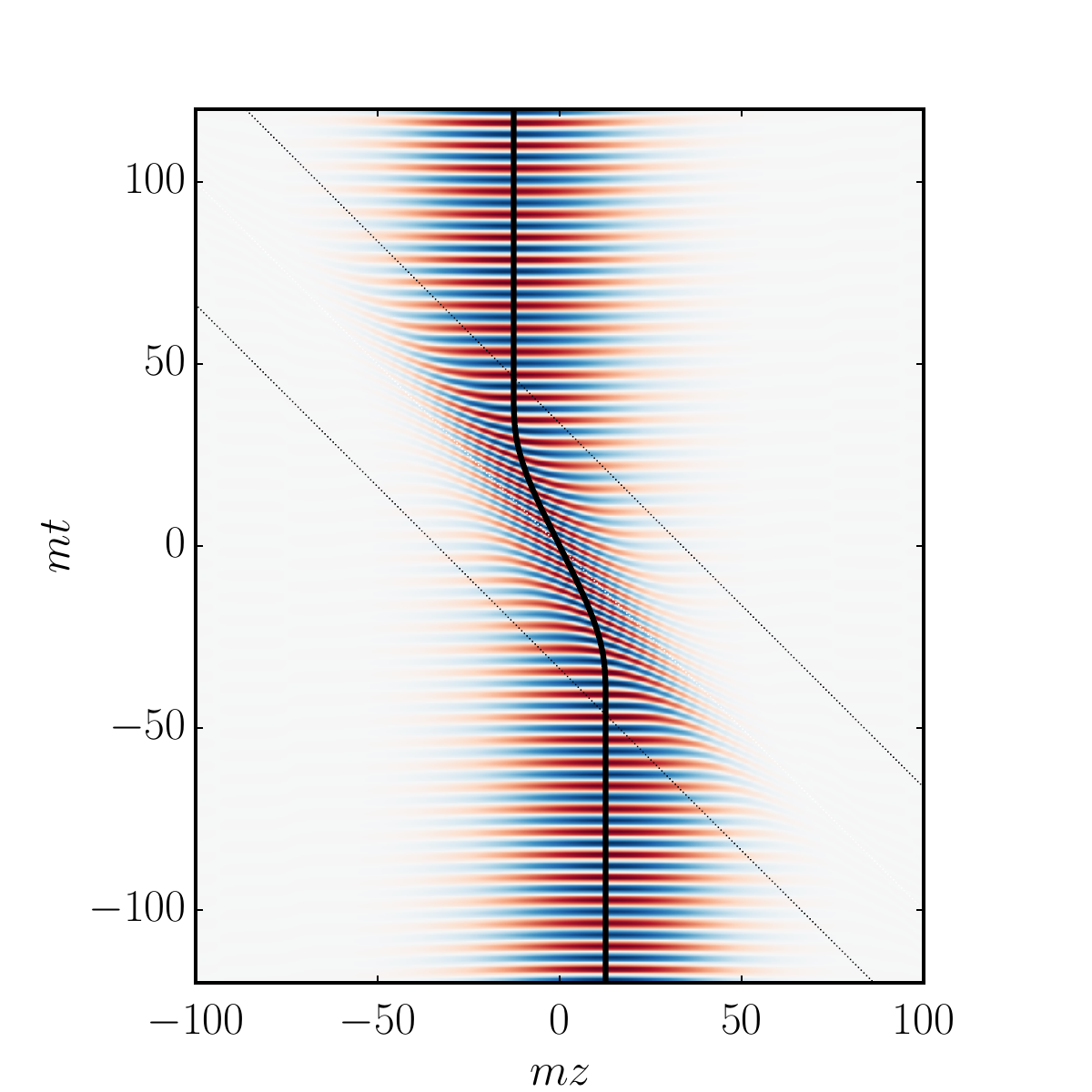}
    \caption{
        Evolution of a Volkov wavepacket (in time $t$ and coordinate $z$) in a laser pulse with normalized amplitude $a_0 = 2$ and duration $\Delta\phi = 20$ {(color scale indicates the wave function).}
        The wavepacket begins with a Gaussian distribution of lightfront momentum (mean $mc$ and standard deviation $0.05 m c$) and zero transverse momentum.
        The centroid closely follows the equivalent classical trajectory, which is indicated by a solid black line.
        The laser pulse propagates between the two dashed lines.
        Reproduced from \citet{seipt.2017}.}
    \label{fig:VolkovWavepacket}
    \end{figure}

The dynamics of free electrons/positrons in QED is described via the Dirac equation, which takes the form $[\gamma^\mu(p_\mu-eA_\mu)-m c]\psi=0$ in the presence of an external EM field with potential $A_\mu$.
If this EM field is a plane wave with wavevector $\kappa_\mu$, where $\kappa^2=0$, then the Dirac equation can be solved analytically. We note that the EM field of a plane wave depends on the coordinates only through phase $\varphi= (\kappa  x)$, so that the 4-potential is also a function of $\varphi$ only, $A_\mu=A_\mu(\varphi)$.
The solution of the Dirac equation in such a background is called a \emph{Volkov state}~\cite{volkov.zp.1935}:
    \begin{align}
    \begin{split}
    \psi_{pr}(x) &=
        \left[ 1 + \frac{e \slashed{\kappa} \slashed{A}(\varphi)}{2 (kp)} \right]
        \exp\!\left[ -\frac{i S(\varphi)}{\hbar} \right]
        u_{pr},
    \\
    S(\varphi) &=
        p x + \int_{-\infty}^\varphi \! \rmd\varphi^\prime \, \frac{2 e (pA)(\varphi') - e^2 A^2(\varphi')}{2 (k p)},
    \end{split}
    \label{eq:Volkov}
    \end{align}
where $u_{pr}$ is the free spinor, normalized to $\overline{u}_{pr}u_{pr^\prime}=2m\delta_{rr^\prime}$,
and the slashes indicate a contraction with the Dirac matrices $\gamma^\mu$, i.e. $\slashed{X} = \gamma^\mu X_\mu$.
Note that the argument of the exponent in \cref{eq:Volkov} coincides with the classical action $S$ for a particle with asymptotic momentum $p$ moving in the EM field of a plane wave, multiplied by the imaginary unit.
These states have a phase-dependent momentum that corresponds closely to the classical kinetic momentum, as can be seen in \cref{fig:VolkovWavepacket}, where the motion of the centroid of a Volkov wavepacket is compared to the classical trajectory~\cite{seipt.2017}.

In the Feynman rules of SFQED, Volkov states are used to describe external electrons: see \citet{mitter.1975} and the tutorial by \citet{seipt.2017}.
This is indicated by double fermion lines in a Feynman diagram (see \cref{fig:Nonperturbativity}).
The scattering matrix elements for processes of interest are then constructed from the standard diagrammatic representation.
In \cref{sec:ComptonBreitWheeler}, we illustrate these for the tree-level, i.e. lowest order, processes shown in \cref{fig:ComptonBreitWheeler}: multiphoton Compton scattering and Breit-Wheeler electron-positron pair creation. It is not always guaranteed that the tree-level processes dominate: nonperturbativity can also arise in the coupling to the radiation field, if the field is sufficiently strong (e.g. $\chi_e > 1$) or long in duration. In these cases, the usual hierarchy, that the probability to emit $N$ photons is larger than the probability to emit $N+1$, can be reversed.
The importance of high-order processes and the link to radiation reaction are discussed in \cref{sec:HigherOrderProcesses} and \cref{sec:QuantumRR} respectively.
{The Volkov solution assumes that the EM wave propagates in vacuum, but it is possible to account for the presence of a background plasma by similar means~\cite{raicher.pra.2013,mackenroth.pre.2019}.}

\subsubsection{Multiphoton Compton and Breit-Wheeler processes}
\label{sec:ComptonBreitWheeler}

	\begin{figure}
	\centering
	\includegraphics[width=0.8\linewidth]{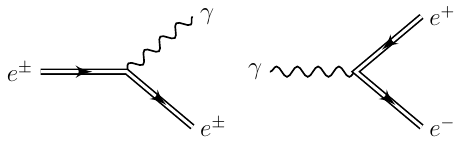}
	\caption{
		The Feynman diagrams of the Compton ($e\rightarrow e\gamma$)
		and Breit-Wheeler ($\gamma\rightarrow ee$) processes.
		Double fermion lines indicate that the process occurs in
		an external field.}
	\label{fig:ComptonBreitWheeler}
	\end{figure}

In order to calculate the probability of the multiphoton Compton and Breit-Wheeler processes, which are $e(p)\rightarrow e(p^\prime)\gamma(k^\prime)$ and $\gamma(k)\rightarrow e(p^\prime)e(q^\prime)$ respectively (see \cref{fig:ComptonBreitWheeler}), Volkov states are used as in and out states for electrons and positrons in the matrix elements. Here $p$, $q$ and $k$ are momenta of initial electron, positron and photon respectively and primed momenta refer to the same particle but in the final state. In what follows we discuss the Compton process, but since the analytical structure of the matrix element for Breit-Wheeler process is similar, these results can be translated to Breit-Wheeler process too.
{The matrix element for the Compton process is}
\begin{equation}
    M=-ie\int d^4x \overline{\psi}_{p^\prime r^\prime}(x)(\gamma^\mu\epsilon^\prime_\mu) e^{ik^\prime x} \psi_{pr}(x).
    \label{eq:MatrixElement}
\end{equation}
Having integrated the above expression over light-front variables $x^\perp$ and $x^-$, we obtain:
\begin{equation}
    M=-ie(2\pi)^4\int\frac{ds}{2\pi}\delta^{(4)}(p^\prime+k^\prime-p-sk)\text{{\cirth O}}(s),
\end{equation}
\begin{multline}
 \text{{\cirth O}}(s)= \text{\cirth  G}_0\text{\cirth  g}_{(s)}+\text{\cirth  G}_+\text{\cirth  g}_+(s)\\
 +\text{\cirth  G}_-\text{\cirth  g}_-(s)+\text{\cirth  G}_2\text{\cirth  g}_2(s),  
\end{multline}
where $\text{\cirth  G}_{0,\pm,2}$ are Dirac current structures and $\text{\cirth  g}_{0,\pm,2}$ are integrals over plane-wave field phase, determining the amplitude of the process {(see \citet{seipt.pra.2013} for details)}. The variable $s$ enters from the 4-momentum conservation, $p+sk=p^\prime+k^\prime$, and can be defined as the amount of 4-momentum absorbed from the background field, $s=(pk^\prime)/(p^\prime k)$. Note, that it is sometimes referred to as a total number of background field photons absorbed. However, this is valid only in the case of infinitely long plane-wave field, when the periodicity of the background field means that $s$ must be integral; thus discrete Fourier components may be interpreted {as} ``photons'' with momentum $k_\mu$. The frequency of the emitted photon can be determined from {energy and momentum} conservation:
\begin{equation}
    \omega^\prime=\frac{s (pk)}{(p+sk)n^\prime},
\end{equation}
The differential probability of the Compton process is 
\begin{equation}
    dP^c=\frac{\alpha\omega^\prime|\text{{\cirth O}}[s(\omega^\prime)]|^2}{16\pi^2 (pk) (p^\prime k)}d\omega^\prime d\Omega,
\end{equation}
where $d\Omega=d\phi d\theta$ is the solid angle element related to the emitted photon energy $\omega^\prime$. The dependence of the Compton process on the plane-wave field parameters: amplitude and temporal profile, is encoded in the $\text{\cirth  g}_{0,\pm,2}$ integrals. These integrals can be evaluated numerically \cite{mackenroth.pra.2011,seipt.pra.2011,krajewska.pra.2012}, or using a saddle point method for highly oscillating phase integrals \cite{narozhny.jetp.1996,mackenroth.prl.2010,seipt.pra.2015}, or analytically \cite{seipt.jpp.2016}.

\subsubsection{Formation length}
\label{sec:FormationLength}

The radiation formation length or the coherence interval is the space-time region where the functions that determine the probability of the processes in strong fields are formed \cite{ritus.jslr.1985}.
This parameter plays an important role in calculation and analysis of the Compton and Breit-Wheeler processes, since the particle interactions over this interval might lead to suppression or enhancement of total intensity or the shape of the spectrum. 

There are several approaches to defining the formation length in the classical theory \cite{uggerhoj.rmp.2005, baier.pr.2005}.
First, one can consider it to be the distance $L_f$ travelled before an electron, moving with velocity $\beta$, and an emitted photon are separated by $\lambda/(2\pi)$, where $\lambda$ is the photon wavelength.
For the ultrarelativistic case, this gives:
\begin{equation}
L_f=\frac{2\gamma^2 c}{\omega}. \label{eq:lf}
\end{equation}
Second, the theory of synchrotron emission in the magnetic field can be used. Taking into account the emission cone of $1/\gamma$, the gyromagnetic curvature, as well as the characteristic synchrotron radiation frequency, one arrives at similar expression for $L_f$: {see \cref{eq:FormationLength}.}

\begin{figure}
	\centering
	\includegraphics[width=0.7\linewidth]{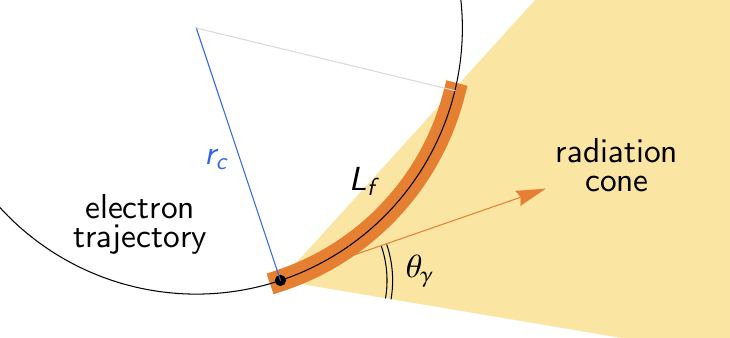}
	\caption{
		Classically, the photon formation length $L_\mathrm{f}$ can be related to the emission angle $\theta_\gamma$ and the instantaneous radius of curvature $r_c$ of the electron trajectory.
		Reproduced from \citet{blackburn.pra.2020}.}
	\label{fig:FormationLength}
	\end{figure}

A more general, geometric, approach was used in \citet{blackburn.pra.2020}, where the formation length of a photon emitted at an angle $\theta$ to the electron momentum is related to the instantaneous radius of curvature of the electron trajectory, $r_c$, by $L_f=2r_c\theta$ (see \cref{fig:FormationLength}).
The radius of curvature is given by
$r_c = \lc (\gamma^2 - 1) [\chi^2-(\vec{E} +\vec{v}\times\vec{B})^2/\Ecrit^2]^{-1/2}$ for a constant, homogeneous field~\cite{seipt.ppcf.2019}.
For $\gamma \gg 1$, $r_c \simeq \lc \gamma^2/\chi$.
We note that for the case of a plane EM wave, where $\chi = 2a_0\gamma[\hbar\omega_0/(mc^2)]$, and taking $\theta=1/\gamma$ as a characteristic value, a well-known result can be recovered, $L_f=c/(a_0\omega_0)$ \cite{ritus.jslr.1985}.
This can also be obtained from \cref{eq:lf} by substituting the characteristic frequency of the emitted photon, $\omega \sim \omega_0 a_0\gamma^2$, in synchrotron radiation.
By taking into account the angular dependence of the photon emission, one may show that the formation length depends on the photon energy~\cite{blackburn.pra.2020}:
    \begin{equation}
    L_f \simeq \frac{c \chi^{1/3}}{a_0\omega_0}\left(\frac{1 - f_c}{f_c}\right)^{1/3}.
    \end{equation}
Thus, even for very strong fields, photons emitted with sufficiently low energy can have a formation length comparable to the wavelength of the background field: see \cref{fig:Map}.
This result is consistent with that obtained from a full theoretical analysis from SFQED, which we now outline.

Calculating the probability of photon emission in a pulsed plane wave, $dP^c$, in QED involves a double integral over phases $\phi_1$ and $\phi_2$, because the amplitude $M$, \cref{eq:MatrixElement}, is squared.
By redefining this as an integral over average and interference phase, $\phi_\text{av} = (\phi_1 + \phi_2) / 2$ and $\phi_\text{in} = (\phi_1 - \phi_2) / 2$, respectively, one can identify the range of $\phi_\text{in}$ that provides the dominant contribution to the `local' rate $d P^c / d \phi_\text{av} \sim \int \cdots \rmd\phi_\text{in}$~\cite{nikishov.jetp.1964,ritus.jslr.1985,harvey.pra.2015,dipiazza.pra.2018,ilderton.pra.2019}.
This size of this region, which can be identified as the `formation phase', depends on both $a_0$ and the lightfront momentum fraction of the emitted photon~\cite{dipiazza.pra.2018}.

\subsubsection{The locally constant, crossed fields approximation}
\label{sec:LCFA}

If the characteristic length or duration of a SFQED process is much smaller than the respective spatial and temporal inhomogeneities of the electromagnetic field in which it takes place, the local probability of the process can be calculated in the framework of the locally constant field approximation (LCFA)~\cite{ritus.jslr.1985,reiss.jmp.1962,baier1998}. This approximation is based on the fact {that} the domain of SFQED includes high energy (ultra-relativistic) particles and high intensity electromagnetic fields (see \cref{fig:Map}). When these fields are transformed into the ultrarelativistic particle's rest frame, they become mutually perpendicular electric and magnetic fields of almost equal amplitude. In most cases these high intensity fields come from laser pulses and are characterized by field amplitude, $a_0\gg 1$, and some wavelength $\lambda$, for which the formation length of the multiphoton Compton and Breit-Wheeler processes are much shorter than the laser wavelength. This leads to an assumption that the probabilities of these processes in an arbitrary field can be obtained as an integral over the time and space of the corresponding probability in a constant crossed field calculated at local value of the field. We note that the same method can be used to calculate the probability of Schwinger pair production in arbitrary electromagnetic field \cite{narozhny.pla.2004}. Though in this case the field does not need to be assumed crossed.
This approximation underpins the numerical tools used to simulate laser-matter interactions in the quantum regime, which we discuss in \cref{sec:Numerics}.
{Benchmarking of the LCFA, the limits of its applicability, and means to go beyond it, are discussed in \cref{sec:OpenQuestions}.}

In what follows we revisit the multiphoton Compton and Breit-Wheeler processes in a constant crossed field~\cite{ritus.jslr.1985}.
These processes can be characterised by differential probability rates, i.e. probability per unit time and $f$, 
{where the latter quantifies the fraction of energy that goes to the created photon ($f = f_c$) or electron/positron ($f = f_b$), respectively.}
These differential probabilities for both processes can be written in the following form:
\begin{equation}
\frac{dP}{df}=-\frac{\alpha}{\lc/c}\frac{1}{l_0} F_\mp\left(\xi,\nu\right),
\label{eq:C&BWRate}
\end{equation}
{where the} function $F_\pm(\xi,\nu)$ is
\begin{equation}
F_\pm\left(\xi,\nu\right)=\frac{2^{-(1\pm 1)/2}}{\sqrt{3}\pi}\left[\int\limits_\xi^\infty K_{\frac13}(z^\prime)dz^\prime\pm \nu K_{\frac23}(\xi)\right].
\end{equation}
{For the Compton process, the relevant subscript is ``+'' and $P=P^c$, $f=f_c$, $l_0=\gamma_e$, $\xi=\xi_c$, $\nu=\nu_c$, where}
\begin{equation}
\xi_c=\frac{2}{3}\frac{f_c}{(1-f_c)\chi_e},~~~\nu_c=1-f_c+\frac{1}{1-f_c}. 
\end{equation}
{For the Breit-Wheeler process, the relevant subscript is ``-'' and $P=P^b$, $f=f_b$, $l_0=\hbar\omega/m c^2$, $\xi=\xi_b$,  $\nu=\nu_b$, where}
\begin{equation}
\xi_{b}=\frac{2}{3}\frac{1}{f_{b}(1-f_{b})\chi_\gamma},~~~\nu_{b}=\frac{1-f_{b}}{f_{b}}+\frac{f_{b}}{1-f_{b}}.
\end{equation}
{We can also express the total power radiated by an electron in the Compton process:}
\begin{equation}
{\Power_\text{rad} = mc^2\gamma \int_0^1 \! f_c \frac{dP^c}{df_c} \, df_c.}
\label{p_rad}
\end{equation}
The explicit forms of these variables allows us to emphasize the similarities and differences between Compton and Breit-Wheeler processes. For $f_{(c,b)}\rightarrow  1$, {\it i.e.}, when almost all energy of an initial particle is transferred to a photon in Compton or to either electron or positron in Breit-Wheeler, $\xi_c$ and $\xi_{b}$, as well as $\nu_c$ and $\nu_{b}$ demonstrate the same behavior, $\xi_{(c,b)}\rightarrow \infty$ and $\nu_{(c,b)}\rightarrow \infty$. In the opposite situation of $f_{(c,b)}\rightarrow 0$, one can see that $\xi_{c}\rightarrow 0$ and $\nu_{c}\rightarrow 2$, while $\xi_{b}\rightarrow \infty$ and $\nu_{b}\rightarrow \infty$. Thus, in the first case the differential probabilities of Compton and Breit-Wheeler processes should be similar. Moreover, we note that the Breit-Wheeler differential probability is symmetric with respect to $f_{b}\rightarrow  1-f_{b}$.

One can express the rates using the synchrotron functions \cite{gonoskov.pre.2015} (see derivation, e.g., in appendix of \cite{harvey.pra.2016}):
\begin{equation}
\frac{d P}{d f_{c,b}} = 
\frac{\sqrt{3}\alpha c \chi}{l_0 \lambdabar}
\begin{cases}
\frac{\left(1 - f_c\right)}{f_c} \left[F_1(\xi_c) + \frac{f^2_c}{1-f_c} F_2(\xi_c)\right],\\ \\
\left(f_b - 1\right)f_b F_1(\xi_b) +  F_2(\xi_b),
\end{cases}
\label{eq:ratesViaSynchFunctions}
\end{equation}
where $F_1(\xi) = \xi \int_\xi^\infty dx K_{5/3}(x)$ and $F_2(\xi) = \xi K_{2/3}(\xi)$ are the first and the second synchrotron functions.
In this expression, the use of synchrotron functions makes apparent the transition to classical synchrotron emission ($f_{c,b} \rightarrow 0$, $\xi_{c}\rightarrow 0$, and $\xi_{b}\rightarrow \infty$) famously expressed by the first synchrotron function (\cref{eq:ClassicalEmissionRate}).

\paragraph{Differential probabilities, $f_{(c,b)}\rightarrow 1$.} In this case the Eq. (\ref{eq:C&BWRate}) can be integrated to yield the result:  
    \begin{equation} \label{dP^C&BW}
    dP=\frac{2}{3 \mp 1}\frac{\alpha c}{\pi \lc l_0}\frac{\chi^{1/2}}{\left(1-f\right)^{1/2}}\exp\left[-\xi\right]df,
    \end{equation}
which highlights the similarities between $dP^c$ and $dP^{b}$ in this limit. Each of these functions has a maximum at 
    \begin{equation}\label{max energy}
    f=1-\frac{4}{3\chi},
    \end{equation}
which corresponds to the enhancement of the high energy electrons/positrons or photons production in $\gamma\rightarrow ee$ and $e\rightarrow\gamma e$ processes respectively. However, for this maximum to exist the following condition should be satisfied: $\chi_e>4/3$ for Compton and $\chi_\gamma>8$ for Breit-Wheeler \cite{bulanov.pra.2013}.

\paragraph{Differential probabilities, $f_{(c,b)}\rightarrow 0$.} The differential probability scales as $f_c^{-2/3}$ in this case: 
    \begin{equation}\label{dPe_low energy}
    \frac{dP^c}{df_c}=-\frac{2\alpha c}{3^{1/3}\pi\Gamma(1/3)\lc}\frac{1}{\gamma}\left(\frac{\chi_e}{f_c}\right)^{2/3}, 
    \end{equation} 
where {$f_{c}\ll 1$ and $\chi_e$}. However the total probability of emission remains finite and the intensity of the radiation emission scales as $f_c^{1/3}$ as $f_c\rightarrow 0$. 

\paragraph{Total probability rates.} The total probability of the Compton process is  
    \begin{equation}
    P^{c}(\chi_e) =
    -\frac{\alpha c}{2\sqrt{3}\pi^2\lc}\frac{\chi_e}{\gamma}\int\limits_0^\infty dy\frac{5+7\zeta+5\zeta^2}{(1+\zeta)^3}K_{2/3}(y),
    \end{equation}
where $\zeta=(3/2)\chi_e y$. For $\chi_e\ll 1 $ and $\chi_e\gg 1$ the integration can be carried out, and
    \begin{equation}
    P^{c}\simeq
    \begin{cases}
    0.92\frac{\alpha}{\lc/c}\sqrt{\frac{I}{I_S}}\left(1-0.92\chi_e+...\right),~~~\chi_e\ll 1 \\ 
    0.93\frac{\alpha}{\lc/c}\sqrt{\frac{I}{I_S}}\chi_e^{-1/3}\left(1-0.58\chi_e^{-2/3}\right),~~~\chi_e\gg 1.
    \end{cases}
    \end{equation}
The probability of the Breit-Wheeler process is
    \begin{equation}\label{gamma probability}
    P^{b}=-\frac{\alpha c \chi_\gamma}{32\sqrt{3}\pi^2\lc}\frac{m_e c^2}{\hbar \omega}\int\limits_{\frac{8}{3\chi_\gamma}}^\infty \frac{(8\zeta+1)K_{2/3}(y)}{\zeta\sqrt{\zeta(\zeta-1)}}dy,
    \end{equation}
where $\zeta=(3/8)\chi_\gamma y$. For $\chi_\gamma\ll 1 $ and $\chi_\gamma\gg 1$ the integration can be carried out, and
    \begin{equation}\label{gamma probability limits}
    P^{b}=\begin{cases}
    0.073\frac{\alpha}{\lc/c}\left(\frac{I}{I_S}\right)^{1/2}\exp\left(-\frac{8}{3\chi_\gamma}\right),~~~\chi_\gamma\ll 1, \\
    0.67\frac{\alpha}{\lc/c}\left(\frac{I}{I_S}\right)^{1/2}\chi_\gamma^{-1/3},~~~\chi_\gamma \gg 1.
    \end{cases}
    \end{equation}
We note that the probability of the Breit-Wheeler process decreases for both small and large values of parameter $\chi_\gamma$. This means that there exists a value of $\chi_\gamma$ that maximizes the probability of pair production. From Eq. (\ref{gamma probability}) we find the value is ${\chi_\gamma^\text{max}}\approx 12$ and the maximum probability is ${P^{b}_{R,\text{max}}}\simeq (\alpha/5) (c/\lc) \sqrt{I/I_S}$.
It is proportional to the square root of intensity. 

\subsubsection{Radiative corrections}
\label{sec:RadiativeCorrections}

\begin{figure}
    \includegraphics[width=0.7\linewidth]{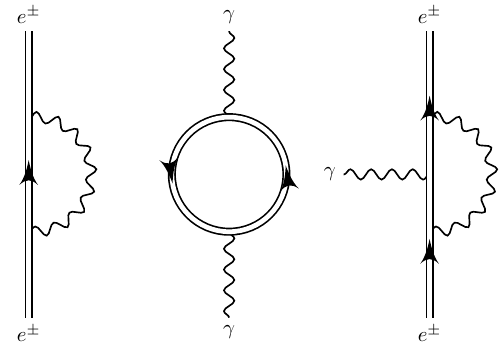}
    \caption{From left to right: the one-loop mass operator, polarization operator and vertex correction.}
    \label{fig:RadiativeCorrections}
    \end{figure}

{Most of the analytical studies of strong field QED phenomena have been carried out at tree-level, mainly due to the exploratory nature of these studies, and because experimental results did not require calculations at high accuracy}. The experiments \cite{bula.prl.1996,bamber.prd.1999,poder.prx.2018,cole.prx.2018} produced only a limited number of events, mostly establishing the understanding that strong fields effects exist and can have a strong influence on the charged particle dynamics in these fields. However, future development of theory and experiments would require higher accuracy of calculations. Moreover, there are several theoretical questions that require such calculations to be resolved. The calculations of radiative corrections at one-loop level, mass and polarization operators and vertex radiative correction  (see \cref{fig:RadiativeCorrections}), as well as at higher orders were reported on in Refs. \cite{ritus.jetpl.1970, ritus.npb.1972, ritus.ap.1972, narozhny.prd.1979, narozhny.prd.1980, morozov.jetp.1981} for crossed EM fields and in plane waves for mass and polarization operators in Refs. \cite{baier.jetp.1975a, baier.jetp.1975b, becker.jpa.1975}. More recent work on the radiative corrections include \cite{akhmedov.jetp.1983,akhmedov.pan.2011, fedotov.jpcs.2017,mironov.prd.2020}, and \cite{dipiazza.prd.2020}, where the vertex radiative corrections and first order radiative corrections to the Compton scattering have been calculated, as well as \cite{edwards.prd.2021}, where the impact of all-orders radiative corrections is studied.

\paragraph{Polarization and mass operators}

The possibility of the Breit-Wheeler process in the constant electromagnetic field leads to the fact that the external field plays a role of a homogeneous anisotropic medium with dispersion and absorption. The {index of refraction} of such a medium can be defined in terms of the photon mass $1-n_{||,\perp}=\mu^2_{||,\perp}/2 k_0^2$, where the photon mass is defined through the probability of the Breit-Wheeler process: $P_{||,\perp}=-(n/k_0)\Im \left(\mu^2_{||,\perp}\right)$. Following Ref. \cite{ritus.jslr.1985}, we obtain the following expression for the polarization operator    
\begin{equation}
\label{eq:mu2}
    \mu^2_{||,\perp}(\chi_\gamma)=\frac{\alpha m^2}{6\pi}\int\limits_1^\infty du \frac{8u + 1\mp 3}{zu\sqrt{u(u-1)}}f^\prime(z),
\end{equation}
where $f(z)=i\int_0^\infty dt \exp{-i(zt+t^3/3)}$ and $z=(4u/\chi_\gamma)^{2/3}$. For $\chi_\gamma\ll1$ and $\chi_\gamma\gg 1$ the asymptotic expressions can be derived from Eq. (\ref{eq:mu2}):
\begin{equation}
\label{E130}
\mu_\pm^2=-\alpha m_e^2\left[
\frac{11\mp 3}{90\pi}\chi_\gamma^2+i\sqrt{\frac{3}{2}}\frac{3\mp 1}{16}\chi_{\gamma}
e^{-\frac{8}{3\chi_\gamma}}\right]
\end{equation}
for $\chi_\gamma\ll 1$ and
\begin{equation}
\label{E131}
\mu_\pm^2=-\alpha m_e^2\left[\frac{10\mp 2}{84\pi^2}\sqrt{3}\Gamma^4\left(1-i\sqrt{3}\right)(3\chi_\gamma)^{2/3}\right]
\end{equation}
for $\chi_\gamma\gg 1$.
Analogously, the electron mass operator in constant crossed electromagnetic field can be obtained from the probability of the multi-photon Compton process \cite{ritus.jslr.1985}: \begin{equation}
    \Delta m=\frac{\alpha m}{6\pi}\int\limits_1^\infty du \frac{5 + 7u + 5u^2}{z(1+u)^3}f^\prime(z),
\end{equation}
where $z=(4u/\chi_e)^{2/3}$. The asymptotic expressions for $\chi_e\ll1$ and $\chi_e\gg 1$ read as follows:
\begin{equation}
    \Re\left(\Delta m\right)=\frac{4\alpha m}{3\pi}\chi_e^2\left(\log\chi_e^{-1}+C+\frac{1}{2}\log 3 - \frac{33}{16}\right)+...
\end{equation}
\begin{equation}
    \Im\left(\Delta m\right)=-\frac{5\alpha m}{4\sqrt{3}}\chi_e\left(1-\frac{8\sqrt{3}}{15}\chi_e + \frac{7}{2}\chi_e^2\right)+...
\end{equation}
for $\chi_e \ll 1$ and 
\begin{equation}
    \Delta m=\frac{7\Gamma(2/3)(1-i\sqrt{3})\alpha m}{27\sqrt{3}}(3\chi_e)^{2/3}.
\end{equation}
for $\chi_e \gg 1$.
Note that both $\mu^2_{||,\perp}$ and $\Delta m$ in the case of $\chi\gg1$ demonstrate the $\alpha\chi^{2/3}$ behaviour, which points at possible breakdown of the semi-classical perturbation theory at high values of $\chi$. 

\paragraph{Cherenkov radiation}
As it was mentioned above, in the presence of strong electromagnetic field, the QED vacuum behaves as a medium with an {index of refraction} larger than unity~\cite{gies.prd.2018,dittrich.stmp.2000,narozhny.jetp.1969,ritus.jetpl.1970,shore.npb.2007}, i.e. the {phase velocity} of the interacting electromagnetic waves is below speed of light in vacuum. One of the consequences of this fact is a possibility of the Cherenkov radiation of the high-energy electrons traversing the electromagnetic field~\cite{ritus.jetpl.1970, dremin.jetpl.2002, macleod.prl.2019, schwinger.annphys.1976, becker.physica.1977,ginzburg.pr.1979}, which is connected to the radiative corrections. The angle and the intensity of the Cherenkov radiation are determined through the real part of the polarization operator \cite{ritus.jslr.1985}:
\begin{equation}
    \theta_\pm=\left[-\frac{\Re\mu^2_{||,\perp}(\chi_\gamma)}{\omega^{\prime 2}}-\frac{m^2}{p_0^2}\right]^{1/2},
\end{equation}
\begin{equation}
    dI_\pm=\alpha\left[-\Re \mu^2_{||,\perp}-\frac{m^2\omega^{\prime 2}}{p_0^2}\right]\frac{d\omega^\prime}{\omega^\prime}.
\end{equation}

In classical electrodynamics, as described by the Maxwell equations, the electromagnetic wave frequency and wavevector are related to each other by $\omega^2 -\pmb{k}^2c^2 = 0$.
In the quantum vacuum, polarization effects result in the dispersion equation 
\begin{equation}\label{E120}
\omega^2-\pmb{k}^2c^2-\mu_\pm^2\left(\chi_e\right)c^2\hbar^{-2}=0.
\end{equation}
In the limit $\chi_\gamma\ll 1$, the difference between the {index of refraction} and unity, $\Delta n_\pm = n_{\pm} - 1$, is
\begin{equation}\label{E140}
\Delta n_\pm=\alpha\frac{11\mp 3}{45\pi}\left(\frac{E}{E_S}\right)^2.
\end{equation}
The introduction of the photon mass leads to the energy conservation equation having two solutions, first, the usual Compton one, and, second, corresponding to the Cherenkov radiation. Thus, kinematic considerations reveal that the emission of a photon by an electron in a strong electromagnetic field is described by a synergic Cherenkov-Compton process \cite{bulanov.prd.2019}. The estimates show that for a 10 PW laser pulse  electrons with an energy more than 10 GeV are needed to observe Cherenkov photons.   

\begin{figure}
    \includegraphics[width=\linewidth]{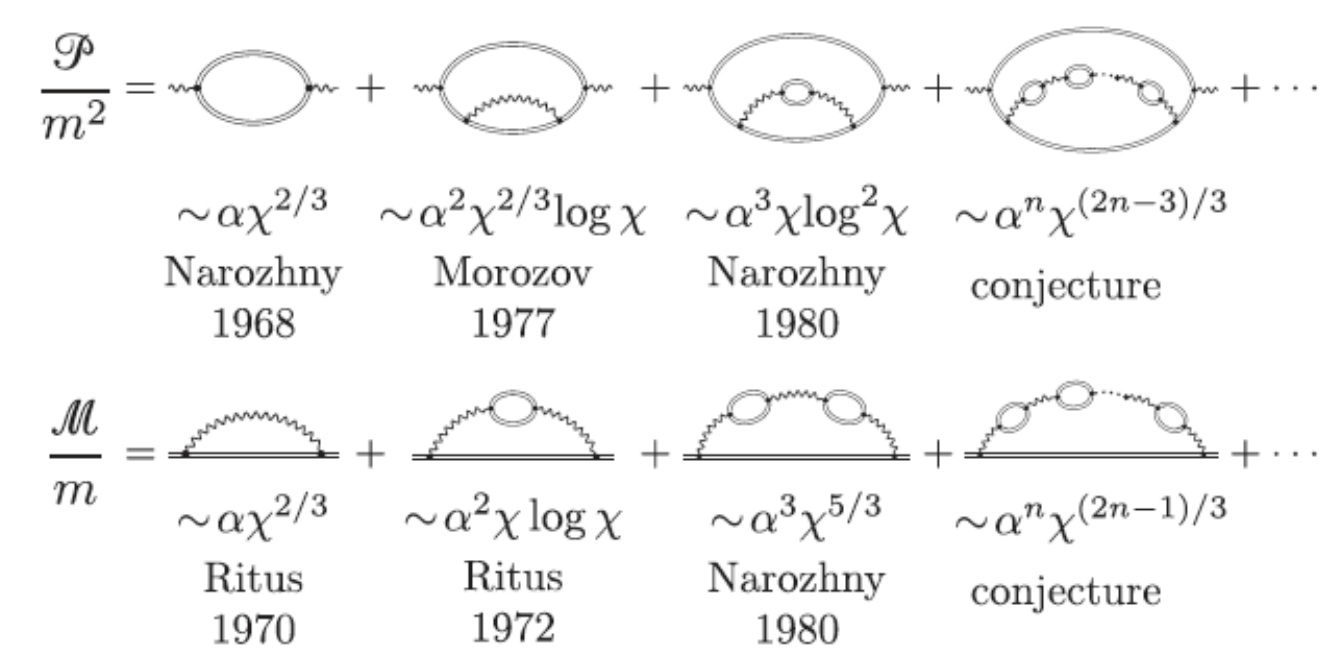}
        \caption{
        Dressed loop expansion of the polarization operator P
(top row) and mass operator M (bottom row). According to the Ritus-Narozhny conjecture, the diagrams shown represent the dominant contribution at n loop and $\alpha\chi^{2/3}$ is the true expansion parameter of SFQED in the regime $\chi\gg 1$.
        Reproduced from \citet{yakimenko.prl.2019}.}
    \label{fig:RadCor}
    \end{figure}

\paragraph{Fully nonperturbative QED}
In most cases considered above, as well as in the cited literature, the radiative corrections are either negligible, or their contribution is much smaller than that of the tree level diagrams. However, in extremely strong electromagnetic fields, characterized by $\chi\gg 1$, this hierarchy is no longer valid. According to the Ritus-Narozhny conjecture \cite{ritus.jslr.1985,narozhny.prd.1980}, the parameter of the SFQED dressed loop expansion is $\alpha\chi^{2/3}$ (see \cref{fig:RadCor}). This means that at $\alpha\chi^{2/3}>1$ each order of  the radiative corrections can no longer be considered small compared to the previous one. So the dressed loop expansion fails and the existing theoretical framework can no longer be used in this case.
Moreover, these results question the applicability of the {semiclassical methodology used in, e.g., QED-PIC simulations (see \cref{sec:Numerics})}, since they are based upon a separation of time-scales: the formation time for quantum processes should be very short when compared to classical propagation time  between incoherent quantum events.
However, calculations (for constant crossed fields) show that in extremely strong fields, $\chi \gg1 $, the mean free paths for electrons and photons are on the order of the Compton wavelength $\lambdabar_C \sim 1/m$ for $\alpha\chi^{2/3} \sim 1$. Of course, the concept of a classical particle and thus classical motion has no meaning on the Compton scale, seriously challenging the applicability of QED-PIC at extreme field strengths \cite{fedotov.jpcs.2017}.
{Experimental configurations that have been proposed to reach such extreme conditions include collisions of dense lepton bunches~\cite{yakimenko.prl.2019}, electron beams and high-intensity lasers~\cite{blackburn.njp.2019,baumann.scirep.2019}, and electron beams and aligned crystals~\cite{dipiazza.prl.2020}.}

The above mentioned calculations were performed in the constant crossed field, which is widely used in QED-PIC, when the separation of scales is assumed to be valid. It was noted in Refs. \cite{ritus.ap.1972,ritus.jslr.1985} that in the case of an alternating field a logarithmic behaviour, more typical for QED processes should take place. More detailed consideration of this problem was carried out in \cite{ilderton.prd.2019, podszus.prd.2019}, where a detailed study of the behaviour of Compton and Breit-Wheeler processes in pulsed EM fields in the limit of $\chi\rightarrow \infty$ was carried out. However, the limit $\chi\rightarrow \infty$ is the combination of particle energy and field strength, whereas the probability of the process is the function of energy and field strength separately. That is why one needs to distinguish the cases when $\chi\rightarrow \infty$ because of $\gamma\rightarrow\infty$ from $a\rightarrow\infty$:
    \begin{align}
    P^c &\sim \frac{\alpha\chi^{2/3}}{\gamma\omega - k p},
    &
    a_0 &\to \infty,
    \\
    P^c &\sim \frac{\alpha a_0^3}{\chi_e}\log\chi_e,
    &
    {\gamma} &\to \infty.
    \end{align}
Thus, the probability of the Compton process increases with $\chi_e$ at high field strength, whereas at high particle energy it decreases with $\chi_e$. Since this probability also determines the magnitude of loop corrections, it means that in one case they can exceed the tree level contribution, whereas in the other case it is not so. These results indicate that there is no universal high $\chi_e$ behaviour.    
We note that $\alpha\chi^{2/3}>1$ corresponds to $\chi>1600$, which is well beyond the reach of current state-of-the art laser facilities and conventional accelerators.

\subsubsection{Higher order processes}
\label{sec:HigherOrderProcesses}

	\begin{figure}
	\includegraphics[width=\linewidth]{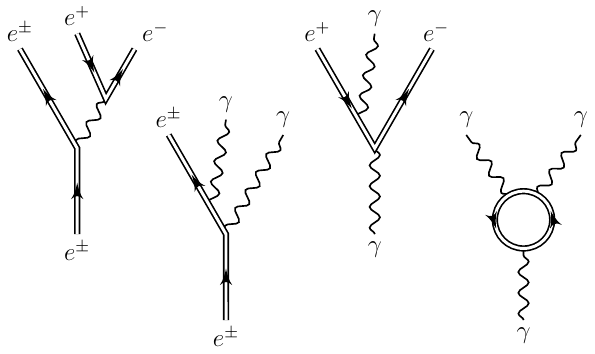}
    \caption{
        Examples of higher order SFQED processes.
        From left to right: trident pair creation, double nonlinear Compton scattering, phototrident pair creation and photon splitting.
    }
	\label{fig:HigherOrderProcesses}
	\end{figure}

The elementary processes shown in \cref{fig:ComptonBreitWheeler} may be combined to form more complex QED interactions.
Each additional vertex reduces the total probability, as compared to a tree-level process, by a factor of $\alpha \ll 1$.
However, this does not necessarily mean that higher order processes are less probable overall, because of additional volume factors which appear in the \emph{cascade} contribution (shortly to be defined);
thus a two-vertex process, scaling quadratically with pulse duration, can {potentially} dominate over a single vertex process, which scales only linearly.
Theoretical calculations of these processes, which involve integrating over the momenta of the intermediate (i.e. unobserved) particles, are much more complicated in SFQED than in vacuum QED, due to the nontrivial spacetime dependence of the background field and basis states.
The most general case amenable to study is a pulsed plane EM wave, with results also available for constant, crossed fields.

Of the two-vertex processes shown in \cref{fig:HigherOrderProcesses}, trident pair creation, the emission of a photon which creates an electron-positron pair, and double nonlinear Compton scattering, the emission of two photons, have attracted most attention.
The former, in particular, represents the first step in the production of an avalanche of secondary particles (see \cref{sec:Cascades}).
The presence of the background field means that it is possible for the intermediate particle to be real, i.e. on the mass shell, leading to resonances in the probability~\cite{oleinik.jetp.1967}.
This raises the possibility that a multivertex process can be factorized into a sequence of first-order processes, which occur to real particles.
The physical basis for this separation is that the intermediate particle can propagate a macroscopic distance if it is on-shell; otherwise its range is limited by the uncertainty principle to a single formation length.
The validity of this factorization, called the `cascade approximation', is essential to numerical simulations of QED processes~\cite{blackburn.rmpp.2020}.

{A two-vertex process can in principle be separated into `two-step' and `one-step' components.
The one-step component encompasses all contributions to the probability that are not accounted for by the cascade, which includes terms where the intermediary is purely virtual, as well as interference and exchange terms~\cite{ilderton.prl.2011}.}
Theoretical calculations of two-vertex processes focus not only on their phenomenology, but also the relative magnitudes of these two contributions.
For double nonlinear Compton scattering this includes both monochromatic~\cite{lotstedt.prl.2009} and pulsed plane waves~\cite{seipt.prd.2012,mackenroth.prl.2013}, and constant, crossed fields~\cite{king.pra.2015}.
Trident pair creation has been studied for pulsed plane-wave backgrounds, to reconsider experimental results~\cite{hu.prl.2010,ilderton.prl.2011} (see \cref{sec:Experiments}) and the cascade and LCFA approximations~\cite{mackenroth.prd.2018,dinu.prd.2018}, as well as for constant-crossed field backgrounds~\cite{king.prd.2013,king.prd.2018}.
Here the name `trident' encompasses all contributions to the probability; it sometimes refers exclusively to the one-step process~\cite{burke.prl.1997}, with the two-step process treated separately as nonlinear Compton scattering followed by nonlinear Breit-Wheeler pair creation.

The two-step contribution is the `product' of \emph{polarized} subprocesses~\cite{king.prd.2013,king.pra.2015,dinu.prd.2020}.
Numerical implementations of SFQED processes in simulation codes have generally employed polarization-averaged rates.
Calculations of the trident pair creation and double Compton probabilities in a constant, crossed field, which compare the use of polarization-averaged and -resolved rates, indicate that this is accurate to within a few per cent~\cite{king.pra.2013,king.pra.2015}.
However, the spin and polarization of intermediate particles may evolve under the action of the external field, due to, e.g. vacuum polarization~\cite{king.pra.2016} and spin precession, enhancing this difference.
We discuss the importance of these dynamics in \cref{sec:Spin} and their possible applications in \cref{sec:PolarizedBeams}.

The dominance of the two-step contribution itself, assuming polarization-resolved subprocesses, is assessed for the trident process in a constant crossed field~\cite{king.prd.2013} and in a pulsed plane wave~\cite{dinu.prd.2018,mackenroth.prd.2018}.
The error in taking only this contribution is one part in a thousand for $a_0 = 50$ and an electron energy of 5~GeV, and becomes substantial for $a_0 < 10$ only if the electron energy exceeds 100~GeV~\cite{mackenroth.prd.2018}.
However, the two-step contribution can be dominant even if $a_0 \simeq 1$, provided that the field duration is sufficiently long~\cite{dinu.prd.2020}: note that in this case, the subprocesses must be described with rates that go beyond the LCFA (see \cref{sec:LCFA}) and the treatment of the polarization of the intermediate particle is more complicated~\cite{dinu.prd.2020}.

An approximation to the one-step trident rate, calculated using the Weizs\"acker-Williams (WW) approximation~\cite{erber.rmp.1966}, has been applied to allow the one-step and two-step yields to be compared~\cite{bamber.prd.1999,bell.prl.2008,kirk.ppcf.2009,blackburn.prl.2014}.
While this is a reasonable approximation to the purely virtual component, once exchange and interference contributions are included, the total one-step rate is negative for $\chi_e \lesssim 20$~\cite{king.prd.2013,king.prd.2018,dinu.prd.2018}.
{In the high-energy limit (at fixed $a_0$), the one-step part gives the dominant contribution and the WW approximation is accurate~\cite{torgrimsson.prd.2020}.}

Further higher order processes include: phototrident pair creation~\cite{morozov.jetp.1977,torgrimsson.prd.2020}, photon splitting~\cite{adler.ap.1971,akhmadaliev.prl.2002,dipiazza.pra.2007}; Delbr{\"u}ck scattering, the scattering of a photon by a vacuum fluctuation~\cite{milstein.pr.1994,dipiazza.pra.2008,koga.prl.2017}; and, in part, triple and quadruple nonlinear Compton scattering~\cite{dinu.prd.2019}.
The interaction where a single electron emits an arbitrarily large number of photons is particularly interesting, because it represents the analogue of radiation reaction in QED~\cite{dipiazza.prl.2010}: see \cref{sec:QuantumRR}.
(Note that QED is a multiparticle theory and therefore there is no guarantee that only a single electron emerges from the interaction; however, pair creation is negligible for $\chi_e \ll 1$.)

\subsection{Quantum radiation reaction}
\label{sec:QuantumRR}

Radiation reaction appears to be more straightforward, at least conceptually, in QED.
The conservation of momentum is applied at the level of the scattering amplitude (see \cref{sec:ComptonBreitWheeler}); therefore when an electron emits a photon in Compton scattering, the self-consistent recoil is automatically accounted for.
However, that recoil is proportional to $\hbar$ and it vanishes in the classical limit.
The resolution to this problem is proposed in \cite{dipiazza.prl.2010}.
A complete calculation of the interaction between an electron and a prescribed, strong electromagnetic field in QED means solving for the scattering amplitude, and final state, for an initial state containing only a single electron~\cite{dipiazza.rmp.2012}.
In the perturbative approach to this problem, cascade processes, which include quantum radiation reaction, manifest themselves as high-order contributions to the $S$-matrix.
However, as discussed in \cref{sec:HigherOrderProcesses}, if $a_0 \gg 1$, the small size of the formation length~\cite{ritus.jslr.1985} means that high-order terms are dominated by \emph{incoherent} contributions, where each elementary QED process occurs in a distinct region of space~\cite{dipiazza.rmp.2012}.
Thus we may identify quantum radiation reaction as the recoil arising from the sequential, incoherent emission of many photons~\cite{dipiazza.prl.2010}.
While the recoil from an individual emission is proportional to $\hbar$, the number of photons emitted $N_\gamma \propto \alpha \propto \hbar^{-1}$, leading to a total momentum change $\Delta p \propto \hbar^0$ that is non-zero in the classical limit.

Provided that $a_0 \gg 1$ and $\chi \lesssim 1$, so that photon emission may be treated as occurring instantaneously, and pair creation is negligible, quantum radiation reaction can be included as a supplement to the classical dynamics. The conditions listed mean that the local constant field approximation is valid and the probability of photon emission is described by the expressions from section \ref{sec:LCFA}. The evolution of the electron distribution function, $\phi_e(t,\vec{r},\vec{p})$, including quantum radiation reaction effects, is governed by the following kinetic equation~\cite{shen.prl.1972,sokolov.prl.2010,elkina.prstab.2011,neitz.prl.2013,bulanov.pra.2013,ridgers.jcp.2014}:
    \begin{multline}
    \frac{\partial \phi_e}{\partial t}
        + \frac{\vec{p}}{\gamma m} \cdot \frac{\partial \phi_e}{\partial \vec{r}}
        - e \left( \vec{E} + \frac{\vec{p} \times \vec{B}}{\gamma m} \right) \cdot \frac{\partial \phi_e}{\partial \vec{p}} =
    \\
    -\phi_e \int\! w_{e \to e\gamma}(\vec{p},\vec{q}) \,\rmd^3\vec{q}
    \\
    + \int\! \phi_e' w_{e \to e\gamma}(\vec{p}', \vec{p}' - \vec{p}) \,\rmd^3\vec{p}',
    \label{eq:KineticEquations}
    \end{multline}
where we have taken $\phi_e = \phi_e(t, \vec{r}, \vec{p})$ and $\phi'_e = \phi_e(t, \vec{r}, \vec{p}')$ for brevity. The right-hand side of \cref{eq:KineticEquations} is a collision operator that accounts for the dynamical effect of photon emission, i.e. radiation reaction. Quantum effects are manifest in the dependence on $w_{e \to e\gamma}(\vec{p},\vec{q}) \,\rmd^3\vec{q}$, the probability rate that an electron with momentum $\vec{p}$ emits a photon with momentum $\vec{q} = \hbar\vec{k}$. If the pair creation is not negligible, it couples the electron and positron distribution functions to the photon distribution function $\phi_\gamma$, and three coupled kinetic equations can be used to describe the interaction of electrons, positrons, and photons with strong EM fields, which is discussed in \cref{sec:Numerics}.

The evolution of the distribution function, including photon emission, at arbitrary $\chi_e$ is generally obtained by numerical solution of \cref{eq:KineticEquations}.
The relevant methods are discussed in \cref{sec:Numerics}.
However, it is possible to obtain analytical insight when considering specific field configurations or if $\chi_e$ is not too large.
In the former case, one can apply the Markov chain formalism to assess the long-term evolution of the distribution function~\cite{bashinov.pre.2015}.
In the latter case, \cref{eq:KineticEquations} can be reduced to a Fokker-Planck equation~\cite{neitz.prl.2013}.
This corresponds to taking the first two terms in a Kramers-Moyal expansion of the collision operator~\cite{niel.pre.2018}.
Following \citet{vranic.njp.2016},
    \begin{equation}
    \frac{\rmd \phi_e}{\rmd t} =
        \frac{\partial}{\partial p_i}
        \left[
            A_i \phi_e + \frac{1}{2} \frac{\partial}{\partial p_j} (B_{ij} \phi_e)
        \right]
    \label{eq:FokkerPlanck}
    \end{equation}
where the drift and diffusion coefficients $A$ and $B$ are, respectively,
    \begin{align}
    A_i &= \int\! q_i w(\vec{p},\vec{q}) \,\rmd^3\vec{q},
    &
    B_{ij} &= \int\! q_i q_j w(\vec{p},\vec{q}) \,\rmd^3\vec{q},
    \end{align}
and the indices $i,j$ denote components of the momentum.
An equivalent expression is obtained in terms of the lightfront momentum $p^-$ in \citet{neitz.prl.2013}.
The validity of the Fokker-Planck approach is limited by the importance of higher orders in the Kramers-Moyal expansion: the third term is larger than 10\% of the diffusion term if $\chi_e \gtrsim 0.25$, and as $\chi_e$ approaches unity the expansion breaks down~\cite{niel.pre.2018}.
Physically, this means that the momentum change from emitting a single photon becomes as important as the cumulative momentum change from emitting many~\cite{bell.prl.2008}.

Further analytical insight can be obtained by considering \emph{moments} of the distribution function.
Results obtained for specific field geometries include the evolution of:
the mean energy of electrons in a rotating electric field, assuming $\chi_e \ll 1$~\cite{elkina.prstab.2011};
and the mean and standard deviation of the lightfront momentum (or energy) of an electron beam in a plane-wave laser pulse, assuming $\chi_e \ll 1$~\cite{neitz.prl.2013,yoffe.njp.2015,vranic.njp.2016}, and extended to arbitrary $\chi_e$~\cite{ridgers.jpp.2017}.
The evolution of the mean Lorentz factor $\mu_\gamma$, for an arbitrary field configuration with electric component $\vec{E}$, is given by~\cite{niel.pre.2018}:
    \begin{equation}
    \frac{\rmd \mu_\gamma}{\rmd t} =
        -\frac{e \avg{\vec{u}\cdot\vec{E}}}{m c}
        - \frac{2\alpha c}{3 \lc} \avg{\chi_e^2 G(\chi_e)},
    \label{eq:MeanEvolution}
    \end{equation}
where $\avg{\cdots}$ denotes the population average.
If the effect of the external field may be neglected, such as in a magnetic field or a plane wave with amplitude $a_0 \ll \gamma$, the evolution of the $n$th order central moment, $\avg{(\gamma - \mu_\gamma)^n}$, may be given analytically~\cite{niel.pre.2018}.

	\begin{figure}
	\centering
	\includegraphics[width=\linewidth]{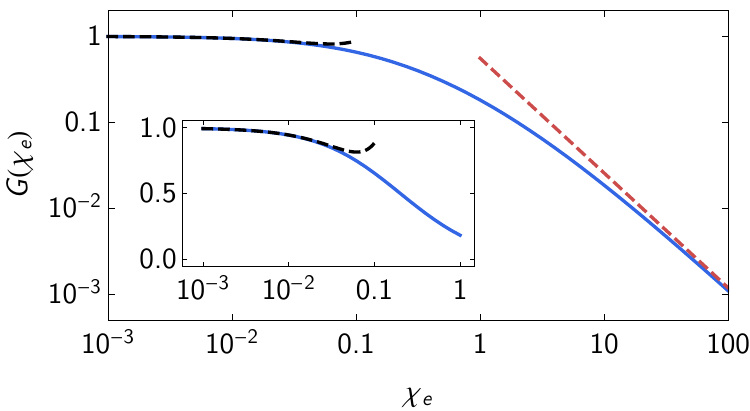}
	\caption{
		$G(\chi_e)$ is the factor by which quantum effects reduce the radiation power from its classically predicted value: full expression (blue) and limiting values at small (black, dashed) and large (red, dashed) $\chi_e$, from \cref{eq:Gaunt,eq:GauntLimits} respectively.}
	\label{fig:G}
	\end{figure}

An important auxiliary function appears in \cref{eq:MeanEvolution}, providing the link to the single-particle emissivity.
The factor $G(\chi_e)$ gives the ratio between the instantaneous radiation powers predicted by QED and by the classical theory~\cite{erber.rmp.1966,SokolovTernov};
it is sometimes referred to as the `Gaunt factor', as it is a multiplicative (quantum) correction to a classical result, first derived in the context of absorption~\cite{gaunt.pt.1930}.
Using the expression for total emitted power, we have $G(\chi_e) = (\lim_{\chi_e \to 0} \int \hbar\omega \,\rmd I)^{-1} (\int \hbar\omega \,\rmd I)$. Expressed as a single integral~\cite{baier1998,SokolovTernov,ritus.jslr.1985},
	\begin{align}
	G(\chi) &=
	    \frac{3\sqrt{3}}{\pi}
	    \int\limits_0^\infty
	        \frac{8 u + 15 \chi u^2 + 18 \chi^2 u^3}{(2 + 3 \chi u)^4} K_{2/3}(u)
	   \,\rmd u,
	\label{eq:Gaunt}
	\\
	&\simeq
	    \begin{cases}
		1 - \frac{55\sqrt{3}}{16}\chi + 48\chi^2 & \chi \ll 1 \\
		\frac{16 \Gamma(2/3)}{3^{1/3} 27} \chi^{-4/3} & \chi \gg 1
		\end{cases}
	\label{eq:GauntLimits}
	\end{align}
where $K$ is a modified Bessel function of the second kind and $\Gamma$ is the gamma function.
These are plotted in \cref{fig:G}, which shows that the reduction becomes significant even at $\chi_e \simeq 0.1$.
The asymptotic expressions given in \cref{eq:GauntLimits} are within 5\% of the true value for $\chi < 0.05$ and $\chi > 200$ respectively~\cite{blackburn.rmpp.2020}.
Analytical approximations to \cref{eq:Gaunt} include $G(\chi) \simeq
[1 + 4.8(1+\chi)\ln(1+1.7\chi) + 2.44\chi^2]^{-2/3}$, which has a fractional error smaller than 2\% across the full range of $\chi$~\cite{baier1998}, and $G(\chi) = (1 + 12\chi + 31\chi^2 + 3.7\chi^3)^{-4/9}$, which is accurate to within 2\% for $\chi < 5$ and $\chi > 1000$~\cite{thomas.prx.2012}.

The classical limit of \cref{eq:FokkerPlanck} allows us to recover a deterministic radiation-reaction force, as discussed in \cref{sec:ClassicalRR}.
Assuming that photons are emitted parallel to the electron momentum, i.e. that $\gamma \gg 1$, the drift and diffusion coefficients are $A = 2\alpha m c^2 \chi_e^2 / (3 \lc)$ and $B = 55\alpha m^2 c^3 \chi_e^3 \gamma / (24 \sqrt{3} \lc)$ to lowest order in $\chi_e$~\cite{neitz.prl.2013,vranic.njp.2016}.
Only $A$ is non-zero in the classical limit $\hbar \to 0$, in which case the characteristics of \cref{eq:FokkerPlanck} satisfy $\frac{\rmd p_i}{\rmd t} = -A_i$~\cite{vranic.njp.2016}, i.e. the {Landau-Lifshitz} equation in the ultrarelativistic limit.
This is also obtained by setting $G(\chi_e) = 1$ in \cref{eq:MeanEvolution}.

Taking its classical limit of QED is also a means of distinguishing between the different classical theories of radiation reaction discussed in \cref{sec:ClassicalRR}.
The evolution of the expectation value of the momentum~\cite{krivitski.spu.1991,ilderton.plb.2013} and the position~\cite{ilderton.prd.2013} of a charge interacting with an external EM field have been shown to obey classical equations of motion in the limit that $\hbar \to 0$.
In particular, it has been shown that, at first order in $\alpha$, only the LAD, Landau-Lifshitz and Eliezer-Ford-O'Connell theories are consistent with QED~\cite{ilderton.prd.2013}.
(Recall that the Landau-Lifshitz equation is itself the result of a perturbative expansion of the LAD equation, in the small parameter $\tau_\text{rad} \propto \alpha$; see \cref{sec:FromLADtoLL}.)
These three equations could be distinguished at second order in $\alpha$, although the much larger number of contributing diagrams at this order significantly increases the difficulty of the calculation~\cite{ilderton.prd.2013}.
{The use of resummation techniques has now allowed \citet{torgrimsson.prl.2021} to recover LL from QED, to all orders in $\alpha$.}

The Landau-Lifshitz equation, being consistent with QED as discussed above, forms the basis for a `semiclassical' (also called `modified classical') theory of radiation reaction.
One may appeal to momentum conservation to argue that, because quantum effects reduce the radiation power by a factor $G(\chi_e)$, the radiation-reaction force must also be reduced by the same factor~\cite{erber.rmp.1966}.
Thus the equation of motion becomes:
    \begin{equation}
	\frac{\rmd p^\mu}{\rmd \tau}
	    = -\frac{e F^{\mu\nu} p_\nu}{m c} + G(\chi_e) g_\text{LL}^\mu,
	\label{eq:ModifiedLandauLifshitz}
	\end{equation}
where $g_\text{LL}$ is the Landau-Lifshitz force given in \cref{eq:LandauLifshitz}.
Its use has also been justified on the grounds that Fokker-Planck formulation, \cref{eq:FokkerPlanck}, is equivalent to a stochastic differential equation for the single-particle momentum~\cite{neitz.prl.2013,niel.pre.2018}, in which the deterministic components are given by \cref{eq:ModifiedLandauLifshitz}.
Therefore the `modified classical' model neglects stochastic effects of quantum radiation reaction, which are discussed in detail in \cref{sec:Nondeterminism}.

An alternative radiation reaction force that accounts for quantum corrections to the radiation power is proposed in \citet{sokolov.jetp.2009}.
A key aspect is that collinearity between the charge's four-momentum and four-velocity is abandoned, in order to renormalize the LAD equation.
However, it has been shown that the momentum evolution predicted by this model is not consistent with the classical limit of QED, unlike the Landau-Lifshitz, Ford-O'Connell and LAD equations~\cite{ilderton.prd.2013}.
It has been suggested that it is the definition of momentum in QED that should be reassessed in light of these results~\cite{capdessus.prd.2016}, as the Sokolov model does produce the correct velocity and trajectory~\cite{ilderton.prd.2013}.
Nevertheless, questions arise as to the model's validity even without appealing to QED, as it is capable of causality violation for sufficiently large fields or particle energies~\cite{burton.cp.2014}: the four-velocity $u^\mu$ satisfies $u^2 = c^2 [1 - (2\alpha\chi_e/3)^2]$ rather than $u^2 = c^2$.
As a practical matter, the differences between the various formulations of classical radiation reaction, including the Sokolov model, are very small in most scenarios of interest~\cite{vranic.cpc.2016}.

\subsection{Spin light and spin dynamics}
\label{sec:Spin}

	\begin{figure}
	\centering
	\includegraphics[width=0.8\linewidth]{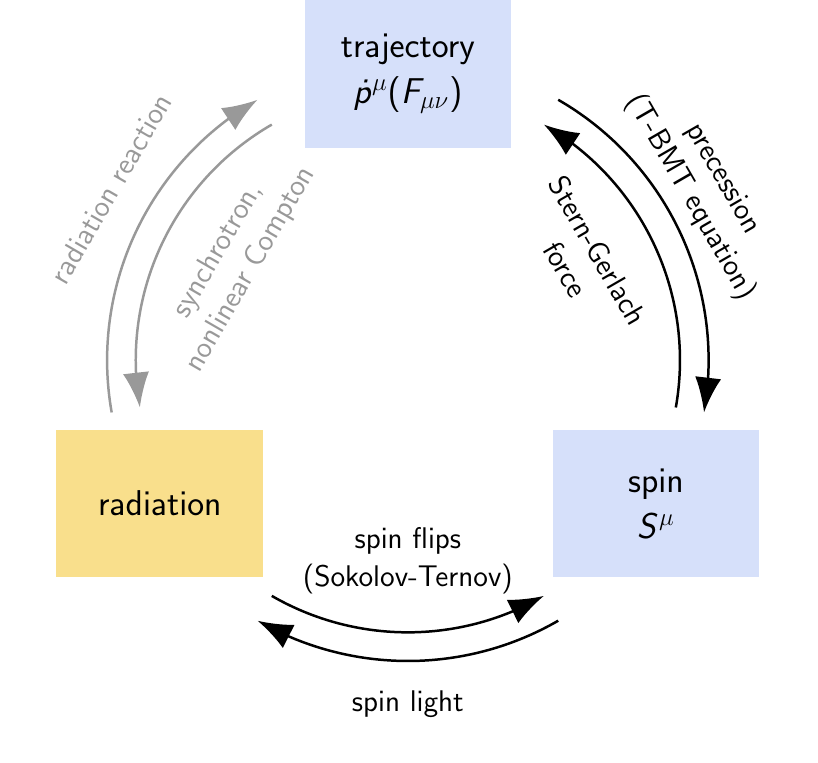}
	\caption{
		Overview of the way spin influences, and is influenced by, particle degrees of freedom (blue) and radiation (yellow).
		Individual couplings are denoted by arrows, with those introduced in \cref{sec:Spin} given in black.
		Adapted from \citet{thomas.prab.2020}.}
	\label{fig:SpinEffects}
	\end{figure}

Electrons are spin-$1/2$ fermions, with an intrinsic magnetic moment $\vec{\mu}_e = -\frac{1}{2} g_s \mu_B \vec{\hat{s}}$ antiparallel to the spin axis $\vec{\hat{s}}$, where $g_s \simeq 2$ is the spin $g$-factor and $\mu_B = e \hbar / (2 m c)$ is the Bohr magneton.
This magnetic moment is a dynamical degree of freedom that couples to the radiation field, affecting the emission spectrum, and to external electromagnetic fields, leading to spin precession~\cite{thomas.nature.1926,bargmann.prl.1959} and additional forces on the particle~\cite{Jackson1999,gerlach.zp.1922}.
In this section we review the role of spin in terms of these couplings, which are illustrated in \cref{fig:SpinEffects}.

In classical electromagnetism, the power radiated by a magnetic moment $\vec{\mu}$ that is accelerated by a static magnetic field $\vec{B}$ is given by
$\Power_\mu = (\alpha/6) (mc^3/\lc) (\mu_\perp^2 / \mu_B^2) \chi_e^4$,
where $\mu_\perp$ is the component of the magnetic moment perpendicular to $\vec{B}$ and $\chi_e = \gamma B / \Bcrit$~\cite{bordovitsyn.pu.1995,Jackson1999}.
(For a neutral particle, $\Power_\mu$ would be the only contribution to radiation emission; this has been explored theoretically for neutrinos in particular~\cite{morley.ap.2015,formanek.ppcf.2018}.)
As the spin magnetic moment is a purely quantum effect, so too is its contribution to radiation emission: $\Power_\mu \propto \hbar^2$.
For an electron, $\Power_\mu$ is smaller than the charge contribution by a factor ${\sim}\chi_e^2$ in the classical regime $\chi_e \ll 1$.
It is more significant in the quantum regime $\chi_e \gtrsim 1$, as radiation associated with changes of the electron spin (spin flips) dominates the high-energy part of the spectrum~\cite{sorensen.nimb.1996}.

The importance of the spin to radiation emission, in the quantum regime, may be understood by the following argument~\cite{kirsebom.prl.2001}:
in the electron instantaneous rest frame, there is an energy difference of $\Delta\Energy = 2\mu_B \abs{\vec{B}_\text{rf}}$ between states where the electron spin is aligned parallel and antiparallel to the rest-frame magnetic field $\vec{B}_\text{rf}$.
A photon emitted in a spin-flip transition therefore has a typical energy (in the lab frame) of $\hbar\omega = \gamma \Delta\Energy = \gamma m c^2 \chi_e$.
For $\chi_e \gtrsim 1$, kinematic pileup means that these photons are concentrated in the tail of the emission spectrum.
This may be seen explicitly by taking the difference between the emission rate of a spin-1/2 electron and that for a spinless electron, i.e. \cref{eq:ClassicalEmissionRate} with the recoil correction \cref{eq:RecoilCorrection}~\cite{sorensen.nimb.1996}:
	\begin{equation}
	\Delta \frac{\rmd W}{\rmd \hbar\omega} =
		\frac{\alpha}{\sqrt{3} \pi \hbar \gamma^2}
		\frac{f^2}{1-f} K_{2/3}(\xi_q).
	\label{eq:SpinLightRate}
	\end{equation}
\Cref{eq:SpinLightRate} gives the `spin light' contribution to the emission spectrum; it is peaked at $f \simeq \chi_e /(1 + \chi_e)$, which bears out the argument given above for the typical energy, once corrected for recoil.
Note that \cref{eq:SpinLightRate} is calculated for constant, crossed fields.
The role of spin is further eludicated by exact QED calculations that compare the scattering and creation of Klein-Gordon (spin-0) and Dirac (spin-1/2) electrons in a pulsed plane wave background~\cite{panek.pra.2002,boca.nimb.2012,krajewska.lpb.2013,jansen.prd.2016}.

	\begin{figure}
	\centering
	\includegraphics[width=0.7\linewidth]{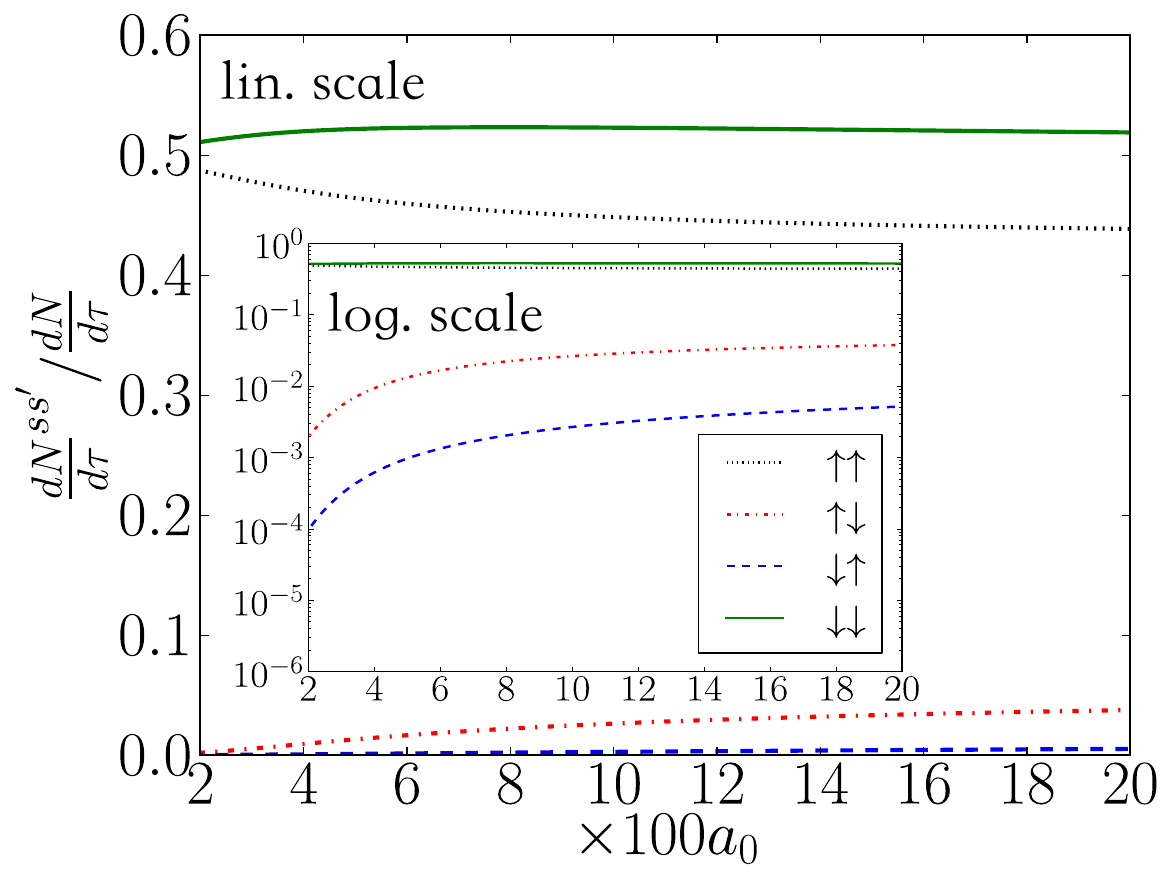}
	\caption{
		The spin-resolved photon emission rates $\rmd N^{s,s'}/\rmd\tau$, normalized to the spin-averaged rate $\rmd N/\rmd\tau$, for electrons orbiting in a rotating electric field of normalized amplitude $a_0$.
		Here $s$ and $s'$ are the projections of the electron initial and final spin on the axis of rotation.
		Reproduced from \citet{delsorbo.pra.2017}.}
	\label{fig:SpinResolvedRates}
	\end{figure}
	
The spin contribution to radiation depends on the orientation of the spin with respect to the external electromagnetic field. This is captured by emission rates that are resolved in the initial and final spin states of the participating particles~\cite{SokolovTernov,seipt.pra.2018}, whereas \cref{eq:SpinLightRate}, like the results in \cref{sec:LCFA}, is calculated for \emph{unpolarized} electrons. A dramatic consequence of spin dependence is the Sokolov-Ternov effect, where lepton beams, orbiting in a magnetic field, self-polarize due to the asymmetric emission probabilities for spin-up and spin-down states~\cite{SokolovTernov,ternov.jetp.1962,jackson.rmp.1976}.

This is also predicted to occur at the magnetic node of the EM standing wave formed by two counterpropagating lasers, but over much shorter timescales: the asymmetric probability of emission for electrons in a rotating electric field, shown in \cref{fig:SpinResolvedRates}, leads to near-total polarization of the electron population within a few femtoseconds at intensities of $10^{23}~\Wcmsqd$~\cite{delsorbo.pra.2017,delsorbo.ppcf.2018}. The prospect of using intense lasers to polarize particles (or to induce their creation) has led to considerable activity in recent years, largely focused on laser collisions with ultrarelativistic electron beams: this is discussed in \cref{sec:PolarizedBeams}.

Spin polarization must be defined with respect to a particular basis.
(Recall that only one component of the spin can be defined at any one time.)
A fixed choice of quantization axis means that polarization is only defined in one dimension; recently a method has been proposed that can simulate the evolution of all three components of the polarization vector~\cite{li.prl.2020pr}.
For a plane EM wave, an appropriate non-precessing quantization axis is given by $\zeta^\mu = \beta^\mu - (p.\beta / \kappa.p) \kappa^\mu$, where $p$ is the electron four-momentum, $\kappa$ is the wavevector, and $\beta = (0, \vec{\beta})$, where $\vec{\beta}$ is a unit vector directed along the wave's magnetic field (as seen in the lab frame)~\cite{king.pra.2015}.
In a static magnetic field, $\vec{\zeta}$ is directed parallel to the field~\cite{SokolovTernov}; in a rotating electric field, it is directed along the electric field's axis of rotation~\cite{delsorbo.pra.2017}.
The occupancies of the spin-up and spin-down states, as defined with respect to this axis, do not change except by interaction with the radiation field.
This can even occur without emission~\cite{meuren.prl.2011}: see \cref{sec:RadiativeCorrections}.

In general, studies of laser--electron-beam or laser-plasma interactions must take also into account the interaction of the spin with the \emph{external} electromagnetic field.
Two such effects are shown in \cref{fig:SpinEffects}: spin precession and spin forces.
The former can be modelled classically using the Thomas--Bargmann-Michel-Telegdi (T-BMT) equation~\cite{thomas.nature.1926,bargmann.prl.1959}.
For an electron with spin four-vector $S^\mu$ and normalized momentum $u^\mu = p^\mu / (m c)$ in an electromagnetic field $F_{\mu\nu}$, this reads:
    \begin{equation}
    \frac{\rmd S_\mu}{\rmd \tau} =
        -\frac{e}{m} \left[
            \frac{g_s}{2} F_{\mu\nu} S^\nu
            + \frac{g_s - 2}{2} (S^\nu F_{\nu\rho} u^\rho) u_\mu
        \right]
    \label{eq:TBMT}
    \end{equation}
where $S^2 = -1$.
According to \cref{eq:TBMT}, a plane EM wave causes the spin of an electron with $g_s = 2$ to precess, but does not change its asymptotic value.
{(The precession of the spin vector inside the wave is necessary to maintain the relation $S_\mu p^\mu = 0$, because the electron momentum is time-dependent.)}
The same is recovered from the quantum description by taking expectation values of bilinears of Volkov states~\cite{seipt.pra.2018}.
The T-BMT equation has been used in particle-in-cell codes to study spin evolution in laser-driven wakefields, for example~\cite{vieira.prstab.2011,wen.prl.2019,an.pop.2019}.

The spin modifies the particle momentum indirectly, by affecting the radiation power~\cite{delsorbo.pra.2017} and thereby the strength of radiation reaction.
Signatures of spin-dependent radiative deflection in tightly focused lasers are studied in~\citet{geng.njp.2020}.
However, if the external electromagnetic field is non-homogeneous, its interaction with the spin also leads directly to a force on the particle; this contribution to the equation of motion is often referred to as the `Stern-Gerlach force', after the seminal experiment that demonstrated the intrinsic quantum nature of the electron magnetic moment~\cite{gerlach.zp.1922}.

This force is proportional to the gradient of the electromagnetic field, which means that it is much weaker than the Lorentz force~\cite{vieira.prstab.2011} and radiation reaction~\cite{tamburini.njp.2010} in interactions with intense optical lasers (see also \citet{thomas.prab.2020}).
On the other hand, it plays a more significant role in fields with high frequency components, as found astrophysically~\cite{mahajan.mnras.2015}, in wavebreaking plasmas~\cite{flood.pla.2015}, or in collisions with intense X rays~\cite{wen.sr.2016}.
While spin-dependent dynamics are captured without approximation by the Dirac equation~\cite{dirac.prsa.1938}, it is useful to consider classical formulations of the same in scenarios where solution of the Dirac equation is numerically unfeasible.
The first classical, covariant formulation of the Stern-Gerlach force was proposed by \citet{frenkel.zp.1926};
this has been compared to the `Foldy-Wouthuysen' model, which is based on the classical limit of the given representation of the Dirac equation~\cite{foldy.pr.1950,silenko.pra.2008}.
The two could be distinguished by measuring the ponderomotive deflection of a spin-polarized electron beam by a tightly focused, ultraintense laser~\cite{wen.sr.2016,wen.pra.2017}.

Finally, we note that the QED emission rates depend on the polarization of the photon, as well as the spin of participating leptons.
This is of particular importance in higher-order QED processes, including cascades, and is discussed in \cref{sec:Cascades}.
The application of spin/polarization-dependent emission to the production of polarized particle beams is discussed in \cref{sec:PolarizedBeams}.

\bib

%% file: numerics.tex
\section{Numerical methods}
\label{sec:Numerics}

Due to the highly nonlinear and geometrically complex nature of the effects caused by RR and SFQED, theoretical investigations are tightly connected with the development of numerical methods.
As these are a key facilitator of this field, we briefly outline the main numerical approaches and some important methodological aspects of their use.

\subsection{Classical radiation emission and reaction}

Computing the radiation of moving charges is an important numerical problem, even in the absence of significant radiation reaction effects.
If this is the case, one can just track or calculate one (or several) representative trajectories using any suitable numerical approach.
The radiation produced can then be obtained by the numerical integration of the emission integral, \cref{eq:SpectralIntensity}, over the trajectories~\cite{martins.spie.2009}.
If information about the trajectory is only available at a sequence of time steps, this implies a limit on the highest frequency that can be resolved, by the Nyquist-Shannon sampling theorem; this can be overcome by the use of high-order interpolations~\cite{thomas.prstab.2010}.
The emission integral accounts fully for the interference between radiation emitted from different elements of the trajectory.
Considerable speedup is possible if this interference can be neglected and contributions summed according to local synchrotron rates~\cite{reville.apj.2010,wallin.pop.2015,martins.ppcf.2016, martins.spie.2009}.
In many studies of laser-driven environments, doing this is reasonable because the formation length $L_f$ is much smaller than the laser wavelength (see \cref{fig:Map} and \cref{fig:a0_wavelength}).

    \begin{figure}
    \includegraphics[width=0.8\linewidth]{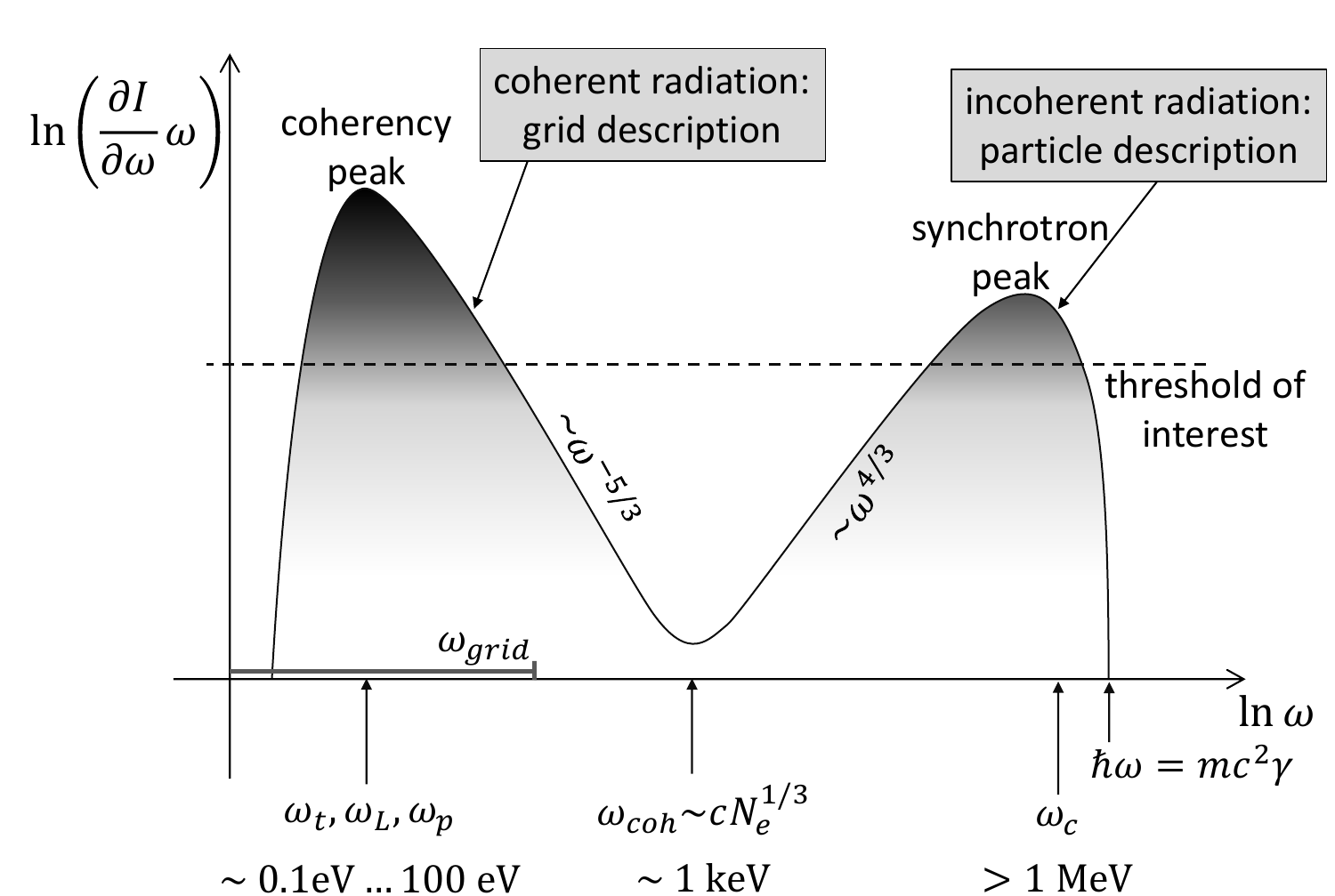}
    \caption{
        Schematic representation of the energy spectrum of the electromagnetic field in a high-intensity laser-matter interaction.
        {The frequencies shown are characteristic of (from left to right): the target ($\omega_t = c/L$, target size $L$), the laser ($\omega_L$), the plasma density ($\omega_p$), the upper limit of coherent emission processes ($\omega_\text{coh}$), and incoherent synchrotron emission ($\omega_c$).}
        Reproduced from \citet{gonoskov.pre.2015}.
    }
    \label{fig:DualFields}
    \end{figure}

The interference between the emission of different particles is routinely accounted in particle-in-cell (PIC) codes via the charge current weighted at the spatial grid nodes.
At high intensity, when radiation emission and RR are strongest, the characteristic frequency $\omega_c$ of synchrotron radiation is far from being resolved on the grid ($\omega_c \gg \omega_\text{grid} = 2 \pi c / \Delta$ for grid spacing $\Delta$) and thus it is accounted for separately. This does mean double counting of particles’ emission, because the expressions for synchrotron emission account for all frequencies including those resolved. Fortunately, in laser-matter interactions the double counting is negligible because of a vast separation of scales.
The characteristic frequency grows with the particle energy (scaling as $\gamma^3$) and when RR effects are notable, this frequency is much higher than the plasma frequency (see \cref{fig:DualFields} and \citet{gonoskov.pre.2015} for more details).

Within the classical domain, radiation reaction can be accounted for by the incorporation of an additional force on the particle, which depends on the instantaneous momentum and the local values of the electromagnetic fields \cite{tamburini.njp.2010, sokolov.pop.2011, capdessus.pre.2012, kravets.pre.2013, green.cpc.2015,vranic.prl.2014}.
Most implementations have focused on the {Landau-Lifshitz} equation, which eliminates the pathological solutions of the Lorentz-Abraham-Dirac (LAD) equation (see \cref{sec:FromLADtoLL}).
(Numerical solution of the LAD equation via backward integration in time is discussed in \citet{koga.pre.2004}.)
The dominant term in the LL equation is proportional to $\gamma^2$ and corresponds to the synchrotron energy loss rate; other terms involving field derivatives are smaller and, in any case, subleading with respect to spin forces (see \cref{sec:Spin}) and quantum corrections~\cite{tamburini.njp.2010}.
Note that in a PIC simulation, where macroparticles represent many individual particles, the `real' charge and mass must be used to compute the magnitude of the RR force~\cite{vranic.cpc.2016}: contrast this with the Lorentz force, which depends only on the charge-to-mass ratio and therefore may be computed using the macroparticle charge and mass.

The radiation and radiation reaction can be calculated together, self-consistently, in the synchrotron regime:
one can {determine} the total energy emitted in synchrotron radiation within a single time step and deducts this amount from the particle's momentum.
This corresponds to the leading term of the LL force and thus gives, to first order, a self-consistent model, which one can implement as an extension of the PIC method: see, for example, \citet{kostyukov.pop.2003, rousse.prl.2004}.
It is also possible to extend these calculations to sample the energy and/or angular distribution of local synchrotron radiation with one or several delta-function-like contributions to the resulted distribution~\cite{sun.prstab.2011, chen.prstab.2013, wallin.pop.2015}.
Note that the use of Monte Carlo method not only facilitates the computations but also gives a natural way to account for probabilistic/stochastic nature of recoils in the quantum domain we now discuss.

\subsection{QED-PIC}

When the particle energy and field strength are sufficient that the quantum parameter $\chi \sim 1$, a classical treatment is insufficient.
As discussed in \cref{sec:ClassicalLimits}, the total energy loss is overestimated, because a significant part of the emission spectrum exceeds the particle kinetic energy.
Correcting the magnitude of the RR force by the Gaunt factor $G_e(\chi_e)$, \cref{eq:Gaunt}, provides a way to account for this overestimation~\cite{kirk.ppcf.2009, thomas.prx.2012, blackburn.prl.2014, yoffe.njp.2015, zhang.njp.2015}.
However, if the typical energy of an emitted photon, whether from the classical or quantum-corrected spectrum, becomes comparable to the particle kinetic energy, an individual particle can lose a significant part of its energy in a single transition.
Since, in many cases of interest, the formation length $L_f$ is much smaller than the laser wavelength (the typical minimal spatial scale of the interaction processes, see \cref{fig:Map}), the radiation losses can be viewed as a sequence of instantaneous events that occur with certain probability.

This concept provides the basis for an extended PIC model, which is commonly referred to as QED-PIC~\cite{duclous.ppcf.2011, nerush.prl.2011, elkina.prstab.2011, sokolov.pop.2011, ridgers.jcp.2014, gonoskov.pre.2015, arber.ppcf.2015, lobet.jpcs.2016}.
The model splits the electromagnetic field into two parts: a coherent classical field, sampled at points on a mesh, and a population of high-frequency photons, which originate from the incoherently summed synchrotron emission of particles.
This is motivated by the separation of energy scales and the consequent negligible role of double counting of the particles' emission~\cite{gonoskov.pre.2015}: in case of significant radiation losses, the energy of incoherent photons ($\hbar\omega \gg 1$~keV) is much larger than that of photons in coherent field-plasma interactions (see fig.~\ref{fig:DualFields}).
In much the same way that charged particles are treated in ordinary PIC codes, the distribution of high-frequency photons is sampled with an ensemble of so-called macrophotons, which propagate ballistically at the speed of light.
A key element for the QED-PIC method is the so-called event generator, which is an algorithm that controls the emission of photons and the creation of electron-positron pairs, according to the locally computed probabilities and characteristics of the QED processes in question.
This methodology can be used to account many SFQED process, but in many cases, a sufficient extension of the PIC model is provided by the inclusion of nonlinear Compton scattering (photon emission) and Breit-Wheeler pair production.

In the simplest case, the event generators for nonlinear Compton scattering and Breit-Wheeler pair production are based on spin- and polarization-averaged rates (as given in \cref{sec:LCFA}) and the assumption that, at high energy, daughter particles are collinear with their parent particles.
Radiation reaction is accounted for by subtracting the momentum of the photon from the emitting electron (positron), whereas in pair production the photon is replaced by an electron and a positron.

The event generators decide pseudorandomly when these events occur and how the momentum should be partitioned.
Unless using a method which can cope with multiple events in a single time step (see \cref{sec:Subcycling}), one has to choose a sufficiently short time step ($\Delta t \ll L_q, L_p$ for all particles/photons), such that the probability of multiple events occurring within a single time step is negligible, i.e. $\Delta t \int_0^1 P^\prime_f (\chi, f^\prime) df^\prime \ll 1$, where $P^\prime_f = \partial P/\partial f$ is the differential rate (per unit time) of the relevant process.
The maximum time step $\Delta t_\text{QED}$ that satisfies this condition is set by the nonlinear Compton rate~\cite{ridgers.jcp.2014}.
It may be compared to maximum time step permitted by the Courant-Friedrichs-Lewy condition for finite-difference-time-domain (FDTD) methods, $\Delta t_\text{CFL}$, and the need to the resolve the Debye length, $\Delta t_\text{D}$, as follows~\cite{ridgers.jcp.2014}:
    \begin{align}
    \frac{\Delta t_\text{QED}}{\Delta t_\text{CFL}} &\simeq
        \frac{10 N}{a_0}
    &
    \frac{\Delta t_\text{QED}}{\Delta t_\text{D}} &\simeq
        \frac{100}{a_0} \sqrt{\frac{n_e m c^2}{\ncrit k_B T} },
    \end{align}
where $N$ is the number of {grid cells} used to resolve the laser wavelength, $n_e$ is the electron number density, $\ncrit$ is the critical density and $T$ is the plasma temperature.

The generation of photons/particles can be handled in various ways.
A method based on cumulative path computation and inverse transform sampling is developed in~\citet{duclous.ppcf.2011, ridgers.jcp.2014}. For a given particle/photon and the corresponding rate $P(t) = \int_0^1 P'_f[\chi(t), f'] d f'$, the probability of no event happening over the time interval $[t, t + \Delta t]$ decreases as $P(\Delta t) = \exp\left(-\int_t^{t+\Delta t} P(\tau) d\tau\right)$. Using a pseudorandom value from the unit interval $r_1 \in [0, 1)$, we can sample an effective path interval between subsequent events $l_e = - \ln(r_1)$ and use it to determine the instant of the next event, by requiring that the cumulative integral $\int_t^{t+\Delta t} P(\tau) d\tau \geq l_e$ (one can handle several events within one time step).
The momentum fraction $f$ is determined by solving $r_2 = C(f,\chi) = \int_0^f P'_f(\chi, f') d f'$, where $r_2$ is another pseudorandom value from $ r_2 \in [0,1]$ and $C$ is the cumulative distribution function.
The inverse (quantile) function $C^{-1}(f,\chi)$ can, for example, be precalculated and tabulated.
For nonlinear Compton scattering, the differential rate $W'_f \to \infty$ as $f \to 0$ as $f^{-2/3}$, which is integrable.
A low $f$ cutoff is commonly applied to exclude this region: a cutoff equivalent to $2mc^2$ is sufficiently small not to affect subsequent pair production and does not affect the magnitude of radiation emission too much, if $\gamma$ is sufficiently large.
The need for a cutoff can be avoided by augmenting the tabulated values of $C(f,\chi)$ (or its inverse) with the asymptotic analytical expressions~\cite{wallin.pop.2015}.

One may instead use rejection sampling~\cite{nerush.prl.2011,elkina.prstab.2011}, which bypasses the need to calculate the cumulative distribution function.
One uniformly distributed random value $r_1 \in [0, 1]$ defines the candidate value of $f$ and another $r_2 \in [0, 1]$ defines whether this event happens: the event is accepted if $r_2 < N'_f = \Delta t P^\prime_f(\chi, r_1)$.
This procedure requires that $N^\prime_f < 1$ for all values of $f \in [0, 1]$ and for all possible values of $\chi$.
Choosing a sufficiently short time step can ensure this for the Breit-Wheeler process.
However, the infrared divergence for nonlinear Compton scattering means that this will be violated for some vicinity of $f = 0$.
This sets an effective cutoff and moreover imposes an additional time step restriction.
Both problems can be eliminated, following \citet{gonoskov.pre.2015}.
For the former, the candidate $f$ can be generated as $f = r_1^3$, i.e. with small values, more often, such that the corresponding acceptance probability $3 r_1 ^2 P^\prime_f(\chi, r_1^3)$ becomes bounded from above. With this modification, the restriction on the time step can be estimated and used to subdivide the time step (see \cref{sec:Subcycling}).

More advanced treatment of nonlinear Compton scattering can include the angular dependence of the emission spectrum \cite{sun.prstab.2011,blackburn.pra.2020}, as well as the spin and polarization properties: see \cref{sec:Spin}.

\subsection{Advanced components}

\subsubsection{Subcycling}
\label{sec:Subcycling}

The growth of the nonlinear Compton rate at high intensity means that it becomes the most important limiting factor on the time step.
Subcycling, i.e. dividing the global time step into smaller fractions and testing for emission at each stage, is an efficient means of scaling simulations of photon emission~\cite{volokitin.jpsc.2020}.
This assumes that the EM field is approximately constant over a single time step, which is generic requirement for a simulation to be appropriately resolved. Note that subcycling permits a single particle/photon to cause cascaded generation of many new particles (if $\chi$ is large) in the course of a single global time step.
Subcycling can also ensure accurate prediction of particle trajectories in strong fields~\cite{arefiev.pop.2015,tangtartharakul.jcp.2021}.

\subsubsection{Resampling}

In QED-PIC codes particle and photon distributions are sampled by an ensemble of so-called macroparticles, each having a weight that quantifies the number of real particles or photons it represents.
In general, macroparticles created by QED processes inherit the weight of their parent macroparticle.
(The distinction between macroparticles and macrophotons is not important here and therefore we use the former term only.)
The prolific generation of new particles in, for example, an electromagnetic cascade (see \cref{sec:Cascades}) can significantly increase the computational demands of running a simulation.
Managing this growing demand requires a procedure that repeatedly decreases the number of macroparticles, while increasing the weight of individual macroparticles accordingly.
This is referred to as ensemble \textit{resampling} or \textit{down-sampling}.
Two main approaches are used: \textit{merging} (coalescing), in which closely placed macroparticles are collapsed, and \textit{thinning}, in which some macroparticles are deleted, while others have their weight increased.

    \begin{figure}
    \includegraphics[width=\linewidth]{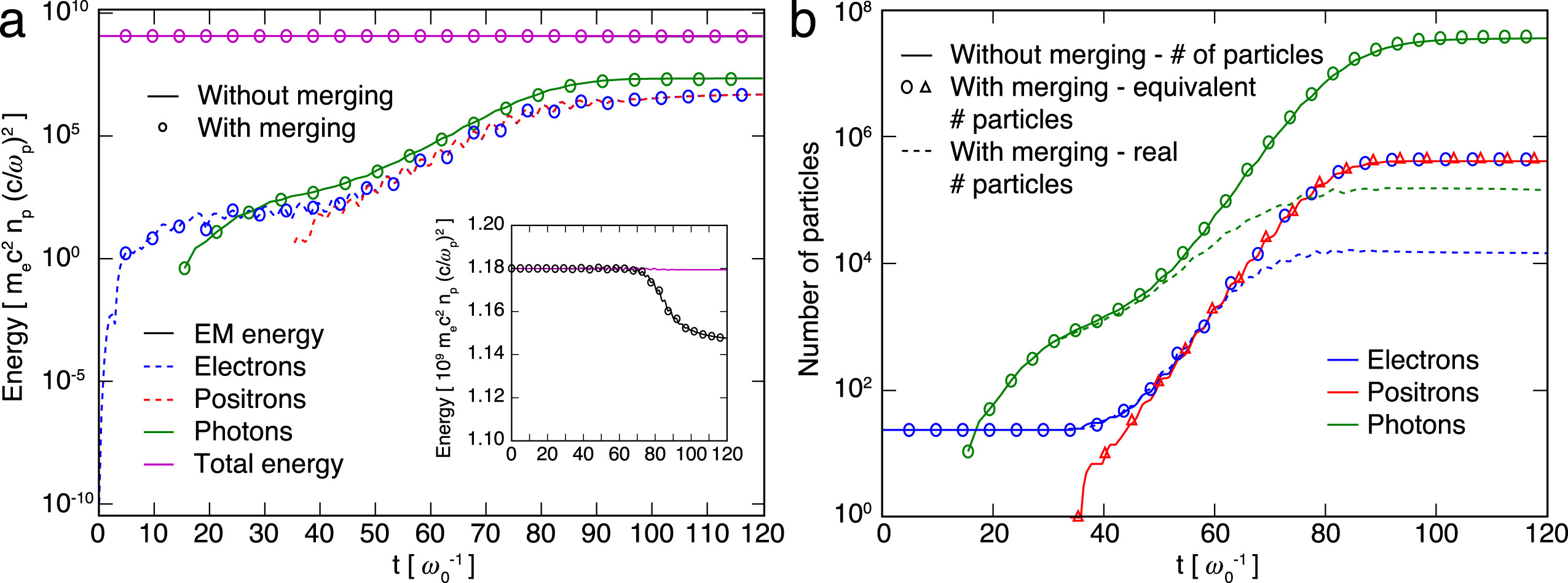}
    \caption{
        Particle merging in a simulation of a laser-driven QED cascade: energy conservation and and growth in the number of particles over the course of the simulation.
        Reproduced from \citet{vranic.cpc.2015}.
    }
    \label{fig:Resampling}
    \end{figure}

Preserving conserved quantities (energy, momentum and charge) and macroparticles' contributions to grid quantities (weighted charge and current density) is recognised an important issue in the development of resampling algorithms, as small errors can accumulate over the growth of a cascade.
The thinning procedure can be realized by removing randomly selected macroparticles and redistributing their weight uniformly among others of the same kind~\cite{timokhin.mnras.2010}; one can redistribute their energy as well~\cite{nerush.prl.2011}.
A method of coalescing two macroparticles, while preserving their contributions to the charge density at grid nodes, was proposed in \citet{lapenta.jcp.1994}.
Preserving the energy and momentum of a merged subset of macroparticles can be accomplished by replacing the subset with a \emph{pair} of macroparticles with appropriately chosen momenta~\cite{rjasanow.jcp.1996, rjasanow.jcp.1998, vranic.cpc.2015}.
The application of the latter to simulations of a QED cascade is shown in \cref{fig:Resampling}, where good agreement is found  for the evolution of the energy and the number of particles between the merging and no-merging simulations.
The approach of replacing local subsets with smaller subsets, with improved uniformity and conservation properties, is developed in \cite{lapenta.cpc.1995, assous.jcp.2003, welch.jcp.2007, pfeiffer.cpc.2015, faghihi.jcp.2020}.

Even if the procedure is not completely conservative, errors can be reduced by applying the merging procedure only to subsets of macroparticles that are \textit{close} in coordinate and momentum space~\cite{rjasanow.jcp.1996, rjasanow.jcp.1998}.
These clusters can be obtained by octree binning~\cite{martin.jcp.2016}, a Voronoi algorithm~\cite{luu.cpc.2016}, or by identifying highly populated cells in coordinate-momentum space~\cite{vranic.cpc.2015, chang.njcp.2017}.
Choosing the size of cells or clusters, in coordinate-momentum space, to be merged involves a tradeoff: small sizes mean a high density threshold to trigger merging, whereas large sizes may flatten out small-scale features (abrupt changes, peaks etc.) that could be of importance.
In the latter case, the problem is caused by adding new macroparticles and assuming a uniform (or smoothed-out) distribution of real particles across the cell/cluster. 

One can avoid this by choosing, probabilistically, one of the several possible weight modifications, such that the weight of each macroparticle remains unchanged on average (with account for probabilities of modifications); in each modification one or more macroparticles are assigned with zero weight and are therefore removed~\cite{gonoskov.arxiv.2020, muraviev.cpc.2021}.
This thinning approach, referred to as \textit{agnostic} resampling, preserves all the distribution functions (coordinate, momentum, angular, etc.) of particles on average and thus does not require any prior knowledge about the scales of features in these distributions.
In addition, an agnostic resampling algorithm that provides a way to preserve conservation laws, contributions to grid values and also central moments of particle distribution across the cell/cluster is proposed in \citet{gonoskov.arxiv.2020}.
The efficiency of agnostic resampling has been demonstrated in \citet{muraviev.cpc.2021}, where several algorithms for agnostic resampling have been compared with two merging algorithms.
The study showed that reducing the variance of weight values across the ensemble (provided by some versions of agnostic resampling) can further increase performance (as characterized by the smallness of numerical errors for a given number of macroparticles used).

\bib

%% file: dynamics.tex
\section{Charged particle dynamics}
\label{sec:Dynamics}

While the effect of radiation reaction (RR) on particle dynamics has been studied theoretically for decades, the development of large-scale numerical simulations has revealed complex new dynamics, which arise in strong-field environments from the competition between acceleration -- by the action of the external EM field -- and deceleration -- by RR.
In this section we classify and review different types of RR-dominated motion, which include dissipation (\cref{sec:Dissipation}), nondeterminism (\cref{sec:Nondeterminism}), trapping (\cref{sec:Trapping}), chaos (\cref{sec:Chaos}), cascades (\cref{sec:Cascades}) and depletion (\cref{sec:Depletion}).
How these effects could be leveraged in the creation of new radiation and particles sources will be discussed in \cref{sec:Applications}.
We begin with an overview of the interaction geometries which characterise laser-matter interactions.

\subsection{Interaction geometries}
\label{sec:Geometries}

    \begin{figure}
    \includegraphics[width=0.7\linewidth]{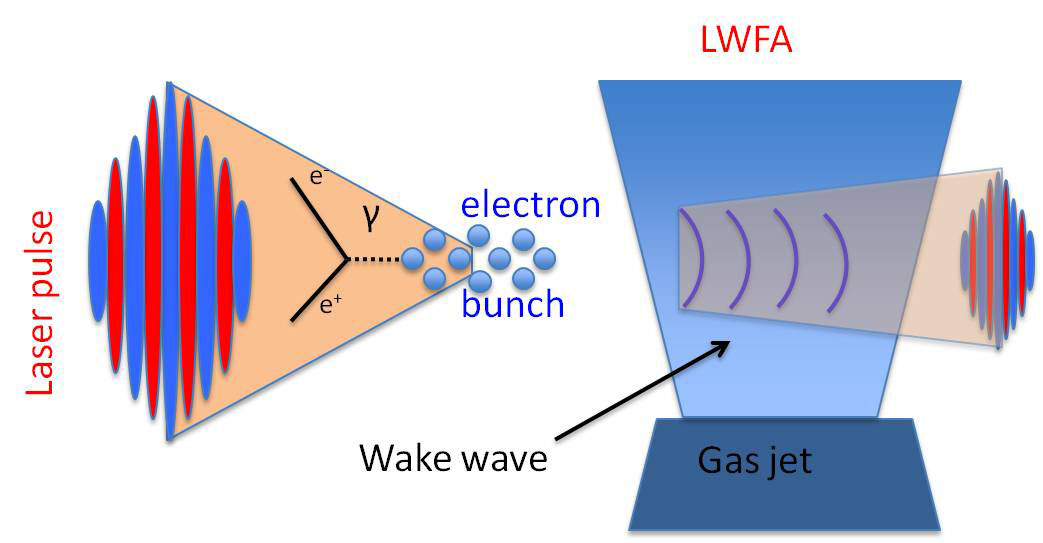}
    \includegraphics[width=0.7\linewidth]{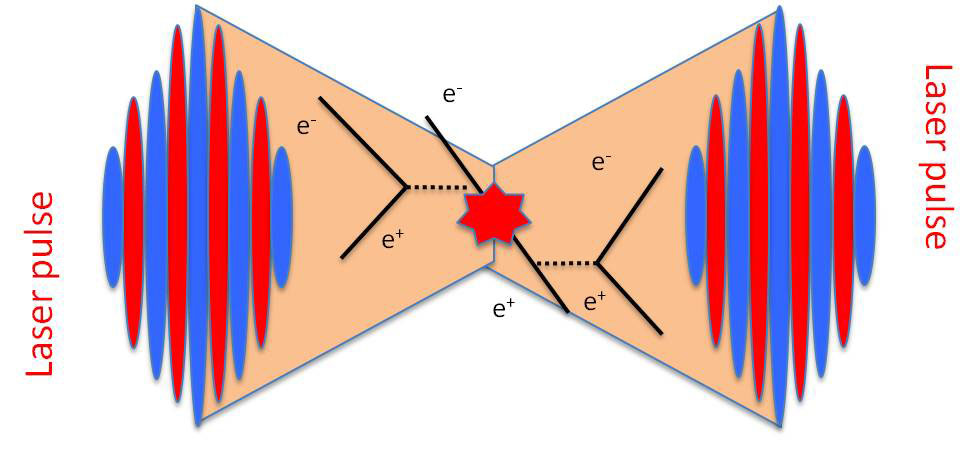}
    \caption{
        Principal experimental schemes aimed at the study of
        nonlinear QED:
        (upper panel) laser--electron-beam interactions (all optical setup).
        (lower panel) colliding laser pulses;
        Reproduced from \citet{bulanov.aip.2012}.}
    \label{fig:LaserSchemes}
    \end{figure}

Let us briefly consider the interaction of a single electron with a plane electromagnetic wave in terms of the parameter $\chi_e$, in order to demonstrate the importance of the interaction geometry.
The first general question for many applications is the energy coupling, i.e. the conversion of electromagnetic energy of laser pulse into the kinetic energy of particles. 
Although the collective effects are central for this question, it is instructive to start from considering the acceleration and deceleration of an individual electron in plane waves. 

Two principal interaction schemes can be used to categorize charged particle behaviour in strong EM waves.
The particle motion is dominated by either transverse or longitudinal motion (defined with respect to the laser propagation axis), with characteristic Lorentz factors $\gamma \sim a_0$ or $\gamma \gg a_0$, respectively.
[See \cref{fig:LaserSchemes}(a) and \cref{fig:LaserSchemes}(b).]
In both cases, the electron's radiation losses manifest themselves as a strong decelerating force.
While this loss completely dominates the particle motion in the longitudinal case, in the transverse case, the particle is continuously reaccelerated such that it can reach a kind of equilibrium.
Thus, in the longitudinal case the field serves as a `target' only, whereas in the transverse case it is not only a target but also an accelerator. 

Let us now characterise these two schemes from the point of view of the parameter $\chi_e$. For an electron propagating at angle $\theta_\kappa$ to {a plane EM wave with wavevector} $\kappa$:
\begin{equation}
    \chi_e=\gamma\frac{E_0}{E_{cr}}\left(1-\beta \cos\theta_\kappa\right),
\end{equation}
where $E_0 / \Ecrit \simeq 2.1 \times 10^{-4} I_0^{1/2} [10^{22}~\Wcmsqd]$ for peak intensity $I_0$, {assuming circular polarization (CP)}. Thus, the parameter $\chi_e$ is maximised for $\theta_k=\pi$, i.e., when an electron collides head-on with the EM wave\footnote{We note that $\chi_\gamma=(\hbar\omega/mc^2)(E_0/E_{cr})(1-\cos\theta_\kappa)$ is also maximised at $\theta_\kappa=\pi$. In the case of co-propagation, $\theta_\kappa=0$, $\chi_e\sim1/\gamma$, and $\chi_\gamma=0$. The latter is due to the vanishing of the terms corresponding to photon-photon scattering in the EM field Lagrangian.}.
For an electron propagating in a pure electric field $E_0 \vec{e}$ at an angle $\theta_e$ to the field vector $\vec{e}$:
\begin{equation}
    \chi_e=\gamma\frac{E_0}{E_{cr}}\sqrt{1-\beta^2\cos^2\theta_e}.
\end{equation}
The parameter $\chi_e$ is maximised when the electron momentum is perpendicular to the electric field.
This can be realized, for example, at the magnetic nodes of a standing wave formed by counterpropagating pulses with parallel linear polarization.

We now discuss the threshold intensities for when electron motion is dominated by classical radiation reaction or quantum effects for these two schemes of interaction. To do so, it is useful to introduce the dimensionless parameter 
    \begin{equation}
    \erad =
        \frac{4 \pi r_e}{3 \lambda} =
        \frac{2\alpha}{3} \frac{\hbar\omega_0}{m c^2},
    \label{eq:erad}
    \end{equation}
which appears when comparing the magnitude of the Lorentz and RR forces~\cite{bulanov.ppr.2004}.

\emph{Laser-electron beam.}---
The electron-laser collision geometry is characterized by the electron maintaining the longitudinal type of motion over the entire course of interaction (despite the radiation losses).
Such motion permits analytical calculations in both quantum (see \cref{eq:DiPiazza}) and classical cases. The related experimental scheme is shown for a head-on collision by the green lines in \cref{fig:Map}. 

We may define the threshold for the radiation reaction effects {to be} significant by requiring that the electron lose half of its energy after the interaction \cite{bulanov.aip.2012}\footnote{We note that there are other definitions of this threshold in the literature, e. g., in Ref. \cite{thomas.prx.2012} it is a loss of 10\% of electron energy per laser wavelength}:
\begin{equation}
    a_0>\acrr=\left(\erad \omega_0\tau_L\gamma_0\right)^{-1/2}
    \simeq 24 \lambda^{1/2}[\micron] / N^{1/2}
\end{equation}
where $\gamma_0$ is the initial Lorentz factor, $\tau_L = N \lambda/c$ is the laser pulse duration, $N$ is the equivalent number of cycles, and $\lambda$ is the wavelength.
For a 10-GeV electron beam colliding with a ten-cycle laser pulse ($\lambda = 0.8 \mu$m), the radiation reaction limit is $\acrr \approx 7$ and the corresponding peak intensity is $I_{crr} \approx 2\times 10^{20}$ W/cm$^2$.

The threshold for quantum effects is when a single photon of the emitted radiation has energy comparable to that of the electron. The characteristic photon energy in the case of laser-electron collision is $\hbar \omega_m\approx\hbar\omega a_0\gamma_e^2$, which corresponds to the condition $\chi_e\sim 1$. Thus, the threshold for quantum effects is reached when \cite{bulanov.aip.2012}
\begin{equation}
    a_0>\aqrr=\frac{2 \alpha}{3 \erad \gamma_0} \simeq 21\lambda [\micron].
\end{equation}
For a 10 GeV electron beam colliding with a 0.8 $\mu$m laser pulse, the quantum limit is {$\aqrr=17$} and the corresponding intensity is $I_{qrr}=8\times 10^{20}$ W/cm$^2$. Note that the hierarchy $\aqrr>\acrr$ is reversed if $\lambda<0.16$ $\mu$m.

\emph{Colliding lasers.}---
Here two counter-propagating laser beams are used to form
an electromagnetic standing wave.
Several types of standing waves are possible depending on the polarization of beams (see \cref{sec:Trapping}) and the setup can be extended to more than two beams, as well as to include a target or self-generated cascade (see \cref{sec:DipoleWaves}). Nevertheless, characteristic thresholds can be obtained from considering the case of two counterpropagating, {CP} beams, which form a standing wave with electric and magnetic field vectors that rotate in the transverse plane.
This case permits analytical calculations under both classical and quantum treatment of radiation reaction. At the electric-field antinode 
the electron orbits with large Lorentz factor as in the single-pulse case, but without the longitudinal drift. A qualitatively similar scenario is expected in the interaction of a single laser pulse with an overdense plasma, due to reflection of the laser from the critical surface~\cite{bell.prl.2008}: see, for example, \citet{kostyukov.pop.2016}.

The threshold for the dynamics to be dominated by radiation reaction is given by the $a_0$ at which the energy loss to radiation is balanced by the energy gain from the electric field. The energy loss is given by $\left. \frac{\rmd \Energy}{\rmd t} \right|_\text{rad} = \erad \omega_0 \gamma^2 p_\perp^2 / m$. The rate of energy gain from the external field is $\left. \frac{\rmd \Energy}{\rmd t} \right|_\text{ext} \simeq \omega_0 m c^2 a_0$, neglecting RR effects. Equating the two, and setting $\gamma \simeq p_\perp / (mc) = a_0$, we find that the classical radiation-dominated regime is reached for \cite{bulanov.nima.2011}
    \begin{equation}
    a_0 > \acrr = \erad^{-1/3} \simeq 4.4 \times 10^2 \lambda^{-1/3} [\micron].
    \label{eq:CRRThreshold}
    \end{equation}

The threshold for quantum effects is when a single photon of the emitted radiation has energy comparable to the electron, i.e. $\chi_e \simeq 1$.
For optical wavelengths, the radiation-dominated regime is reached before the quantum regime (see \cref{fig:a0_wavelength}), at an amplitude of~\cite{bulanov.nima.2011}:
    \begin{equation}
    a_0 > \aqrr =
        \frac{4 \alpha^2}{9 \erad} \simeq 2.0 \times 10^3 \lambda [\micron].
    \label{eq:QRRThreshold}
    \end{equation}
The hierarchy of $\aqrr > \acrr$ is reversed if $\lambda \lesssim 0.3~\micron$.
Shorter wavelengths therefore provide an opportunity to enter the quantum regime before large-scale radiation losses are expected: see \cref{fig:a0_wavelength}.
For an electron moving in the focus of two $0.8~\micron$ laser pulses, $\acrr$ corresponds to an intensity of $3.5 \times 10^{23}~\Wcmsqd$ and $\aqrr$ to an intensity of $5.5 \times 10^{24}~\Wcmsqd$.
Above this latter intensity, it would be important to note that the QED radiation power scales as $\chi_e^{2/3}$ rather than $\chi_e^2$.

\subsection{Energy exchange}
\label{sec:Dissipation}

From the theoretical considerations discussed in \cref{sec:ClassicalRR,sec:QuantumRR}, one might expect radiation reaction (RR) to manifest itself primarily as dissipation, i.e. energy loss.
However, the strength of RR effects depends on the strength and orientation of the external, accelerating electric and magnetic fields.
The interplay between the Lorentz and RR forces can lead to both deceleration and acceleration of charged particles, depending on the field geometry.

Consider the interaction of an electron with a pulsed, plane EM wave.
In this scenario the LL equation, \cref{eq:LandauLifshitz}, may be solved analytically~\cite{dipiazza.lmp.2008}.
Let the electric-field components in the $x$- and $y$- directions be $E_0 f'_1(\phi)$ and $E_0 f'_2(\phi)$ respectively, where $E_0 = a_0 m c \omega_0 / e$, {$f_{1,2}(\phi)$ are arbitrary functions of phase $\phi$,} and primes denote differentiation with respect to phase.
The lightfront momentum of the electron $p^- = \gamma m c - p_z$ (assuming the laser propagates towards positive $z$), as a function of phase is~\cite{dipiazza.lmp.2008,harvey.prd.2011}:
    \begin{align}
    \begin{split}
    p^-(\phi) &= \frac{p^-_0}{1 + \erad a_0^2 p^-_0 I(\phi) / (mc)},
    \\
    I(\phi) &= \int_{-\infty}^\phi \left[ f_1'(\psi)^2 + f_2'(\psi)^2 \right] \,\rmd\psi.
    \end{split}
    \label{eq:DiPiazza}
    \end{align}
The lightfront momentum always decreases under RR (it would be constant under the Lorentz force alone).
Whether the \emph{energy} decreases or increases, however, depends on its initial value.
If $\gamma$ is initially large, then $p^- \simeq 2 \gamma m c$ and RR manifests itself as an energy loss: measuring this energy loss, whether in the classical or quantum regimes, is an object of experimental investigation (see \cref{sec:Experiments}).
If the charge is initially at rest, however, then it is accelerated to an asymptotic Lorentz factor of $\gamma' \simeq 1 + \erad a_0^2 I(\phi)$.
In fact, there is a net energy transfer from the wave of~\cite{dipiazza.plb.2018}:
    \begin{multline}
    \Delta\Energy = \frac{m c^2 \erad a_0^4}{2}
    \\
    \times \int_{-\infty}^\infty \left[f_1'(\phi)^2 + f_2'(\phi)^2\right] \left[f_1(\phi)^2 + f_2(\phi)^2 + \frac{2}{a_0^2}\right] \,\rmd\phi.
    \end{multline}
In the absence of RR, this would be zero, in accordance with the Lawson-Woodward theorem~\cite{palmer.aip.1995}.

One can qualitatively understand the origin of this energy gain in the following way~\cite{zeldovich.spu.1975}:
in the interaction with the laser, the electron absorbs photons that carry energy $\hbar \omega$ and momentum $\hbar \omega / c$ along the direction of the pulse propagation ($\omega$ is the frequency of one of the photons).
This means that the electron receives the maximum possible longitudinal momentum, $\Delta \Energy/c$, per unit of absorbed energy $\Delta \Energy$.
However, an electron cannot absorb photons without emission, by energy-momentum conservation.
If the electron were to emit photons only along the longitudinal direction, there would be no accumulation of momentum.
However, the emission is not restricted to this direction: thus the energy loss is $\Delta\Energy$ but the loss of longitudinal momentum is $\Delta\Energy/c \cos\theta < \Delta\Energy/c$ (where $\theta$ is the emission angle).
As a consequence, the electron will have non-zero longitudinal momentum after it is overtaken by the pulse. 

Thus RR plays a significant role in altering the energy coupling between fields and particles.
This has dramatic consequences in field configurations in which more work can be done, such as standing waves.
To gain insight into energy coupling in the laser-laser geometry, we consider two indicative cases, theoretically: the rotating and oscillating electric field.

Let us start with the rotating electric field. 
Under the classical treatment of RR, one finds that the electron performs circular orbits, with constant Lorentz factor $\gamma$ given by the root of the following equation~\cite{zeldovich.spu.1975,bulanov.pre.2011}:
    \begin{equation}
    a_0^2 = (\gamma^2 - 1) (1 + \erad \gamma^6).
    \label{eq:Zeldovich}
    \end{equation}
Thus the Lorentz and RR forces are in balance and the energy transferred from field to particle is seamlessly lost as radiation.\footnote{
{Analytical solutions are also available for the case of a rotating electric field, which has a static magnetic field superimposed along its axis of rotation~\cite{nakamura.pre.2020}.}}
\Cref{eq:Zeldovich} has limits $\gamma \simeq a_0$, if $a_0 \ll \acrr$, and $\gamma \simeq (a_0 / \erad)^{1/4}$, if $a_0 \gg \acrr$, where $\acrr$ is given by \cref{eq:CRRThreshold}.
The weakened scaling at high intensity is caused by radiation losses, which decrease the angle between the instantaneous momentum and the electric field vector:
    \begin{equation}
    \tan \varphi = \frac{1}{\erad \gamma^3}.
    \label{eq:ZeldovichAngle}
    \end{equation}
The quantum parameter $\chi_e$ goes from $a_0^2/\acrit $ at $a_0 \ll \acrr$ to $(3/2\alpha)^{1/2 }a_0^{1/2} \acrit^{-1/2}$ at $a_0 \gg \acrr$ (see, e.g., \citet{chang.pre.2015, kostyukov.pop.2016, bashinov.pra.2017,bulanov.nima.2011,bulanov.pre.2011}).
Thus classical radiation losses delay the onset of the quantum regime~\cite{popruzhenko.njp.2019}.
An interesting consequence of \cref{eq:ZeldovichAngle} is that, at large $a_0$, the electron motion is parallel to the electric field and therefore optimized to absorb incident radiation \cite{zhang.njp.2015}.
It should be noted that orbits at the magnetic node are unstable: charged particles will tend to migrate towards the electric nodes, where the magnetic field is strongest~\cite{bashinov.pra.2017} (see \cref{sec:Trapping}).
In the limit of $\chi \gg 1$, quantum corrections alter the scaling of radiation losses, from $\sim\chi^2$ to $\sim \chi^{2/3}$, which changes the trends to $\gamma \approx 84 \acrit^{-1/4} a_0^{3/4}$, $\chi_e \approx 7.1 \times 10^3 \acrit^{-3/2} a_0^{3/2}$, $\varphi \approx 84 \acrit^{-1/4} a_0^{-1/4}$ (see, e.g., \cite{chang.pre.2015, kostyukov.pop.2016, bashinov.pra.2017}).

The second indicative case is the linearly polarized {(LP)} standing wave. This wave has the property that the electric field at the magnetic node oscillates in magnitude, but not direction.
As the electron is driven parallel to $\vec{E}$, radiation losses are suppressed and the energy can reach the characteristic quiver value of $\gamma = a_0$. 
Reduced production of high-energy photons and the consequent suppression of cascade development (see \cref{sec:Cascades}) means that this configuration has been proposed as means to generate critically strong electromagnetic fields~\cite{bulanov.prl.2010s}.
The dynamics are however, more complicated, due to migration around this node.
While it might be expected to be unstable, as in the CP case, there is the surprising result that, in the radiation-dominated regime~\cite{bulanov.ppr.2004}, electrons can become trapped here~\cite{gonoskov.prl.2014}.
We discuss this `anomalous' trapping in \cref{sec:Trapping}.

    \begin{figure}
	\centering
	\includegraphics[width=0.7\linewidth]{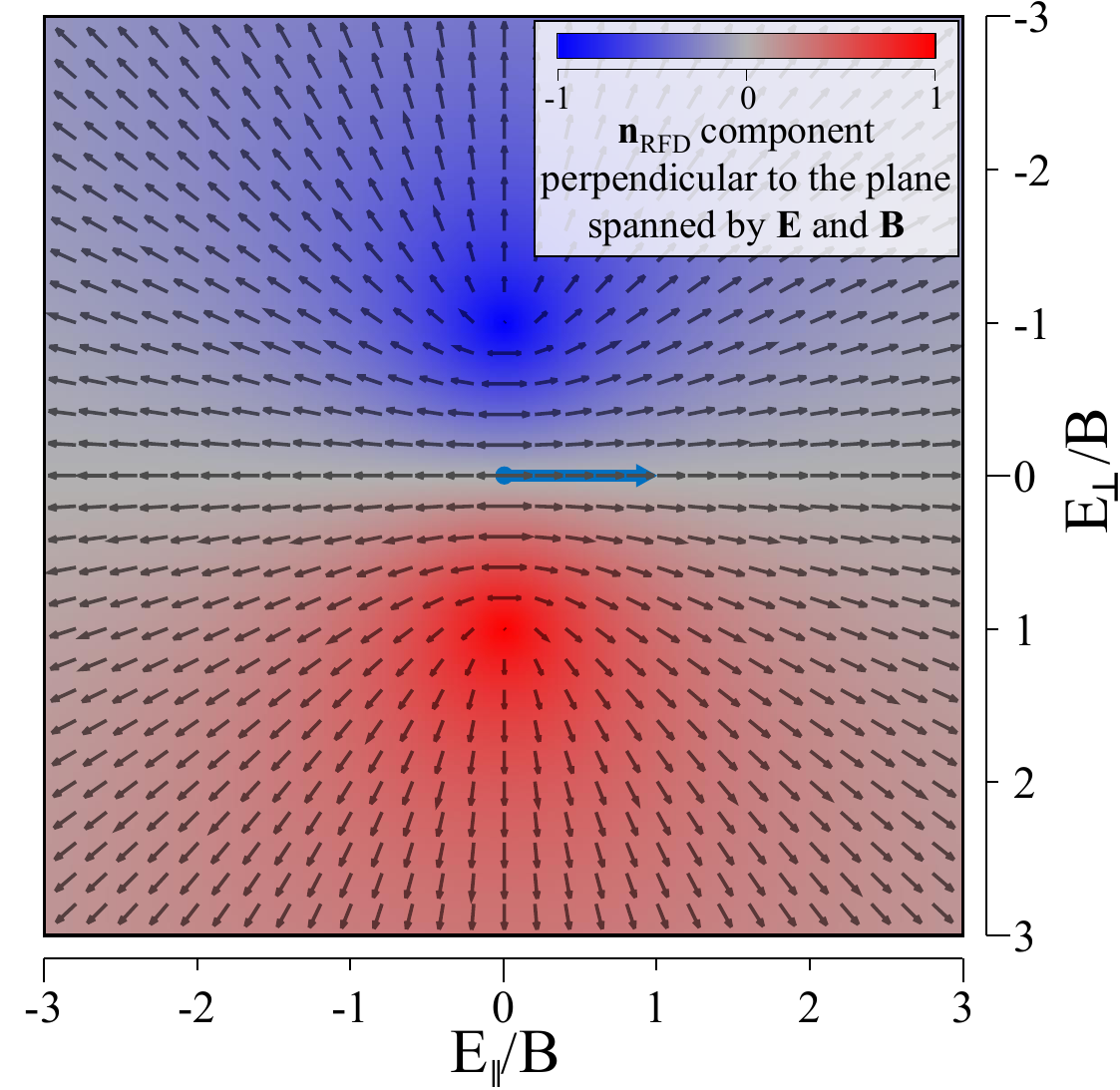}
	\caption{
	    The radiation-free direction $\vec{n}^{+}_\text{RFD}$, as a function of the end point of the electric field vector $\vec{E}$, given a fixed magnetic field vector $\vec{B}$ (blue arrow).
	    Arrows denote the projection of $\vec{n}^{+}_\text{RFD}$ onto the plane spanned by $\vec{E}$ and $\vec{B}$ and the color scale the magnitude of the perpendicular component.
	    Reproduced from \citet{gonoskov.pop.2018}.
    }
	\label{rfd}
    \end{figure}

From the above two cases, we see that energy exchange has different scalings for CP ($\gamma \sim a_0^{3/4}$) and LP ($\gamma \sim a_0$) standing waves.
To understand this observation, we turn to the discussion of the RR's generic influence on the dynamics of charges.
Consider that, for ultrarelativistic particles, the magnitude of radiation losses is determined by the magnitude of the transverse acceleration.
In a general electromagnetic field, with electric and magnetic components $\vec{E}$ and $\vec{B}$, there is always one (unit) direction for which the transverse acceleration vanishes~\cite{gonoskov.pop.2018}:
    \begin{equation}
    \vec{n}_\text{RFD}^{\pm} =
        \frac{\sqrt{ u (1 - u) }  (\vec{E} \times \vec{B}) \pm \left[(1 - u) B \vec{E} + u \left(\vec{E} \cdot \vec{B}\right) \hat{\vec{B}}\right]}{\left[ E^2 B^2  - u \left(\textbf{E} \times \textbf{B} \right)^2 \right]^{1/2}},
    \label{eq:rfd}
    \end{equation}
where
    \begin{align}
    u &=
        \frac{2 c^2 B^2}{E^2 + c^2 B^2}
        \frac{1 - \sqrt{1 - w}}{w},
    &
    w&=
        \frac{4 (\vec{E}\times c\vec{B})^2}{(E^2+ c^2 B^2)^2}.
    \end{align}
Here the superscript denotes the sign of the charge.
This vector is named the \emph{radiation-free direction} (RFD) by virtue of the fact that radiation is suppressed when the particle has momentum parallel to $\vec{n}_\text{RFD}$.
If this is not the case, momentum in any other direction is rapidly exhausted by radiation emission; reacceleration by the external field then brings the particle momentum towards the RFD, which is the only stable point~\cite{gonoskov.pop.2018}.
\Cref{rfd} shows that $\vec{E}$ has a non-zero projection on the RFD, making it the only stable attractive point, in all cases except $\vec{E}\cdot\vec{B} = 0$ and $E < c B$.
\Citet{samsonov.pra.2018} analyze particle dynamics in the strong-radiation-damping regime and find the velocity is indeed determined by the local and instant EM field.
This causes trajectories to be periodic in a wide variety of oscillating EM fields.
It is also shown that the angle between the RFD and the Poynting vector is always less than $\pi/2$, leading to particles being attracted to regions of EM-field absorption~\cite{samsonov.pra.2018}. The tendency of charges to align the motion along the local RFD has been rigorously demonstrated in exact analytical solutions for longitudinal and transverse waves~\cite{ekman.njp.2021}: both cases are special in the context of the arguments given above.

The tendency to the RFD makes it clear that the rotating and oscillating electric fields {represent} limiting cases. In the former, the particles are constantly chasing the RFD, which points along the electric-field vector (see \cref{rfd}), while the deviation angle (\cref{eq:ZeldovichAngle}) remains relatively large. This facilitates energy loss to radiation and leads to a weaker energy scaling $\gamma \sim a_0^{3/4}$ (at $\chi \gg 1$).
In the latter, the RFD flips direction twice per wave period so that charges can quickly adjust their motion to move predominantly along the RFD. This minimises energy losses and so the energy scaling is stronger: $\gamma \sim a_0$.
As these represent worst- and best-case scenarios, we may conclude that the energy coupling for various types of particle dynamics in the laser-laser geometry is likely to be found between these two limiting cases.
This is highlighted by the shaded red region in \cref{fig:Map}.

In both the classical and quantum regimes, the magnitude of radiation reaction is controlled by the transverse acceleration.
While the dynamics are therefore qualitatively similar, the stochastic, discrete nature of losses in the latter leads to unique dynamics which we describe in the following section.

\subsection{Stochasticity, straggling and quenching}
\label{sec:Nondeterminism}

In the quantum regime, $\chi \gtrsim 1$, the typical energy of the emitted photons is comparable to the kinetic energy of the emitting electrons and positrons.
As the photon formation length is much shorter than the characteristic time scale of the particles dynamics (see \cref{fig:Map}) for $a_0 \gg 1$, RR in the quantum regime can be viewed as a sequence of discrete, instantaneous, recoils that occur along the particle trajectory at probabilistically determined intervals.
This has effects on particle motion not captured by a classical treatment, in which RR manifests itself as a continuous force.

As discussed in \cref{sec:QuantumRR}, the probabilistic change of momentum can be described as a stochastic force that can be treated with the Fokker-Planck equation: the drift term describes the rate of radiation losses, whereas the diffusion term describes the discreetness and probabilistic nature of the losses.
The energy loss rate increases with the energy of particles and therefore -- on average --- the most energetic particles lose energy more quickly than others.
This causes contraction of the particle phase space~\cite{tamburini.nima.2011, yoffe.njp.2015, yoffe.nima.2016} and in particular, to the shrinking of energy distribution when an electron beam passes through an intense laser pulse.
However, the diffusion term, which governs stochastic broadening, can dominate for sufficiently high values of $\chi$.
This dominance has been analyzed and numerically demonstrated for the case of a beam of 1 GeV electron passing through a 10-cycle laser pulse of peak intensity $2 \times 10^{22}$~W/cm$^2$ (yielding $\chi = 0.8$)~ \cite{neitz.prl.2013}.
The change of the energy variance can be used as a direct measure of the quantum stochasticity of RR already at intensities of $10^{21}$~W/cm$^2$~\cite{ridgers.jpp.2017}.
If the electrons continue losing energy for a sufficiently long time, their reduced energies start to yield lower values of $\chi$ and the broadening term no longer dominates.
The maximum energy width can be calculated for the given initial beam parameters~\cite{vranic.njp.2016}.
The analysis of this transition and of the applicability of the Fokker-Plank approach, the classical RR treatment and the discrete, stochastic treatment is carried out in \citet{niel.pre.2018}.
In other field geometries, such as linearly polarized standing waves, the Markov chain formalism can be applied to analyse when and how the classical contraction of phase space volume breaks down due to the onset of quantum stochasticity~\cite{bashinov.pre.2015}.

Another notable consequence of the probabilistic nature of RR is the following possibility, for an electron entering a region of strong field that ramps up over a finite duration (or length).
Instead of losing its kinetic energy continuously, as would occur classically, there is a chance that electron penetrates the strong-field region without having radiated.
The quantum parameter of such electrons is therefore larger than expected and, subsequently, there is a greater probability of emitting very high energy photons.
This effect, referred to as \emph{straggling}~\cite{shen.prl.1972}, causes a significant enhancement of the emission of sub-GeV photons by GeV electrons interacting with a laser pulse having intensity of $\sim 10^{22}$~W/cm$^2$~\cite{blackburn.prl.2014}. In addition, straggling particles, which have higher energy, are less susceptible to deflection by gradients in the laser fields: this can result in the narrowing of the emission angle~\cite{harvey.pra.2016}.

In more general terms, the straggling effect implies the appearance of electrons in regions of phase space that are not accessible for trajectories governed by continuous RR forces (either in classical LL or quantum corrected form). This can be used to enhance the production of pairs in various field configurations~\cite{duclous.ppcf.2011},
or to distinguish classical from quantum RR~\cite{geng.cp.2019}.

\emph{In extremis}, an energetic electron can pass through a region of strong field without having radiated even a single photon.
This effect, referred to as quantum \emph{quenching} of radiation losses, can be observable at intensities $\sim 10^{22}$~W/cm$^2$ with 10-cycle pulses (few \% penetrate) and becomes crucial with sub-two-cycle pulses ($> 30$~\% penetrate)~\cite{harvey.prl.2017}. This suggests that highly compressed laser radiation can deflect energetic particles without synchrotron radiation emission, an otherwise inevitable loss mechanism.

\subsection{Particle trapping}
\label{sec:Trapping}

Unlike frictional losses due to, e.g., viscosity, the energy dissipation caused by RR is critically dependent on the local orientation of the electric and magnetic fields.
The strong directional dependence of RR can manifest itself in particle trapping, which we discuss in terms of two main electromagnetic field configurations: travelling and standing waves.

Even in the absence of RR, relativistic effects can cause particles to be `trapped' within travelling electromagnetic waves.
Consider the dynamics of an electron that starts at rest.
If the wave has amplitude $a_0 \gtrsim 1${, the} electric field causes relativistic motion of an electron in the transverse direction, while the magnetic field converts this oscillation into a secular longitudinal drift.
The particle is generally carried in the direction of the wave propagation.
The higher amplitude of the pulse, the closer the longitudinal velocity gets to the speed of light and the longer the electron co-propagates with the pulse.

    \begin{figure}
	\includegraphics[width=\linewidth]{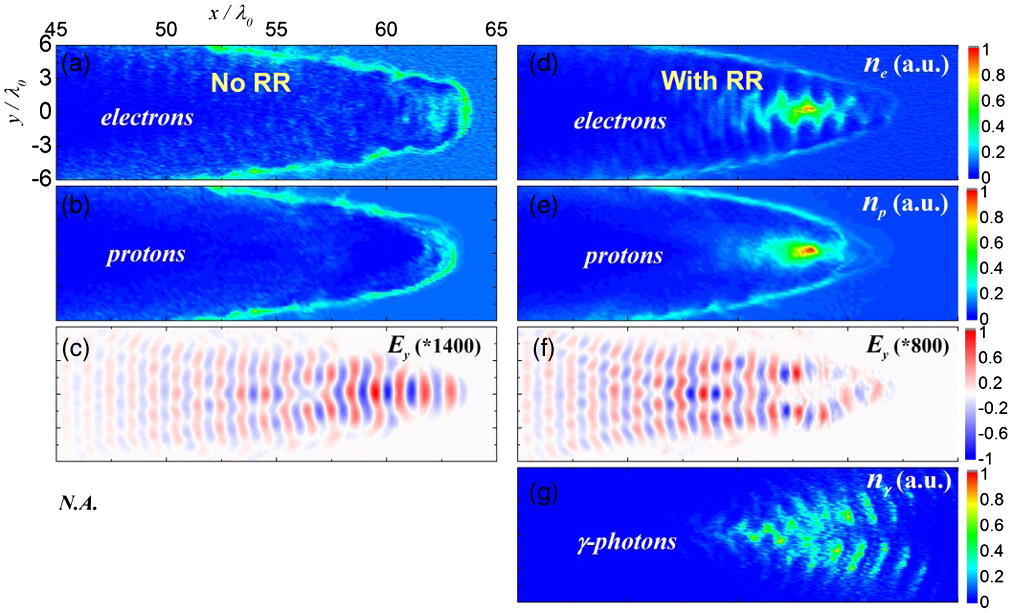}
	\caption{
	    Radiation-reaction trapping in simulations of laser-wakefield acceleration.
	    Panels show the densities of electrons (a, d), protons (b, e) and $\gamma$-photons (g), as well as the transverse component of the electric field (c, f) obtained using 3D PIC simulations with (right column) and without (left column) radiation reaction.
	    Reproduced from \citet{ji.prl.2014}.
    }
	\label{fig:rrt}
    \end{figure}

This can be enhanced by the effect of RR.
\emph{RR-induced trapping} (RRT) is observed in the simulations of laser wakefield acceleration (LWFA)~ \cite{ji.prl.2014}.
In the {considered} `bubble' regime of LWFA, an intense laser pulse propagates through a low-density plasma, displacing the plasma electrons and creating a region of uncompensated positive charge: the charge-separation electric field within this `bubble' then accelerates a trailing population of electrons~\cite{esarey.rmp.2009}.
The displacement of plasma electrons is caused by the ponderomotive force of the laser~\cite{gaponov.jetp.1958,quesnel.pre.1998}, which itself arises from the transverse gradient in intensity of the focused laser pulse.
By comparing simulations with and without radiation losses, as shown in \cref{fig:rrt}, one observes a distinct difference: in the former case, a population of electrons is trapped \emph{inside} the laser pulse and copropagates with it over a long distance.
Simulations show that for a laser pulse with peak intensity of $4 \times 10^{24}~\Wcmsqd$, a population with charge greater than one nC can be carried over a few hundreds of micrometers~\cite{wallin.jpp.2017}.

To explain these observations, note that the RR force is directed antiparallel to the electron's motion in the ultrarelativistic regime.
It does not therefore deflect the electron \emph{directly}; it does, however, reduce the electron $\gamma$ such that the deflection by the laser fields, i.e. the Lorentz force, is enhanced.
One may view such a situation as a manifestation of the electron's tendency to align itself with the local RFD, which points along the laser propagation axis in this case: see \cref{eq:rfd}.

{It is natural to ask whether a trapped electron can gain energy while co-moving with the laser pulse.} In the absence of RR, the oscillatory nature of the fields in an EM wave means that periods of acceleration and deceleration follow each other in sequence, almost compensating for each other (up to a longitudinal drift).
While this would normally preclude net gain of energy, it can be overcome in particular field configurations.
{One possible} idea is to deflect the electron out of the strong-field region, during the deceleration phase, and then return it in time for the next acceleration phase.
This can be realized by laser propagation in a dense plasma channel, where a self-generated, azimuthal magnetic field provides the suitable deflection.
{In this case} RR is essential in facilitating the coalescence of electrons around the laser propagation axis and their optimal phasing, leading to enhanced, directed radiation emission and acceleration of electrons~\cite{zhu.njp.2015,stark.prl.2016, zhu.ncomm.2016,vranic.ppcf.2018,gong.sr.2019}.

    \begin{figure}
	\centering
	\includegraphics[width=0.8\linewidth]{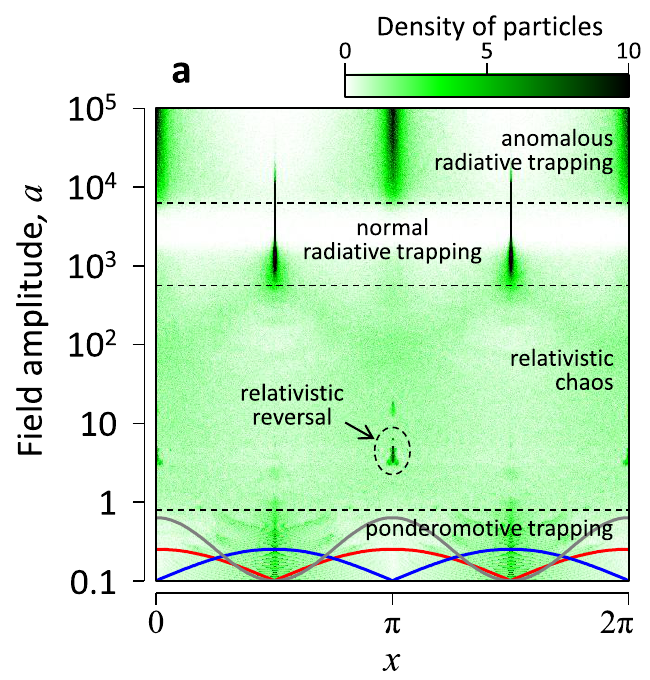}
	\caption{
	    The electron number density as a function of the wave amplitude (here called $a = a_0$) in a linearly polarized, standing electromagnetic wave.
	    As $a$ increases, the ponderomotive, normal and anomalous radiative trapping regimes are accessed.
	    {The curves in the lower part indicate the ponderomotive potential (gray), as well as the electric (red) and magnetic (blue) field amplitudes, in arbitrary units.}
	    Reproduced from \citet{gonoskov.prl.2014}.
    }
	\label{art1}
    \end{figure}

A standing EM wave can be realized in a variety of ways: from the collision of {two counterpropagating} laser pulses to the creation of a dipole wave~\cite{gonoskov.pra.2012}.
We begin with the case of a standing wave formed by two, linearly polarized lasers which have parallel electric-field vectors.
The electron dynamics depend crucially on the field amplitude $a_0$.
Consider the evolution of the density distribution of a population of electrons that are initially uniformly distributed along the collision axis.
After a transient phase that lasts several laser periods, the distribution stabilizes (for a certain phase): the final configuration is shown as a function of $a_0$ in \cref{art1}.

For nonrelativistic amplitudes, $a_0 < 1$, the electrons are trapped around minima in the laser's ponderomotive potential, which are located at the electric-field nodes~\cite{gaponov.jetp.1958}.
In the range $1 < a \lesssim 500$, where radiation losses are not sufficiently large to affect particle dynamics significantly, the dynamics are stochastic~\cite{bauer.prl.1995,lehmann.pre.2012} and, with the exception of relativistic reversal~\cite{kaplan.prl.2005,dodin.pre.2008}, trapping effects are not evident.

Once $a_0 \gtrsim 500$, RR induces two kinds of trapping phenomena.
The first manifests itself as the localization of electrons in the vicinity of electric-field nodes~\cite{kirk.ppcf.2009}, as would be expected in ponderomotive trapping (see \cref{art1}).
{This is therefore referred to as} \emph{normal radiative trapping}.
The cause is that radiation losses quickly drain the energy of electrons and make them gyrate, without any significant drift. Since both the radiation losses and the gyration are enhanced by a magnetic field, this happens in a more prominent form near the electric field nodes where the magnetic field is maximized.

    \begin{figure}
	\centering
	\includegraphics[width=0.8\linewidth]{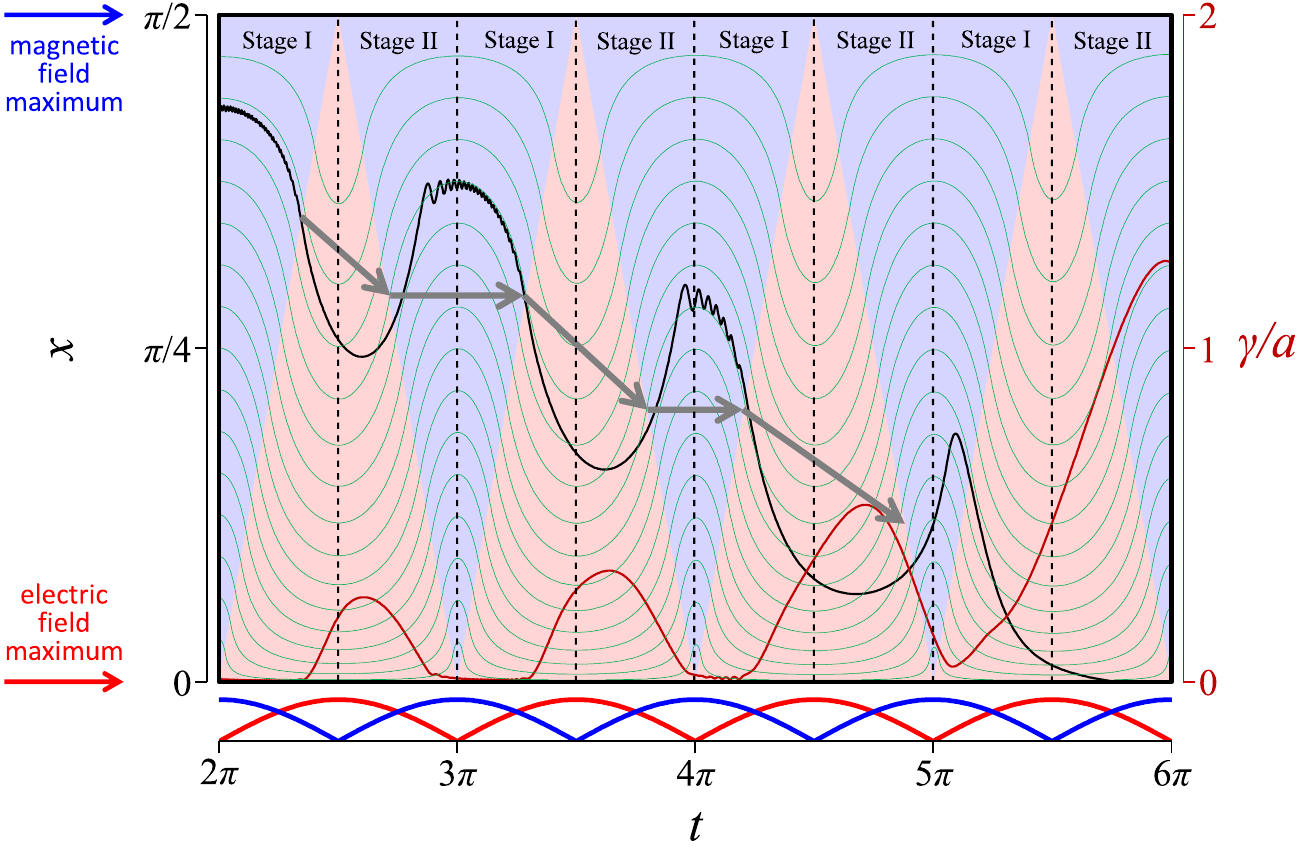}
	\caption{
	    {Explanation of ART through the net migration of electrons towards electric field maxima. The particle trajectory (black curve), its gamma factor (red curve, axis to the right), asymptotic trajectories (green curves) are shown together with the regions of electric (red) and magnetic (blue) field dominance (see details in the text).}
	    Reproduced from \citet{gonoskov.prl.2014}.
	}
	\label{art2}
    \end{figure}

At even higher field strengths, $a_0 \gtrsim 5000$, \cref{art1} shows that the trapping takes a completely different form.
Electrons are instead confined at electric-field \emph{antinodes}, where the ponderomotive potential is maximized.
{This counterintuitive phenomenon is called} \emph{anomalous radiative trapping} (ART)~\cite{gonoskov.prl.2014}.

The explanation of the ART phenomenon involves several logical steps, which we outline here: see details in \citet{gonoskov.prl.2014}.
First, note that in strong fields, RR starts to act as a strong viscosity, suppressing particles’ inertia such that they oscillate predominantly along, and in phase with, the oscillating electric field.
The motion along the $x$-axis (or the collision axis, i.e. perpendicular to both $\vec{E}$ and $\vec{B}$) is characterised by two limiting cases (in which the particle is ultrarelativistic):
gyration with drift velocity $\vec{E} \times \vec{B} / B^2$, if $cB > E$ {(blue regions in \cref{art2})}; or motion along the RFD with projected velocity $c \vec{E} \times \vec{B} / E^2$, if $cB < E$ {(red regions in \cref{art2})}.
As $E$ and $c B$ oscillate, the particle motion switches between these cases.
From stitching these phases together (which is shown by the green lines in \cref{art2}), one might expect that the particle experiences no net migration.
The key is that the particle only approaches these limiting cases and the approach is asymmetric:
the particles have higher energy (during stage II, as labelled in \cref{art2}) after the electric field peaks than after the magnetic field peaks, when radiation losses are maximized.
Thus, during stage II, they are less affected by the magnetic field, which deflects them away from the electric-field antinode.
This causes a net migration of particles towards those electric field antinodes (see black curve in \cref{art2}). 

Even though RR is discrete in nature at the relevant amplitudes, the basic idea still holds.
Furthermore, the threshold to achieve ART is significantly lowered (from 60 to 8 PW) by the onset of cascades, which are triggered at the center of an electric-dipole wave~\cite{gonoskov.prx.2017}.
This provides exactly a 3D standing-wave-like structure, with the antinode at the center.
See \cref{sec:DipoleWaves} for details.

Trapping along the \emph{transverse} direction (along $\textbf{E}$ and perpendicular to the $x$-axis) has been analysed in \citet{esirkepov.pla.2017}. They consider a dissipative dynamical system, $\ddot x + k (F) \dot x=F$, driven by a periodic force, $F$, and a strong nonlinear friction, $k (F) \dot x$. Sufficiently strong friction makes the system drift antiparallel to the spatial gradient of the driving force, causing apparently paradoxical stabilization near local maxima of $F$.

In a circularly polarised standing wave, by contrast, ART does not occur.
Normal radiative trapping does:
numerical simulations show that the radiation losses cause particles to migrate towards field nodes for all $a_0 \gtrsim 70$~\cite{bashinov.pra.2017}.

The tendency of particles to align themselves along the local radiation-free direction may also be viewed from the perspective of how energy is transferred from the external field~\citet{samsonov.pra.2018}.
As the RFD points predominantly in the direction of the local Poynting vector, regions of EM energy absorption cause asymmetry and eventually attract particles: as such, it is dubbed \emph{absorption-induced trapping}~\cite{samsonov.pra.2018}.

It is instructive to compare the trapping cases we have discussed, where particles begin inside the region of strong field, with the case that they begin outside it.
In this case, one would expect ponderomotive expulsion of charged particles from the region of highest intensity.
However, in the RR-dominated regime, this expulsion can be suppressed, as shown in \citet{fedotov.pra.2014}, where the following field geometry is studied: the electric field rotates around an axis, remaining perpendicular to it; the magnetic field oscillates along the axis; and the amplitude of fields decreases with increasing transverse distance.

We conclude this section by observing that trapping phenomena are generally robust against discrete, probabilistic effects that are a feature of RR in the quantum regime.
This is because at the typical wavelength of high-intensity lasers, $\lambda \simeq 1~\micron$, laser-accelerated electrons enter the radiation-dominated regime before quantum effects become essential.
Nevertheless, for longer wavelengths the radiation-dominated regime can be reached when discreteness does not play an important role (see \cref{fig:a0_wavelength}). In this case, the continuous RR force enables a variety of trapping-like phenomena, including large longitudinal oscillations in CP standing waves (called NRT$^+$ in \cite{bashinov.pra.2017}) and attractors in LP standing waves~\cite{esirkepov.pla.2015, bulanov.jpp.2017, esirkepov.pla.2017}, which we discuss in the next section.

\subsection{Chaos}
\label{sec:Chaos}

The fact that there is only an oscillating electric field, and no magnetic field component, at the nodes of a standing wave provides a great simplification from the theoretical point of view.
However, charged particles originating at other phases of the standing wave experience much more complicated field configurations.
Since both the radiation reaction and Lorentz forces depend on the particle’s momentum, the strength of the EM field acting on a particle and on their mutual orientation, nontrivial particle dynamics arise even in the simplest case of two counterpropagating plane waves~\cite{mendonca.pra.1983, bauer.prl.1995, sheng.prl.2002}, demonstrating stochastic heating.
The inclusion of the radiation reaction makes the particle dynamics even richer for high field intensities, including random walk trajectories, Lévy flights, limit circles, attractors and regular patterns~\cite{lehmann.pre.2012, gonoskov.prl.2014, bashinov.pre.2015, esirkepov.pla.2015, jirka.pre.2016, kirk.ppcf.2016, bulanov.jpp.2017, esirkepov.pla.2017}.

   \begin{figure}
 	\centering
 	\includegraphics[width=1\linewidth]{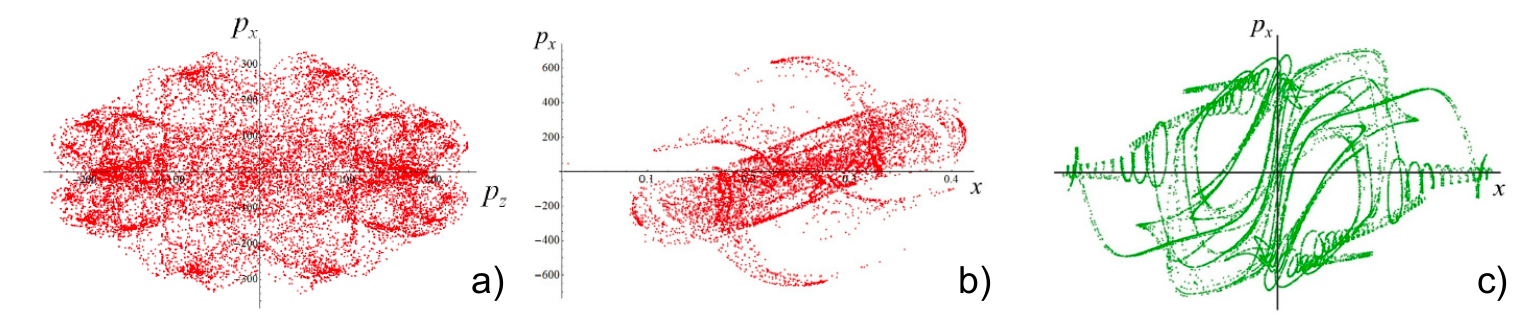}
 	\caption{
         Poincaré sections showing particle coordinates in the space of position $x$ and momentum $p_x$, at time intervals equal to the period of the driving field. a) $a_0=617$, $\varepsilon_{rad}=1.2\times 10^{-8}$. b) $a_0=778$, $\varepsilon_{rad}=6\times 10^{-9}$. c) $a_0=1996$, $\varepsilon_{rad}=1.2\times 10^{-9}$
         Reproduced from \citet{bulanov.jpp.2017}.
     }
 	\label{poincare}
   \end{figure}

It was first noted in \citet{lehmann.pre.2012} that as the strength of the EM field increases, stochastic heating is reduced and attractors begin to appear.
This effect was mainly attributed to phase-space contraction~\cite{tamburini.nima.2011}.
This transition from chaos to regular dynamics can be illustrated by Poincaré sections, which show the particle coordinates in position and momentum space $(x,p_x)$ at intervals of time equal to the period of the driving field.
\Cref{poincare} shows such sections for three different values of $a_0$.
Analysis of the charged particle motion in terms of Lyapunov exponents reveals the existence of attractors, including strange attractors, in the phase space~\cite{esirkepov.pla.2015,bashinov.pre.2015}.
This is corroborated by the positivity of the maximum Lyapunov exponent
    \begin{equation}
    \Lambda=\lim_{t \to \infty} \lim_{|\delta x_0| \to \infty} \frac{1}{t} \ln\frac{|\delta x(t)|}{|\delta x_0|},
    \end{equation}
where $\delta x_0$ is the initial distance between two particle trajectories \cite{eckmann.rmp.1985}.

\begin{figure}
	\centering
	\includegraphics[width=1.0\linewidth]{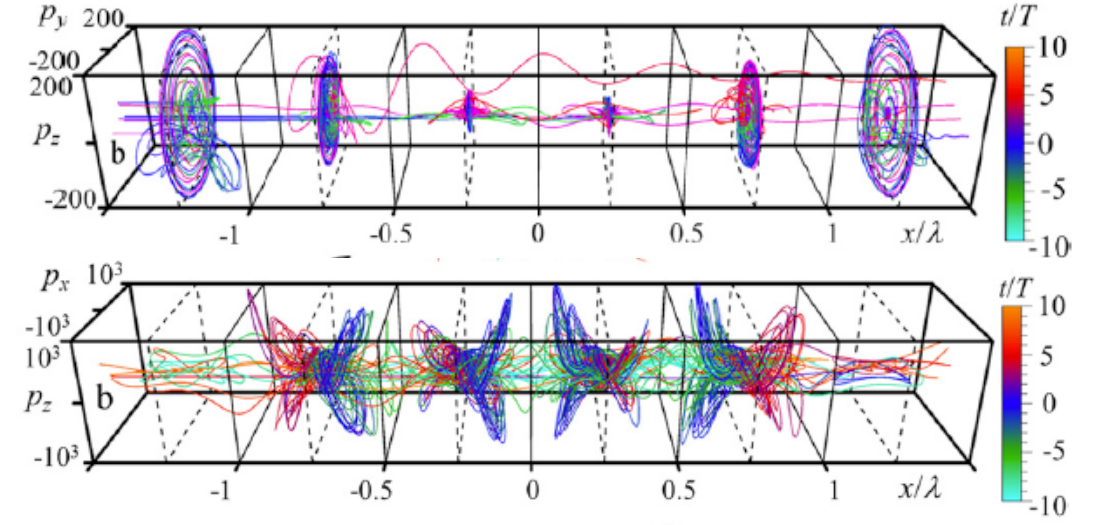}
    \caption{
        Electron trajectories in the standing wave formed by two colliding circularly (upper frame) and linearly (lower frame) polarized laser pulses propagating along the $x$-axis, with $I=1.37 \times 10^{24}$ W/cm$^2$, $\lambda =1~\mu$m, duration 33 fs and focal spot 3 $\mu$m.
		Modelled with semiclassical radiation reaction, \cref{eq:ModifiedLandauLifshitz}.
		Reproduced from \citet{esirkepov.pla.2015}.
	}
	\label{attractors_cl}
    \end{figure}

\begin{figure}
	\centering
	\includegraphics[width=1.0\linewidth]{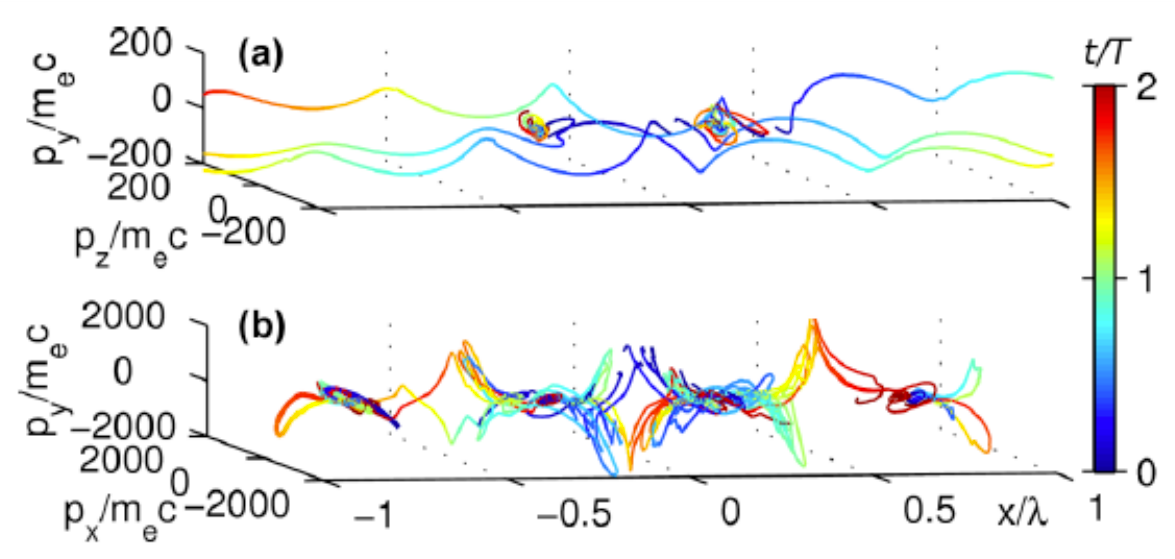}
		\caption{Electron trajectories in the standing wave formed by two colliding circularly (upper frame) and linearly (lower frame) polarized laser pulses propagating along the $x$-axis, with $I=1.11 \times 10^{24}$ W/cm$^2$, $\lambda =1~\mu$m, duration 30 fs and focal spot 3 $\mu$m.
		Modelled with quantum (stochastic) radiation reaction.
		Reproduced from \citet{jirka.pre.2016}.
	}
	\label{attractors_q}
    \end{figure}

Particle trajectories in two colliding EM waves have mainly been studied in the framework of classical physics, using the Landau-Lifshitz equation, \cref{eq:LandauLifshitz}.
One might expect that quantum effects would prevent the formation of regular structures in the particle phase space.
However, \citet{jirka.pre.2016} have shown that particle trajectories still demonstrate strange attractors at electric-field nodes and loops at antinodes: see \cref{attractors_cl} and \cref{attractors_q}.
The stochastic nature of photon emission means that trajectories are no longer smooth, but fluctuate around that predicted by classical equations of motion.
Nevertheless, important questions concerning the connection between radiative trapping, Lévy flights, limit circles and attractors remain to be addressed.

\subsection{Pair cascades}
\label{sec:Cascades}

When high-energy electrons, positrons, or photons interact with a strong EM field, their energy is converted into secondary particles in a process commonly referred to as a cascade.
While the cascade can take different forms, depending on the initial conditions and field configurations, there are two basic cases: the \emph{shower} and the \emph{avalanche}.
In a shower-type cascade, it is the initial energy of particles (or photons) entering the strong field region that drives the production of new particles, i.e., the energy, summed over all particles, remains roughly constant as the cascade develops.
In the avalanche-type cascade, by contrast, the energy of the participating particles is continuously replenished through their repeated acceleration by the EM field.
In this case, the total particle energy grows over time.

The avalanche-type cascade was first identified by \citet{bell.prl.2008}, where it was realized that, although a single laser pulse cannot be both accelerator and target for electrons and positrons, two colliding laser pulses can.
Electrons accelerated by the rotating electric field at an antinode emit high-energy photons via multiphoton (nonlinear) Compton scattering (see \cref{sec:ComptonBreitWheeler}).
This energy loss is compensated by reacceleration, such that a steady-state energy is maintained~\cite{zeldovich.spu.1975,bell.prl.2008,bulanov.pre.2011}.
The high-energy photons, which are embedded in a strong background EM field, can subsequently decay into electron-positron pairs via the multiphoton (nonlinear) Breit-Wheeler process.
These pairs are similarly accelerated by the EM field and, as a result, generate additional photons and pairs.
As the avalanche cascade develops, the laser energy is transformed into the energy of electrons, positrons, and photons.

Prolific production of secondary particles raises the question of
energy balance.
It was noted in \citet{bell.prl.2008} that {creating and accelerating} a certain number of electron-positron pairs per joule of laser energy would itself lead to the complete depletion of the laser.
{This question was further investigated in \citet{fedotov.prl.2010}, where this depletion was argued to 
prevent reaching the critical field with focused laser pulses.
As a consequence, the observation of the Schwinger effect (pair creation from the vacuum) would require the concentration of stray particles in the experimental chamber to be extremely low, so as not to seed a cascade, e.g. as low as $10^5$~cm$^{-3}$ in the case of dipole-wave focusing geometry \cite{gonoskov.prl.2013}. 
In \citet{bulanov.prl.2010s} the authors identified a strong effect of radiation polarization on the development of cascade.
In particular, the initial stage of cascade development was found to be suppressed in the case of two linearly polarized, colliding laser pulses due to the field geometry, which might lead to a significant number of Schwinger pairs to be born before the laser is depleted by the cascade.}
We should also note that depletion of the laser energy is caused not only by pair creation \emph{per se}, but also by collective, i.e. plasma, effects, when the density of pairs becomes sufficiently large.
Heating of the generated electron-positron plasma also leads to rapid laser absorption~\cite{nerush.prl.2011,grismayer.pop.2016}.

The basic features of the avalanche-type cascade in a rotating electric field can be described by the following simple model~\cite{bashmakov.pop.2014,grismayer.pop.2016,grismayer.pre.2017}.
If the probability of the Breit-Wheeler process is $P^b$, then the pair density grows as $n_\pm=n_\gamma[1-\exp(-P^b t)]$, {where $n_\gamma$ is the photon density}.
If the photons are generated by the same pairs, and assuming that their energy is the same, we have $n_\pm=\int_0^t dt^\prime P^c[1-\exp(-P^c(t-t^\prime))]$ and
\begin{equation}
    \frac{dn_\pm}{dt}=\int\limits_0^t n_\pm(t^\prime)P^c P^b\exp[-P^b(t-t^\prime)].
\end{equation}
where $P^c$ is the photon-emission probability.
The solution of this equation has the form $n_\pm\sim \exp[\Gamma t]$, where $\Gamma\sim P^b$ for the most relevant case of $\chi \sim 1$~\cite{fedotov.prl.2010,bashmakov.pop.2014,grismayer.pop.2016,grismayer.pre.2017}.
As pair creation is exponentially suppressed for $\chi < 1$, the onset of the avalanche-type cascade is intrinsically connected with the dominance of the quantum effects~\cite{fedotov.prl.2010,bulanov.prl.2010s,bulanov.nima.2011}.
Recall that $\chi = 1$ corresponds to the intensity of $\sim 5\times 10^{24}$ W/cm$^2$ for a 1-$\mu$m laser, as detailed in \cref{sec:Geometries} (see \cref{eq:QRRThreshold}).

However, numerical studies of avalanche-type cascades, using multidimensional QED-PIC codes, show that the onset of the cascades is observed at lower intensities $I\sim 7\times 10^{23}$ W/cm$^2$~\cite{bashmakov.pop.2014,grismayer.pop.2016,grismayer.pre.2017}.
This is partially due to the fact that the classical prediction for the radiated power, which is used in deriving \cref{eq:QRRThreshold}, is an overestimate;
thus the electron energy is damped more strongly than it should be, according to QED calculations.
The threshold is also lowered by stochastic effects, because it is possible for electrons to radiate less than expected and therefore to have higher energy~\cite{duclous.ppcf.2011}.

The question of how to \emph{initiate} an avalanche has received a great deal of attention. The use of seed high-energy electron or photon beams has been studied~\cite{tang.pra.2014,king.pra.2013}, as well as the role of the detailed spatiotemporal structure of the laser fields ~\cite{tamburini.sr.2017,sampath.pop.2018}. The role of spin and polarization of the electrons, positrons, and photons in multiphoton Compton and Breit-Wheeler processes in relation to the avalanche development was studied in \cite{seipt.njp.2021}. The study of dual-beam cascades has been extended to the case of multiple colliding laser pulses (MCLP)\footnote{We discuss the properties of multiple colliding laser pulses in section \ref{sec:DipoleWaves}.} \cite{gelfer.pra.2015,vranic.ppcf.2017,gonoskov.prx.2017,magnusson.pra.2019}, and to investigations of single-particle and collective dynamics~\cite{jirka.pre.2016,gong.pre.2017,bashinov.pra.2017,luo.sr.2018,efimenko.sr.2018}.

In parallel, a comprehensive study of QED effects on laser interactions with solid- and near-critical density targets has been conducted using simulations. Initially, this effort was justified by considering how they would affect laser-driven ion acceleration~\cite{tamburini.njp.2010}\footnote{We discuss the effect of role of RR and QED in ion acceleration in \cref{sec:Ions}.}, but it was soon realized that such interactions could serve as an efficient source of high-energy photons and electron-positron pairs. The reflection of laser radiation from a plasma surface, and therefore the standing-wave-like structure generated nearby, means that the dynamics bear a close resemblance to the avalanche-type cascade~\cite{ridgers.prl.2012, ridgers.pop.2013, zhang.njp.2015, zhu.njp.2015, chang.pre.2015, liu.oe.2016, zhu.ncomm.2016, wang.pre.2017, slade-lowther.njp.2019, luo.sr.2018, luo.ppcf.2015}.

The shower-type cascade is characterized by the fact that the EM field serves only as a target and does not affect particle trajectories to a significant extent. In this case the particle motion is completely determined by QED processes and the dynamics can be formulated in terms of a system of kinetic equations for electrons, positrons, and photons~\cite{baier.prd.1999,sokolov.prl.2010,duclous.ppcf.2011,bulanov.pra.2013}:
    \begin{align}
    \frac{d \phi_\pm}{dt} &=-\phi_\pm P^c+\int\limits_0^1\left[\phi_\pm\frac{dP^c}{d\epsilon^\prime}+\phi_\gamma\frac{d P^b}{d\epsilon^\prime}\right]d\epsilon,
    \\
    \frac{d \phi_\gamma}{dt} &=-\phi_\gamma P^b+\int\limits_0^1\left[\phi_++\phi_-\right]\frac{dP^c}{d\epsilon^\prime}d\epsilon.
    \end{align}
All variables are as defined in \cref{eq:KineticEquations}, in which electron-positron pair creation is neglected.

The results of numerical studies of shower-type cascades are in good agreement with the results of the simple analysis of 1D kinetic equations.
They demonstrate fast depletion of electron-beam energy and the development of a spectrum with exponentially decaying shape, as well as the generation of electron-positron pairs~\cite{sokolov.prl.2010, bulanov.pra.2013, blackburn.prl.2014, tang.pra.2014, mironov.pla.2014, wang.pop.2015}.
However, we should note that the significant loss of electron beam energy eventually invalidates the 1D approximation usually employed when analyzing shower-type cascades.
If the transverse momentum of particles is no longer much smaller than the longitudinal momentum, this should be taken into account.
Transverse motion of charged particles in EM fields might lead to the development of an avalanche-type cascade in specific EM field configurations, e.g., electron-beam interactions with two colliding laser pulses~\cite{mironov.pla.2014} and electron-beam interaction with {multiple colliding laser pulses}~\cite{magnusson.prl.2019,magnusson.pra.2019}.

\subsection{Electromagnetic field depletion}
\label{sec:Depletion}

At extremely high intensities, cascades can be accompanied by the depletion of the EM field itself~\cite{meuren.prd.2016,seipt.prl.2017}.
Over the course of the multiphoton Compton and Breit-Wheeler processes, which are responsible for transforming the initial electron beam energy into photons and electron-positron pairs, a significant number of photons are absorbed from the EM field. The interaction of an electron beam of sufficiently high charge with an intense EM field can therefore lead to the depletion of the field energy. 

It was shown in \citet{seipt.prl.2017} that the multiphoton Compton process in strong EM fields can lead to the energy loss of the laser sufficiently large to invalidate the background-field approximation. It happens when the number of photons absorbed during the process is of the order of the number of photons in the laser pulse. {This leads to the condition on the field strength, number of electrons per laser wavelength cubed and their energy: $N_e \gamma_e^{-0.92} a_0^{1.08} \gtrsim 6.8\times 10^{11}$, which for $a_0 \sim 10^3$, $\gamma_e\sim a_0$ gives $N_T\approx 10^{11-12}$.} The number of laser photons absorbed during the Compton process is obtained as follows: in a monochromatic plane wave laser field, taken to be circularly polarized for simplicity, the kinematics of the photon emission is governed by four-momentum conservation, $q + sk = q' + k'$, where $s$ is the number of laser photons. Analyzing how $dP^c/ds$ depends on $s$ shows that the most probable number of photons absorbed is:
    \begin{align}\label{s}
    s(\chi_\gamma) = \frac{a_0^3}{\chi_e} \frac{\chi_\gamma}{\chi_e-\chi_\gamma}.
    \end{align}
This is valid for $a_0 \gg 1$ and reproduces the leading order of the related result (18) in \citet{narozhny.jetp.1965}. Thus, when a Compton photon with a given value of $\chi_\gamma$ is emitted, the number of laser photons drawn from the laser field can safely be estimated using \eqref{s} within the model of one-photon incoherent emission\footnote{{A formula completely analogous to \eqref{s} holds for the Breit-Wheeler process, $\gamma + s\gamma_L \to e^+ e^-$, which becomes possible {in a monochromatic wave} above a threshold in photon number, $s \geq s_0 = 2a_0(1+a_0^2)/\chi_\gamma$.}}. We note that $s(\chi_\gamma)=(\omega_0 L_f)^{-3}$, which means that the shorter the formation length $L_f$, the greater the number of photons that must be absorbed from the laser, to make the emission possible.

{In \citet{meuren.prd.2016}, where electron-positron pair production via the multi-photon Breit-Wheeler process is discussed, the energy absorbed from the laser in the interaction is divided into two parts: ``classical'' (absorbed in accelerating the electron and positron out of the laser pulse) and ``quantum'' (absorbed over the formation length).
It is shown that these two parts scale differently with $a_0$: the classical part scales as $a_0^3/\chi_\gamma$, whereas the quantum part scales as $a_0/\chi_\gamma$.
As such, the classical component dominates at high intensity.
The same applies to the multi-photon Compton process~\cite{blackburn.pop.2018}.}  

\subsection{{QED plasmas}}
\label{sec:QEDplasma}

	\begin{figure}
	\centering
	\includegraphics[width=0.8\linewidth]{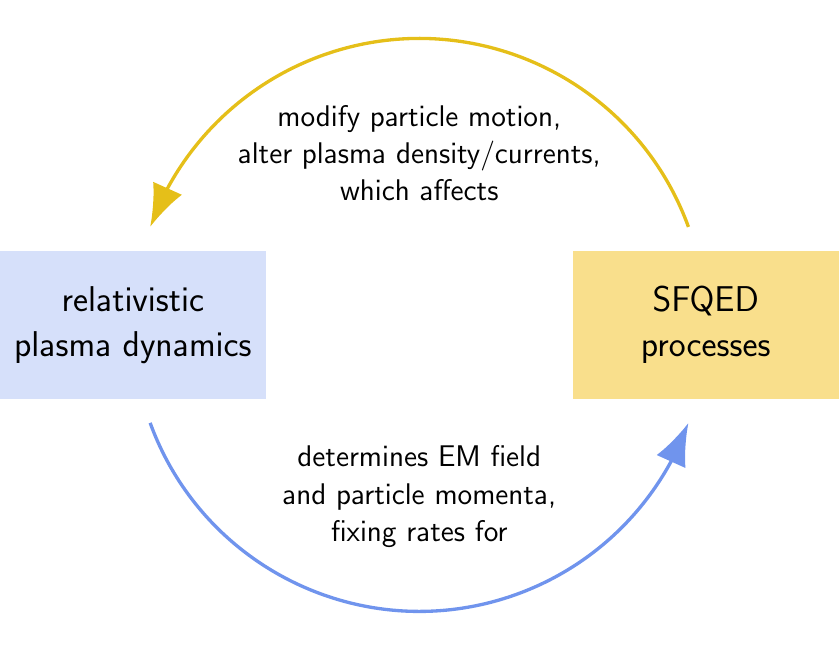}
	\caption{
		Feedback between SFQED processes and (classical) relativistic plasma dynamics in a `QED plasma'.}
	\label{fig:QEDPlasmas}
	\end{figure}

{
In \cref{sec:Cascades} we reviewed how the development of a shower- or avalanche-type cascade led to the prolific production of photons and electron-positron pairs.
When this occurs in a plasma environment, or the density of the produced particles is high enough that a plasma environment is created, it is necessary to consider the coupling between SFQED processes and classical plasma dynamics, as illustrated in \cref{fig:QEDPlasmas}.
This includes the modification of particle trajectories, due to radiative energy losses, and the plasma density, due to pair production, with the consequence that basic plasma properties are continuously changed.
As these properties control the rates for SFQED processes, the interconnection between SFQED effects and plasma behavior gives rise to a unique dynamical system, which is usually referred to as \emph{QED plasma}~\cite{zhang.pop.2020} and is of particular interest for astrophysics (see \cref{sec:AstrophysicalEnvironments}).}

{
The behavior of the QED plasma is expected to differ dramatically from that of the well-studied classical plasma.
These dynamics are primarily studied with numerical simulations (see \cref{sec:Numerics}), but one-dimensional analytical models that include classical radiation reaction are available for a thin foil~\cite{bulanov.pop.2013} or a surface pair plasma formed during hole-boring of a thick target~\cite{kirk.ppcf.2013}.
Prolific electron-positron pair production, whether by means of the Schwinger mechanism~\cite{bulanov.pre.2005} or an avalanche-type cascade of photon emission and nonlinear Breit-Wheeler pair creation~\cite{nerush.prl.2011,grismayer.pop.2016}, leads to the backreaction on the incident laser pulse and its fast depletion (and possible change of polarization).
Radiation reaction effects are expected to dominate in laser-driven ion~\cite{tamburini.njp.2010,delsorbo.njp.2018} and electron~\cite{stark.prl.2016, vranic.ppcf.2018} acceleration, by reducing the available electron energy.
However, the combination of RR, laser and plasma fields can lead to net energy \emph{gain} by electrons (see \cref{sec:Dissipation}), as discussed for the case of a magnetized plasma channel by \citet{gong.sr.2019}.
Energy dissipation affects not only the particles: simulations show that, in hole-boring of a thick target with a CP laser, RR leads to the absorption of angular momentum and then to the generation of a quasistatic, axial magnetic field of GG strength~\cite{liseykina.njp.2016}.
}

{The feedback between classical plasma, i.e. collective, dynamics and SFQED processes may be viewed from a global or local perspective.
Consider the effect of \emph{relativistically self-induced transparency}, when light of sufficient intensity is transmitted through a plasma that has electron density above the classical critical density and is therefore expected to be opaque.
In a global approach, this can be explained by the fact sufficiently intense light drives electron oscillations at relativistic velocity, which increases their effective mass and thereby reduces the plasma frequency from $\omega_p = \sqrt{n_e e^2/(\epsilon_0 m)}$ to $\omega_p / \avg{\gamma}$, where $\avg{\gamma} \sim a_0$ is the average Lorentz factor of the electron population~\cite{kaw.pf.1970}.
A reduction of $\avg{\gamma}$ because of radiative energy losses would therefore counter the relativistic transparency of a QED plasma~\cite{zhang.njp.2015}.
Similarly, exponentially rapid electron-position pair production (see \cref{sec:Cascades}), which increases the current density, would increase $\omega_p$ and cause the plasma to become highly reflective.
Once critically dense, the self-generated plasma would then shield the interior from the incident radiation~\cite{fedotov.prl.2010,bulanov.prl.2010s}.
}

{
While a global perspective allows for general appreciation of SQED effects, it is not sufficient to explain the complex dynamics that occur in the presence of strong dissipation and pair production.
One of the central questions is how this affects the pair cascade itself, which has been studied for laser-plasma~\cite{brady.pop.2014,slade-lowther.njp.2019} and laser-laser~\cite{grismayer.pop.2016} interactions, as well as multiple colliding laser pulses~\cite{gonoskov.prx.2017} and propagating laser-plasma interfaces~\cite{samsonov.sr.2019}.
Numerical studies demonstrate the formation of nonlinear, highly structured plasma as the self-generated current feeds back on the incident light, including the formation of transverse~\cite{muraviev.pre.2021} and radial current sheets~\cite{efimenko.sr.2018} in dual- and multiple-laser interactions, respectively, as well as plasma pinching around the axis of laser-generated dipole waves~\cite{efimenko.pre.2019} (see \cref{sec:DipoleWaves}).
}

\bib

%% file: experiments.tex
\section{Experiments}
\label{sec:Experiments}

In this section we discuss recent and upcoming experimental campaigns in the regime where radiation reaction and strong-field QED effects become important.
Thus far, those experiments have employed electron beams that have been accelerated to ultrarelativistic energies before they encounter an electromagnetic field.
This allows $\chi_e \gtrsim 1$ with fields that are subcritical in the lab frame.
We focus here on experiments that use a high-intensity laser as the target electromagnetic field.
A comprehensive review of crystal-based experiments is given in \citet{uggerhoj.rmp.2005}, with results since then discussed briefly in \cref{sec:AlignedCrystals}.

To identify the interaction regime, we model the interaction as that of a high-energy electron beam and a single, travelling plane EM wave.
The threshold for the energy loss to radiation to exceed the work done by the external field is given by $\gamma_0 (\gamma_0^2 - a_0^2/4) > (4 \erad)^{-1}$, or $\gamma_0 > 2.8 \times 10^2 \lambda^{1/3} [\micron]$, assuming that $\gamma_0 \gg a_0$.
However, because the instantaneous work done by a travelling EM wave $\propto a_0^2 / \gamma_0$ is rather small, this condition's being satisfied does not necessarily mean that radiation losses are large in an absolute sense.
A better way to quantify the latter is to compare the energy lost over a single cycle of the wave to the electron's initial energy.
Thus radiation-reaction effects are significant when
    \begin{align}
    \frac{4\pi}{3} \alpha a_0 \chi_e &> 1
    &
    &\iff
    &
    a_0 &\gtrsim 59 \frac{\lambda^{1/2} [\micron]}{\Energy_0^{1/2} [\text{GeV}]},
    \label{eq:RRCollision}
    \end{align}
where $\lambda$ is the laser wavelength and $\Energy_0$ the electron's initial energy.
This is equivalent to the condition $L_i < \lambda$, where the depletion length $L_i$ is defined in \cref{eq:InertiaLength}.
Quantum effects are significant when
    \begin{align}
    \chi_e &> 1
    &
    &\iff
    &
    a_0 &\gtrsim 105 \frac{\lambda [\micron]}{\Energy_0 [\text{GeV}]}.
    \label{eq:QuantumCollision}
    \end{align}
The intensities equivalent to \cref{eq:RRCollision,eq:QuantumCollision} are $2.9$ and $3.8 \times 10^{21}~\Wcmsqd$, respectively, for a 2-GeV electron beam and a laser wavelength of $0.8~\micron$.
Note that RR can be substantial if the fractional energy loss is 10\% per wavelength~\cite{thomas.prx.2012}, as even a 30-fs pulse is approximately eleven cycles long, and similarly that quantum effects, via the suppression of the radiated power, can be evident at $\chi_e \gtrsim 0.1$ (see \cref{sec:QuantumRR} and \cref{eq:RRCollision,eq:QuantumCollision}).

\subsection{Using conventional accelerators}
\label{sec:ConventionalAccelerators}

The first experimental campaign to demonstrate nonlinear quantum effects took place at SLAC~\cite{bula.prl.1996,burke.prl.1997,bamber.prd.1999}, using its conventional, radio-frequency accelerator to produce a 46.6-GeV electron beam that collided with a laser pulse with peak nominal intensity of $1.3 \times 10^{18}~\Wcmsqd$, duration of $1.4$~ps and wavelength of $1054$~nm (or $527$~nm after frequency doubling).
The classical and quantum nonlinearity parameters, $a_0 \simeq 0.4$ and $\chi_e \simeq 0.3$, were sufficient for observation of multiphoton Compton scattering~\cite{bula.prl.1996} and multiphoton Breit-Wheeler pair creation~\cite{burke.prl.1997}.
Measuring the energy spectrum of the scattered electrons over a range of laser intensities (achieved by varying the pulse energy), and comparing to simulations, showed that up to four laser photons could be absorbed in a single Compton event~\cite{bula.prl.1996}.

The emitted photons also drove the creation of electron-positron pairs, via the multiphoton Breit-Wheeler process~\cite{burke.prl.1997}: in total, $106 \pm 14$ positrons were detected over a sequence of 22,000 collisions with the frequency-doubled laser, with a yield per collision $N_{+}$ that scaled with $a_0$ as $N_{+} \propto a_0^{2n}$, where $n \simeq 5.1 \pm 0.2 \text{(stat.)} \vphantom{a}^{+ 0.5}_{- 0.8} \text{(sys.)}$.
This scaling is justified on the grounds that, in the weakly nonlinear regime $a_0 \ll 1$, the rate is expected to scale as $a_0^{2n}$, where $n$, the number of participating photons, must satisfy $n \geq 5$ to overcome the mass threshold~\cite{burke.prl.1997}.
Since the time of this experiment, the development of theoretical frameworks that treat photon emission and pair creation as two steps of a unified `trident' process, and therefore include contributions where the intermediate photon is off-shell, has offered new insight into its results~\cite{hu.prl.2010,ilderton.prl.2011}: see \cref{sec:HigherOrderProcesses} for details.

    \begin{figure}
    \includegraphics[width=\linewidth]{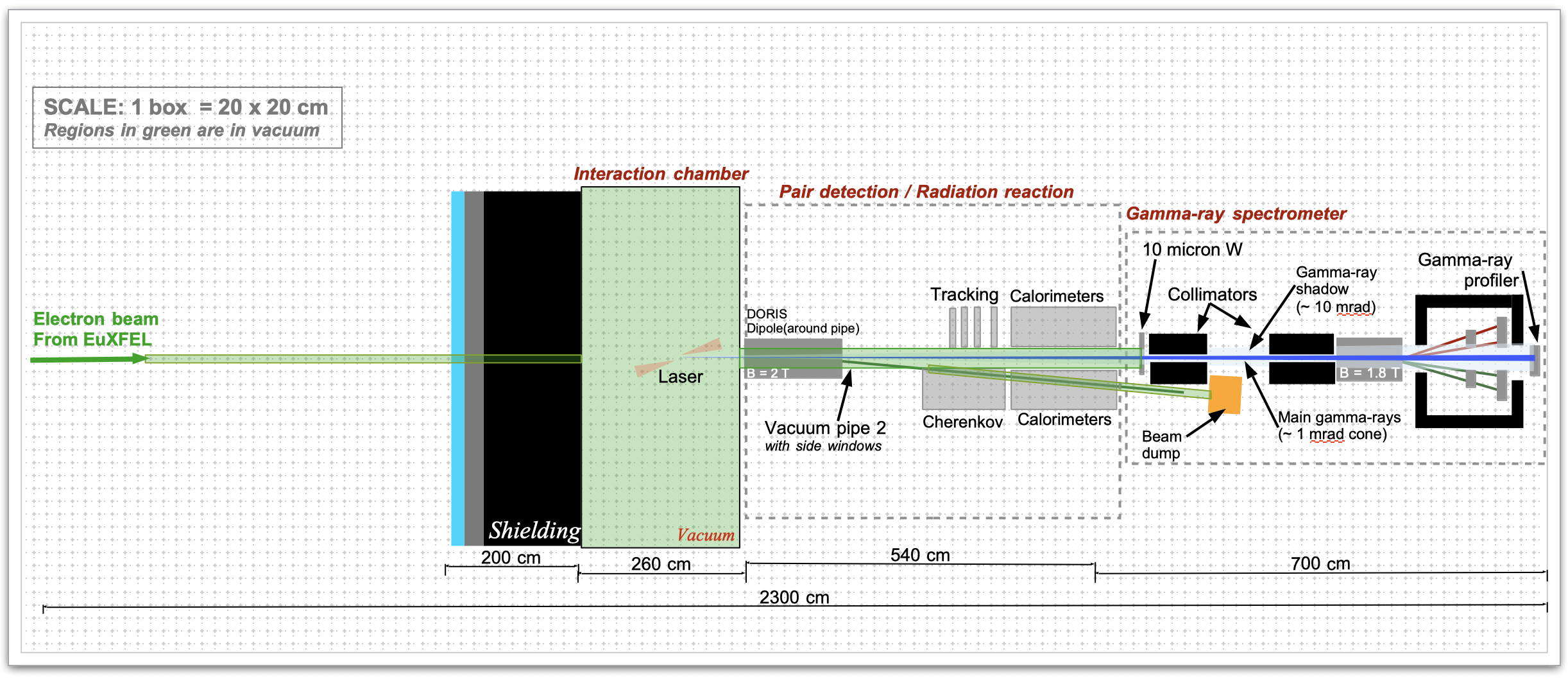}
    \includegraphics[width=\linewidth]{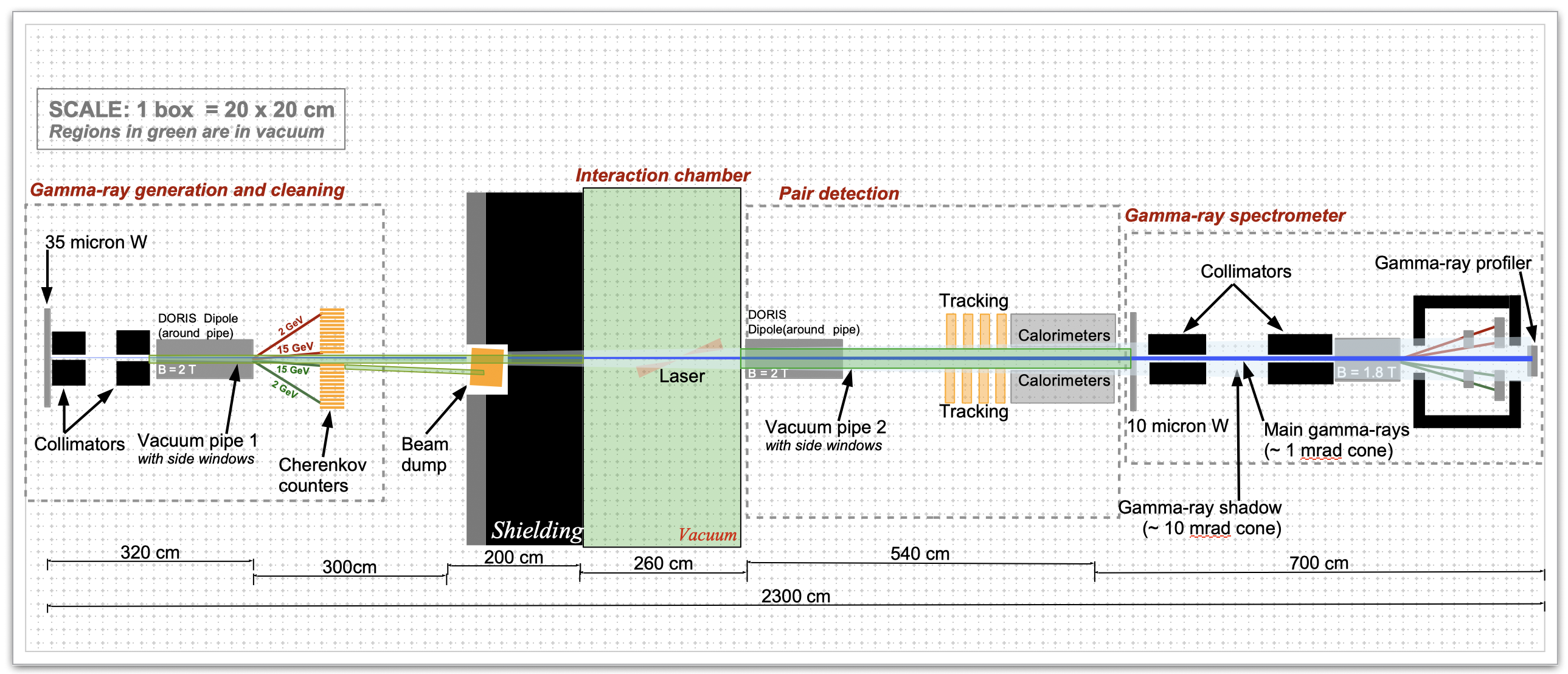}
    \caption{
        Experimental configurations envisaged for laser-electron (top) and laser-$\gamma$ (bottom) collisions at LUXE. Reproduced from \citet{abramowicz.arxiv.2019}.}
    \label{fig:luxe}
    \end{figure}

It is now likely that this experiment will be revisited, more than twenty years later, with a higher intensity laser.
At SLAC~\cite{yakimenko.prab.2019}, the E320 collaboration proposes to collide the 13-GeV FACET-II electron beam with a 10-TW class laser, reaching $a_0$ up to $7$ and $\chi_e \sim 1$, and thereby exploring the ``quantum tunneling regime'' in high-order nonlinear Compton scattering and electron-driven pair creation~\cite{meuren.exhilp.2019}.
The LUXE (Light Und XFEL Experiment) collaboration has proposed a dedicated experiment at the European XFEL~\cite{abramowicz.arxiv.2019,abramowicz.arxiv.2021}, in which one bunch per bunch train of the 17.5-GeV electron beam is extracted into a dedicated beamline and interaction chamber, where it either collides with a 30-TW laser, or is used to create high-energy bremsstrahlung photons which themselves collide with the laser (see \cref{fig:luxe}).
{(There are plans to upgrade the laser up to 300 TW at a later stage.)}
With nonlinearity parameters of $a_0 \simeq 2$ and $\chi_e \simeq 0.4$ expected, scientific goals include exploration of nonlinear Compton scattering and $\gamma$-driven pair creation in the transition from the multiphoton to the tunneling regimes~\cite{abramowicz.arxiv.2019,abramowicz.arxiv.2021}.

\subsection{All-optical setups}
\label{sec:AllOptical}

The experiments described in the previous section depend crucially upon the co-location of a high-power laser system and a conventional, linear accelerator, which usually comes along with a unique set of challenges. However, advances in laser-driven wakefield acceleration have made it possible to accelerate electrons to multi-GeV energies in centimeter-scale gas targets~\cite{kneip.prl.2009,wang.ncomm.2013,leemans.prl.2014,gonsalves.prl.2019} and thereby to realize `all-optical' collision experiments~\cite{bulanov.nima.2011}. Besides the compactness of a laser facility, as compared to a km-scale conventional accelerator, this exploits the high charge (100s pC) and small size (several $\micron$, see \citet{schnell.prl.2012}) of wakefield-accelerated electron beams, which maximises the signal from a collision with a laser pulse focused to a comparable spot size. Collision timing is also aided by the inherent synchronization between the driving laser pulse and the electron beam, and between the two laser pulses, provided they emerge from the same oscillator~\cite{blackburn.rmpp.2020}: compare to the natural alignment afforded by a single-pulse retroreflection geometry~\cite{taphuoc.np.2012}.
{However, laser plasma accelerators suffer from shot-to-shot fluctuations~\cite{samarin.jmo.2017}, which limit the number of observed events reported in \citet{cole.prx.2018,poder.prx.2018} (see discussion in this section).
These fluctuations can be mitigated by advances in laser technology~\cite{maier.prx.2020}.}
We discuss here all-optical laser-electron beam collisions as a means of studying nonlinear effects, radiation reaction and QED. Their role as a source of high-energy radiation is discussed in \cref{sec:RadiationGeneration}.

Since the first demonstration of Thomson scattering in an all-optical laser-electron collision~\cite{schwoerer.prl.2006}, increases in the accessible laser intensity have made it possible to study two, classical, nonlinear effects
\footnote{For completeness, note that the same effects have been observed in other laser-based geometries, including collisions with a 60-MeV electron beam from a linac, using the angular structure of the second and third harmonics at laser amplitudes of $a_0 = 0.3$ and $0.6$ ~\cite{babzien.prl.2006,sakai.prstab.2015}, as well as in a laser-driven gas, using the laser-intensity and electron-density dependence of the second-harmonic light~\cite{chen.nature.1998}.}: if $a_0 \gtrsim 1$, the electron's anharmonic, relativistic motion causes the Thomson edges to be redshifted by a factor $1/(1 + a_0^2)$, as if the mass had increased, and the emission of higher order harmonics~\cite{brown.pr.1964,goldman.pl.1964,sarachik.prd.1970,esarey.pre.1993}. {The nonlinear relativistic Thomson scattering is first identified by spectrally resolved measurement of high-energy (up to 18 MeV) and ultrahigh-brilliance photon beams from all-optical collisions at $\gamma \simeq 1100$, $a_0 \simeq 2$ in \citet{sarri.prl.2014}. The onset of the nonlinear regime is studied in \citet{khrennikov.prl.2015} by tuning the energy of the electron beam in the range $20 < \gamma < 100$ and measuring the spectral peak of the X rays emitted when the beam collides with a laser that has $a_0 \simeq 0.9$.} More recently, \citet{yan.nphot.2017} reported the measurement of photons with harmonic order $>$500 in collisions of electrons with $\gamma = 708 \pm 114$ and a laser pulse with $a_0 \simeq 12$. Further evidence of the transition from the linear to the nonlinear regime is obtained in the asymmetric angular profile of the emitted radiation, as well as in the scaling of the total energy of the radiation with $a_0$~\cite{yan.nphot.2017}.

    \begin{figure}
    \includegraphics[width=\linewidth]{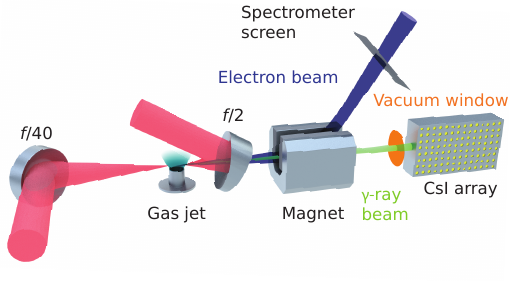}
    \caption{
        An all-optical realization of a laser-electron-beam collision experiment.
        One laser pulse is focused by a short-focal-length optic with a hole in it, to allow for counterpropagation of an electron beam that is accelerated by a laser wakefield in a gas jet.
        Electrons, and the radiation they emit, are transmitted through this hole before being diagnosed.
        The collision is timed so that it occurs close to the edge of the gas jet, where the electron beam is smallest, to maximize overlap with the laser pulse.
        Reproduced from \citet{cole.prx.2018}.
    }
    \label{fig:ColeExpt}
    \end{figure}

All-optical laser-electron collisions have now reached a regime where neither radiation-reaction nor quantum effects are negligible, as the intensities available with tight focussing of petawatt-class lasers, are comparable to the thresholds given by \cref{eq:RRCollision,eq:QuantumCollision}. 
The first experimental campaigns to find evidence of radiation reaction~\cite{cole.prx.2018}, as well as quantum effects on the same~\cite{poder.prx.2018}, have now been reported.
Both were performed at the Rutherford Appleton Laboratory using the Gemini laser system~\cite{hooker.jpf.2006}, which delivers two synchronized pulses with energy ${\sim}10$~J, wavelength $800$~nm and duration $45$~fs.
The design, which is essentially common to both, is sketched in \cref{fig:ColeExpt}.
To produce a high-energy electron beam by wakefield acceleration, one laser pulse is focused with a long-focal-length optic (f/40) onto: the leading edge of a 15-mm, supersonic helium gas jet, in \citet{cole.prx.2018}; or the entrance of a helium-filled gas cell, in \citet{poder.prx.2018}.
The second laser pulse is focused tightly (peak $a_0 \simeq 20$ in both experiments) with a short-focal-length, off-axis parabola (f/2) which has, crucially, an f/7 aperture in its centre.
This allows for counterpropagation of the second, high-intensity laser and the electron beam, which maximizes $\chi_e$, because the first laser, the electron beam and any $\gamma$ rays produced in the collision can pass through the aperture and be blocked or diagnosed.

The advantage of a gas cell, over a gas jet, is the improved stability and higher energy of the electron beam~\cite{poder.prx.2018}.
On the other hand, to avoid the second, higher intensity laser damaging the gas cell itself, the point where it is focused, and where the collision takes place, must be 1~cm downstream, where the electron beam has expanded to a transverse size comparable to that of the laser pulse.
This means that the electrons effectively sample a range of collision $a_0$, which weakens the relevant signals and complicates theoretical analysis.
In \cite{cole.prx.2018}, the second laser is focused at the rear edge of the gas jet, close to where the electron beam leaves the plasma and is therefore much smaller.
A systematic timing offset in both experiments means the $a_0$ at the collision point is $a_0 \simeq 10$, rather than 20.

Since fluctuations and drifts in the collision timing and pointing cause the overlap between the electron beam and the laser pulse, and therefore the strength of any signals of radiation reaction, to vary, access to both electron and $\gamma$-ray data on a shot-by-shot basis is essential for identification of successful collisions~\cite{blackburn.rmpp.2020}.
In both experiments, the total $\gamma$-ray yield deposited in a stack of CsI crystals is used as a criterion to distinguish such collisions (the spectrometer is discussed in detail in \citet{behm.rsi.2018}).

    \begin{figure}
    \includegraphics[width=\linewidth]{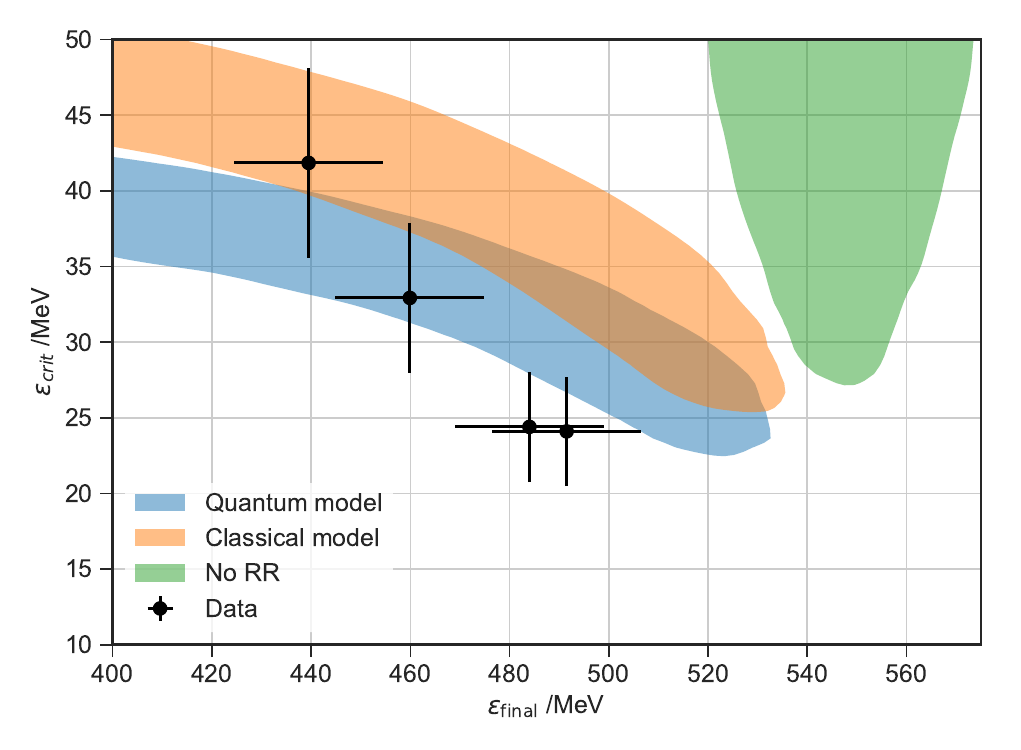}
    \caption{
        Experimental evidence of radiation reaction in the correlation between the energy of the electron beam after the collision and the critical energy of the $\gamma$-ray spectrum, as measured in four successful collisions (points), which is compared to simulations of the interaction that include various models of radiation reaction (no radiation reaction in green, classical in blue and stochastic in orange) and fluctuations in the electron energy and laser intensity.
        Reproduced from \citet{cole.prx.2018}.
    }
    \label{fig:ColeResult}
    \end{figure}

The results from \citet{cole.prx.2018} are shown in \cref{fig:ColeResult}.
Over a sequence of eight shots with the second laser on, and ten shots with it off, four are measured to have a $\gamma$-ray signal significantly above background.
The same four have electron-beam energies (as diagnosed by a sharp feature in the measured spectrum) below 500 MeV, as compared to $550 \pm 20$~MeV for the beam-off shots.
\Cref{fig:ColeResult} shows that this electron energy is negatively correlated with the critical energy of the $\gamma$ rays, a parameter which characterizes the hardness of the photon energy spectrum, and is extracted from measurements of the depth-resolved deposition in the scintillator stack~\cite{behm.rsi.2018}.
The probability for this correlation, and for the electron energies to be below 500~MeV on the same four successful shots, to occur by chance is $0.03\%$ and therefore evidence of radiation reaction is obtained at the three-sigma level~\cite{cole.prx.2018}.
Simulations of the interaction, including fluctuations in the initial electron energy and collision $a_0$, show that the measured energies are inconsistent with the absence of radiation reaction; both classical and quantum radiation reaction are consistent at the two-sigma level.

    \begin{figure}
    \includegraphics[width=\linewidth]{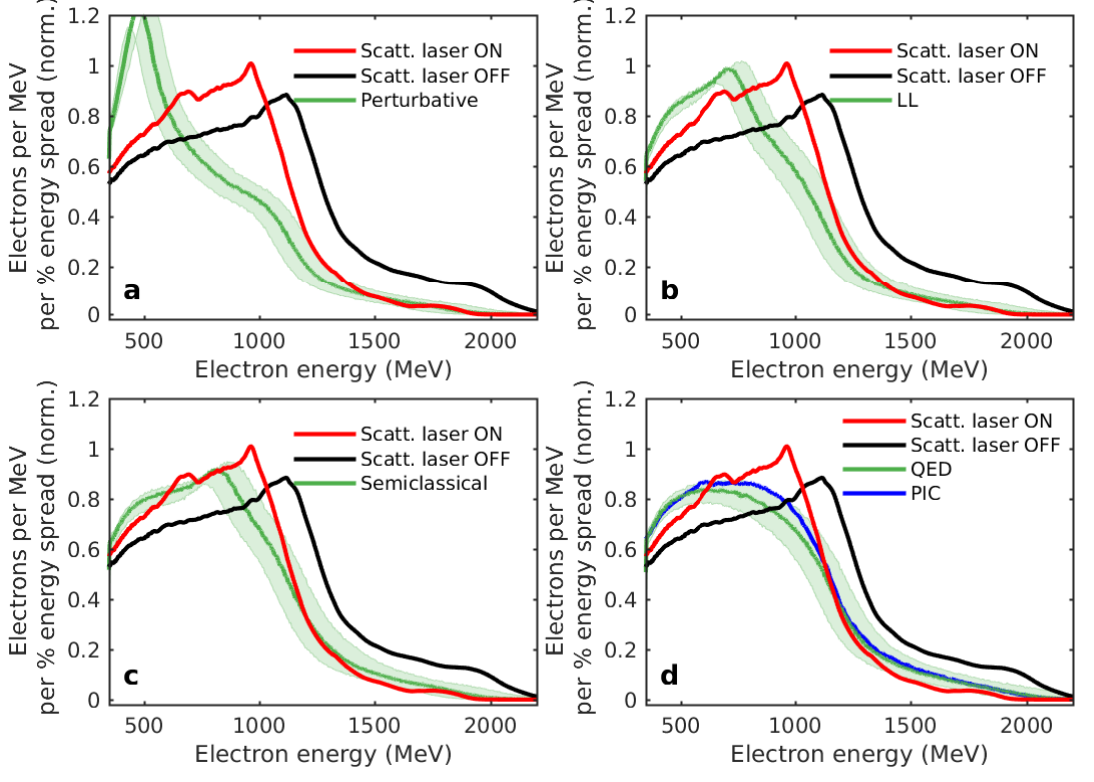}
    \caption{
        Experimental evidence of radiation reaction and quantum corrections:
        electron energy spectra as measured at best overlap, without the scattering laser (black) and with it (red), and from simulations (a) without radiation reaction, (b) classical  (Landau-Lifshitz) radiation reaction, (c) a quantum-corrected Landau-Lifshitz model and (d) stochastic radiation reaction (PIC and single-particle codes in green and blue respectively), all including experimental uncertainties in the initial electron spectrum, the magnetic spectrometer and laser intensity.
        Reproduced from \citet{poder.prx.2018}.
    }
    \label{fig:PoderResult}
    \end{figure}

In \citet{poder.prx.2018}, the total signal in the scintillator, shown to be positively correlated with the energy loss of the electron beam (as compared to a reference, beam-off shot), is used to identify collisions with good and poor overlap.
Energy losses of up to 30\% are measured for the highest energy electrons in the case of best overlap, which is consistent with, but smaller than, that predicted by the classical Landau-Lifshitz equation, \cref{eq:LandauLifshitz}, assuming an initial $\gamma_e \simeq 4000$ and $a_0 = 10$~\cite{poder.prx.2018}.
The best-overlap results are compared in detail against 3D simulations, which use the experimentally measured electron energy spectra and divergence, and model the laser transverse profile by means of a two-component Gaussian fit to the measured focal spot.
These comparisons, with different models of radiation reaction, are shown in \cref{fig:PoderResult}.
Classical radiation reaction is shown to overestimate the energy loss, with a coefficient of determination $R^2 = 0.87$.
The `semiclassical' model, where the strength of the radiation-reaction force is scaled by the quantum correcting factor $G(\chi_e)$, \cref{eq:Gaunt}, gives better agreement ($R^2 = 0.96$) than both the fully stochastic ($R^2 = 0.92$) and classical models.
\Citet{poder.prx.2018} attribute the discrepancy in the former case to failure of the LCFA or to incomplete knowledge of the structure of the laser pulse.
Nevertheless, the failure of the classical model provides strong evidence of quantum corrections to radiation reaction.

\subsection{With high-energy photons}
\label{sec:PhotonPhoton}

Probing regions of strong electromagnetic field with high-energy photons, rather than electrons as described in the previous sections, has certain advantages.
As they are uncharged, photons do not undergo ponderomotive scattering~\cite{gaponov.jetp.1958,kibble.prl.1966,quesnel.pre.1998, narozhny.jetp.2000}, which can prevent electrons penetrating to the point of highest intensity~\cite{fedotov.pra.2014,blackburn.rmpp.2020}, or radiation reaction, which reduces the energy and so the quantum parameter~\cite{yakimenko.prl.2019, blackburn.njp.2019, baumann.ppcf.2019}. Moreover, {signals of Breit-Wheeler pair production can be highly sensitive to the laser pulse shape and duration in the nonlinear regime~\cite{krajewska.pra.2014, titov.prl.2012, titov.pra.2013, titov.epjd.2020}.}
Simulations show, for example, that it is easier to initiate a pair cascade in two counterpropagating lasers by seeding with a photon than an electron, of the same energy~\cite{tang.pra.2014}.

Photons with the desired, GeV-scale, energies can be obtained from multi-GeV electrons in several ways~\cite{albert.ppcf.2016}.
One attractive method is bremsstrahlung of an electron beam in a high-$Z$ target, because photons with energies comparable to the electron itself may be generated.
(Alternative laser-based concepts have been proposed as well: see \cref{sec:RadiationGeneration}).
Denoting the fractional photon energy $f = \hbar \omega' / (\gamma m c^2)$, where $\gamma$ is the Lorentz factor of the incident electron, and expressing the thickness $X$ of the target as a fraction $x$ of the \emph{radiation length} $X_0$, the photon energy spectrum {per incident electron} is, for thin targets $x \ll 1$~\cite{tsai.rmp.1974}
    \begin{equation}
    \frac{\rmd N_\gamma}{\rmd f} =
        \frac{x}{f}
        \left(
            \frac{4}{3} - \frac{4 f}{3} + f^2
        \right),
    \end{equation}
and for moderately thick targets, $0.5 \lesssim x \lesssim 2$,
    \begin{equation}
    \frac{\rmd N_\gamma}{\rmd f} =
        \frac{(1 - f)^{4 x / 3} - e^{-7x/9}}{f \left[ \frac{7}{9} + \frac{4}{3} \ln(1-f) \right]}.
    \end{equation}

Experiments where bremsstrahlung photons collide with a high-intensity laser, envisaged decades ago~\cite{reiss.prl.1971}, now form part of planned experimental programs.
Such experiments would explore electron-positron pair creation and cascade formation in the nonlinear regime~\cite{blackburn.ppcf.2018,hartin.prd.2019}.
The electron beam itself could be sourced from a conventional~\cite{abramowicz.arxiv.2019,abramowicz.arxiv.2021} or laser-wakefield accelerator~\cite{sarri.prl.2013,turcu.rrp.2016}.
{Photon-photon physics platforms can also investigate the linear Breit-Wheeler process~\cite{breit.pr.1934} or real photon-photon scattering, by colliding bremsstrahlung photons with keV X ray photons from a laser-heated target~\cite{pike.natpho.2014,kettle.njp.2021}.}
    
\subsection{Characterization of strong fields}

Precise characterization of the experimental conditions is essential in any experiment that aims for precise comparisons with theory.
In a laser--electron-beam collision, this encompasses factors such as the electron beam energy, energy spread, size, the collision timing, and the degree of overlap~\cite{samarin.jmo.2017}.
While these differ between conventional and laser-wakefield accelerators, there is a common need to determine the peak intensity of the laser, and the spatiotemporal structure of its focus.
The peak intensity is usually calculated using $I_0 = C_s \Energy_L / (\tau A)$, where the total energy of the pulse $\Energy_L$ may be measured at full amplification, the pulse duration $\tau$ may be obtained by intensity autocorrelation (assuming a particular temporal shape) or by frequency resolved optical gating at low power~\cite{trebino.rsi.1997}, and the area $A$ is obtained from imaging of the focal spot at attenuated power ($C_s$ is a shape constant which accounts for the chosen form of spatial and temporal profile, e.g. Gaussian). This provides a useful upper bound, though direct measurements show that the full structure of the focus is more complicated~\cite{kiriyama.ol.2018, jeandet.jpp.2019} and is affected by factors not evident when measured at low power, e.g. spatiotemporal couplings~\cite{pariente.nphot.2016}.

The need to determine the peak intensity at full power, particularly for next-generation lasers where $I_0 \gtrsim 10^{22}~\Wcmsqd$, has motivated a great deal of recent research.
{It is of great importance for dedicated SFQED experiments, which rely on accurate measurement of laser intensity at the interaction point for the interpretation of experimental results. For example, interactions at intensities lower than expected might be wrongly interpreted as a quantum effect, as the electrons lose less energy than classically expected~\cite{samarin.jmo.2017}.}
The intensity measurements rely on exploiting various, intensity-dependent, physical processes, which we summarize here.  

\paragraph{Ionization of heavy atoms}
Strong electromagnetic fields can induce atomic ionization in a number of ways: bound electrons can escape by absorbing multiple photons from the field~\cite{mainfray.rpp.1991} or, in the quasistatic regime, by quantum tunnelling through the Coulomb barrier~\cite{popov.pu.2004}. Using ionization state as an intensity diagnostic is motivated by the fact that the tunneling rate is exponentially sensitive to the external field strength. For a hydrogen atom in its ground state, for example, the ionization rate $\propto (E_a / \abs{\vec{E}}) \exp[-2 E_a / (3 \abs{\vec{E}})]$, where $\abs{\vec{E}}$ is the strength of the external electric field and $E_a = m^2 e^5 / [(4\pi\varepsilon_0)^3 \hbar^4]$ is a characteristic atomic field strength~\cite{landau.1981}. As that characteristic atomic field also grows stronger with increase of $Z$, the maximal ion charge state observed when the laser focuses in, and ionizes, a low-density gas fill provides a way to measure the field strength~\cite{link.rsi.2006,ciappina.pra.2019}.

\paragraph{Charged particle scattering}
The peak intensity may also be determined by measurement of charged particle scattering, either directly or via the emitted radiation.

The laser light scattered from free electrons, produced by field ionization in the focal region, undergoes an intensity-dependent red-shift~\cite{sarachik.prd.1970}, which can be used to determine the field strength in the relativistic regime $I_0 \lambda^2 \sim 10^{19}~\Wcmsqd \micron^2$~\cite{chen.nature.1998,he.oe.2019}.
However, at higher intensities, electrons are likely to be completely expelled from the focal region in advance of the pulse peak, preventing characterization of the highest field strength.
This could be overcome by the use of protons, rather than electrons, due to their higher mass: \citet{vais.njp.2020} propose the distribution of scattered protons, which originate from a low-density hydrogen gas in the region of the laser focus, as an intensity diagnostic in the regime $10^{21} \lesssim I_0 [\Wcmsqd] \lesssim 10^{24}$.

Alternatively, the laser pulse could be probed by an energetic electron beam, because the particles are more `massive' due to relativistic effects.
During its passage through the laser fields, the electron oscillates with characteristic angle $a(\vec{r}) / \gamma$, where $a(\vec{r})$ is the local value of the normalized amplitude ($\max a(\vec{r}) \equiv a_0$) and $\gamma \gg 1$ is its instantaneous Lorentz factor.
The radiation, which is strongly beamed around the electron's instantaneous velocity~\cite{ritus.jslr.1985,blackburn.pra.2020}, is therefore emitted into an angular range of $\abs{\theta} < a_0 / \gamma$.
(For a linearly polarized pulse, the resulting angular profile is elliptical and elongated along the laser polarization axis.)
Measurement of the radiation angular profile therefore gives access to $a_0$ and so the peak intensity~\cite{harshemesh.ol.2012}, as demonstrated experimentally by \citet{yan.nphot.2017}.
In conjunction with measurement of the mean initial and final electron energy, this scheme can be modified to be independent of radiation reaction or quantum effects, which reduce $\gamma$ by an intensity-dependent amount~\cite{blackburn.prab.2020}.

A qualitatively distinct proposal is to use the lower energy radiation emitted \emph{after} the electron has lost most of its energy, when the trajectory is more sensitive to transverse acceleration, as a diagnostic of peak intensity, duration and polarization~\cite{harvey.prab.2018}.
The electrons themselves also carry information about the peak intensity and duration of the pulse, because they are deflected down gradients in the spatial intensity profile as they traverse the focal region: \citet{mackenroth.njp.2019} show that the maximal scattering angle is controlled by the ratio of the laser amplitude and initial electron energy.

Besides the peak intensity, the laser duration (at full amplification) could be determined by measuring the radiation emitted by a chirped electron beam, colliding at right angles to the laser propagation direction~\cite{mackenroth.sr.2019}.
In principle, the radiation pattern is also sensitive to the carrier envelope phase (CEP) of the laser pulse, which can be inferred for a few-cycle pulse by measuring the radiation's angular aperture~\cite{mackenroth.prl.2010}, or in a pulse that is up to ten cycles long by identification of two, asymmetric peaks in the angular distribution of the radiation~\cite{li.prl.2018}.
CEP is conventionally measured using `$f$-to-$2f$' spectral inference~\cite{apolonski.prl.2000}.

\subsection{Future directions}

Future experimental objectives include precision tests of SFQED predictions and studies of radiation-dominated particle and plasma dynamics.
Various methods have been proposed to obtain stronger evidence of radiation reaction or QED effects, than those discussed in \cref{sec:AllOptical}, either by increasing the signal itself, or by increasing sensitivity to the same. For example, the use of comparatively low energy electrons (40 MeV) in a collision with a high-intensity laser means that radiation detected at an angle greater than $90\degree$ to the collision axis would be a sensitive signal to, and positive evidence of, RR effects~\cite{dipiazza.prl.2009}.
The use of ultrashort (few- or sub-cycle) pulses would dramatically increase the impact of quantum effects by raising the possibility of RR quenching~\cite{harvey.prl.2017}.
Similarly, quantum and classical RR could be distinguished by a laser-electron collision at an angle of $90\degree$, where transmission of electrons through the laser would be forbidden classically~\cite{geng.cp.2019}. 

In laser-electron-beam collision experiments, the principal challenges are beam stability (for all-optical geometries) and maintaining the spatial alignment and temporal synchronization of the two beams.
Alignment and synchronization challenges can be overcome by the use of the retro-reflected driver as the strong-field environment~\cite{ji.ppcf.2019,gu.cp.2018} or by using a microchannel target~\cite{he.commphys.2021}.
Fluctuations can be overcome by means of increased statistics, in particular by monitoring the variance of the electron final energy spread over many shots~\cite{ridgers.jpp.2017,arran.ppcf.2019}.
Alternatively, radiation reaction could be diagnosed in a \emph{single} shot by allowing electron beam to diverge before the collision: as only part of the electron beam collides with the laser, both the pre- and post-collision energy spectra could be measured simultaneously~\cite{baird.njp.2019}.

Reaching high values of the quantum parameter $\chi$ is central for such experiments and we therefore review literature devoted to this issue.
One option would be to co-locate a future laser system with a conventional electron accelerator, which can provide lower divergence and energy spread, as well as higher energies and, therefore, larger values of $\chi$.
In this case, optimizing the angle of collision \cite{blackburn.njp.2019} or the localization of the strong-field region by advanced focusing \cite{magnusson.prl.2019, magnusson.pra.2019} can be used to prevent overwhelming radiation losses for electrons entering the strong-field region.

{Another direction is reaching even stronger electromagnetic fields with focused laser radiation.
While this is primarily a matter of advances in multi-PW laser technology (see \cref{fig:LaserHR} and \cite{danson.hplse.2019}), there are alternative methodologies, which have been investigated theoretically.}

The increase of field strength by means of advanced focusing is fundamentally restricted by the diffraction limit. Thus the conversion of laser light into higher frequency radiation is promising route towards achieving even higher intensity.
We note that {the use of} higher frequency radiation makes quantum regime of interaction more accessible, while also reducing the requirements for radiation power (see \cref{fig:a0_wavelength} and discussion in section \cref{sec:PhysicalRegimes}).
Two major approaches are discussed in the literature: the \emph{relativistic flying mirror}, discussed in \cref{sec:FlyingMirrors}, and \emph{high-harmonic focusing}, discussed in \cref{sec:HighHarmonics}.
We discuss another approach, based on the coherent summation of the fields from several laser pulses, in \cref{sec:DipoleWaves}.

\subsubsection{Relativistic flying mirrors}
\label{sec:FlyingMirrors}

The first approach is based on reflection of an EM wave from a mirror that moves with relativistic velocity.
This was first studied in \citet{einstein.ap.1905} as an example of Lorentz transformations.
The frequency of the light reflected by a counter-propagating mirror is upshifted by a factor of $4\gamma_M^2$, where $\gamma_M$ is Lorentz factor of the mirror.
This is sometimes referred to as a `double Doppler effect'.
A mirror of this type could be formed in a plasma by means of relativistically moving density modulations.
In an underdense plasma, the wake generated by a high-power laser pulse forms a density cusp when approaching wavebreaking, which can reflect EM radiation~\cite{bulanov.prl.2003,martins.ps.2004}, similar to the effect of caustics in optics.
An advantage is that the plasma wake has a parabolic shape~\cite{matlis.natphys.2006}, and therefore acts as a focusing optic \cite{jeong.pra.2021}.
However, as this plasma optic moves with relativistic velocity, the radiation is focused in the moving frame with wavelength $\lambda_0/2\gamma_M$, which leads to only $\gamma_M^2$ additional intensification.
The resulting factor $\gamma_M^6$ needs to be modified by the reflection coefficient of this plasma mirror, which scales as $\gamma_M^{-3}$ or $\gamma_M^{-4}$, depending on how the density cusp is modelled~\cite{bulanov.ufn.2013}.
Several other schemes using underdense plasmas have been proposed, including a spherical plasma wave, which focuses the reflected radiation in the laboratory frame, thus, producing an additional $\gamma_M^2$ factor due to focal spot being defined by the wavelength $\lambda_0/4\gamma_M^2$~\cite{bulanov.pop.2012}.
Proof-of-principle experimental results are presented in \citet{kando.prl.2007, pirozhkov.pop.2007, bulanov.ufn.2013, kando.qbs.2018}. 

Alternatively, a relativistic flying mirror can be formed in the laser-solid interactions usually considered for ion acceleration.
In \citet{kulagin.prl.2007, meyer-ter-vehn.epjd.2009, habs.apb.2008, bulanov.pla.2010}, the laser pulse is assumed to be intense enough to drive all the electrons away from the ions in a form of a single high-density sheet (or a series thereof), as would be the case in the `Coulomb explosion' regime.
In the scheme proposed by \citet{esirkepov.prl.2009}, two laser pulses are incident onto a thin foil: one laser accelerates a thin sheet of electrons as described as above and the second laser pulse is reflected from it, increasing its intensity by a factor of $4 \gamma_M^2$.

Experimental verification of the relativistic flying mirror concept in a laser-foil interaction has yet to be achieved, because of the demanding requirements on laser contrast, power, and intensity.
The difficulty in generating dense electron sheets moving with relativistic velocity is that they are susceptible to transverse expansion and instabilities~\cite{macchi.rmp.2013}.

\subsubsection{Focusing of high harmonics}
\label{sec:HighHarmonics}

Another approach is to focus the high-order harmonics of laser light generated by laser-solid interactions.
Here, the rise and fall of the radiation pressure on the electrons at the surface of a solid target causes oscillatory motion of the laser-plasma interface; in the phase of the motion when the surface moves backwards, towards the incident laser, it functions as a moving mirror and thereby upshifts the frequency of the laser light~\cite{bulanov.pop.1994, gordienko.prl.2005}.
The plasma mirror model does not indicate the existence of any preferred interaction conditions, whereas simulations show that under certain conditions, plasma electrons can be bunched into a layer that generates attosecond-duration, coherent synchrotron emission (CSE)~\cite{anderbrugge.pop.2010}.
These observations can be explained by the relativistic electron spring (RES) model~\cite{gonoskov.pre.2011,gonoskov.pop.2018a}, which further predicts the existence of an optimal laser incidence angle [from 50$^\circ$ to 62 $^\circ$, depending on the preplasma scale length~\cite{blackburn.pra.2018}].
The concept of oblique irradiation of a groove-shaped mirror at the optimal angle is considered in \citet{gonoskov.pre.2011}.

The decrease of laser intensity in the transverse directions (from the center of the focal spot to its edges) can be exploited to induce focusing of the high-harmonic radiation,
due to the varying magnitude of the interface oscillations~\cite{naumova.prl.2004}, the varying displacement of the ions~\cite{vincenti.ncomm.2014}, or the varying depth of radiation penetration into a preplasma (plasma denting)~\cite{dromey.nphys.2009}.
Simulations show that the focused radiation has the form of an electromagnetic burst, the short duration of which inhibit the radiation losses that would prevent reaching very high $\chi$~\cite{baumann.scirep.2019}, and furthermore that the mechanism is robust against laser and target imperfections~\cite{vincenti.prl.2019}.
A 3-PW laser could be used to reach an focused intensity of $10^{25}~\Wcmsqd$~\cite{vincenti.prl.2019}; with further developments, it is argued that an intensity equivalent to the critical field strength, $10^{29}~\Wcmsqd$, is a realistic objective~\cite{quere.hplse.2021}.

\subsubsection{Dipole waves}
\label{sec:DipoleWaves}

Ultrahigh intensity can also be achieved by coherent summation of the fields of several laser pulses, which are focused from different directions to the same spot.
Consider a laser system that generates a pulse with peak power $\Power$, which after focusing generates a peak field amplitude of $a_0^{(1)}$.
Splitting the laser into two beams of equal power $\Power/2$ and focusing them to the same point from opposite directions yields an increased {field} amplitude of $a_0^{(2)} = \sqrt{2} a_0^{(1)}${, purely due to how the energy is redistributed in space}. Extending this concept to four or more lasers, which are arranged in counterpropagating pairs in a common plane with polarization along the plane normal, results in an even higher amplitude of $a_0^{(n)} = \sqrt{n} a_0^{(1)}$. Further intensification can be achieved by adding pairs of laser pulse at some angle to this initial plane. This concept of `multiple colliding laser pulses' (MCLP) was formulated in \citet{bulanov.prl.2010m} as a way of lowering the threshold for nonperturbative electron-positron pair production from vacuum. 
The theoretical limit for such a field configuration is given by a `dipole wave', which provides the strongest possible field strength, $a_0^{(d)} \approx 780 (\Power[PW])^{1/2}$, for fixed total power $\Power$~\cite{bassett.oa.1986}, see also \cite{jeong.oe.2020}. It is so named because the field structure can be viewed as the time-reversed emission of a dipole antenna. The concept was revisited in the context of upcoming experiments by \citet{gonoskov.pra.2012}.
Furthermore, the possibility of practically realizing a dipole wave by the use of six or twelve beams, each with a flat-top disk transverse intensity profile, is discussed in the Appendix of \citet{gonoskov.prl.2014}; the intensity boost is also shown in the inset of \cref{fig:LaserHR}.
{Note that we here relate the increase of electric field strength to the increase of \emph{effective} intensity via quadratic dependency neglecting all the geometrical factors for simplicity.}

The fact that the dipole wave achieves the greatest possible field strength, at fixed power, is advantageous for the study of SFQED effects.
It could be used to drive nonperturbative pair production from the vacuum~\cite{gonoskov.prl.2013} or to trigger electromagnetic cascades~\cite{gelfer.pra.2015}.
In addition, the dipole wave naturally provides particle and plasma confinement by the ART effect (see \cref{sec:Trapping}), which can be further enhanced by cascaded pair production in the focus~\cite{gonoskov.prx.2017}.
QED-PIC simulations show that the development of self-sustained cascade can be triggered in a dipole wave even at a total power of 8 PW.
The high degree of localization of the strong-field region facilitates the creation of highly localized quasi-neutral electron-positron plasma structures, with extreme density (up to $\sim 10^{25}$~cm$^{-3}$) at powers 8 PW $< \Power < $ 15 PW~\cite{efimenko.sr.2018}.
For powers $\Power > $20~PW, simulations indicate that a plasma pinching effect occurs and drives an increase in plasma density up to at least $10^{28}$~cm$^{-3}$ and in the electric (magnetic) field strength up to 1/20 (1/40) of $\Ecrit$~\cite{efimenko.pre.2019}.
Quasistatic magnetic fields generated during the interaction could be of interest for astrophysical connections~\cite{muraviev.jetpl.2015}.

    \begin{figure}
    \includegraphics[width=\linewidth]{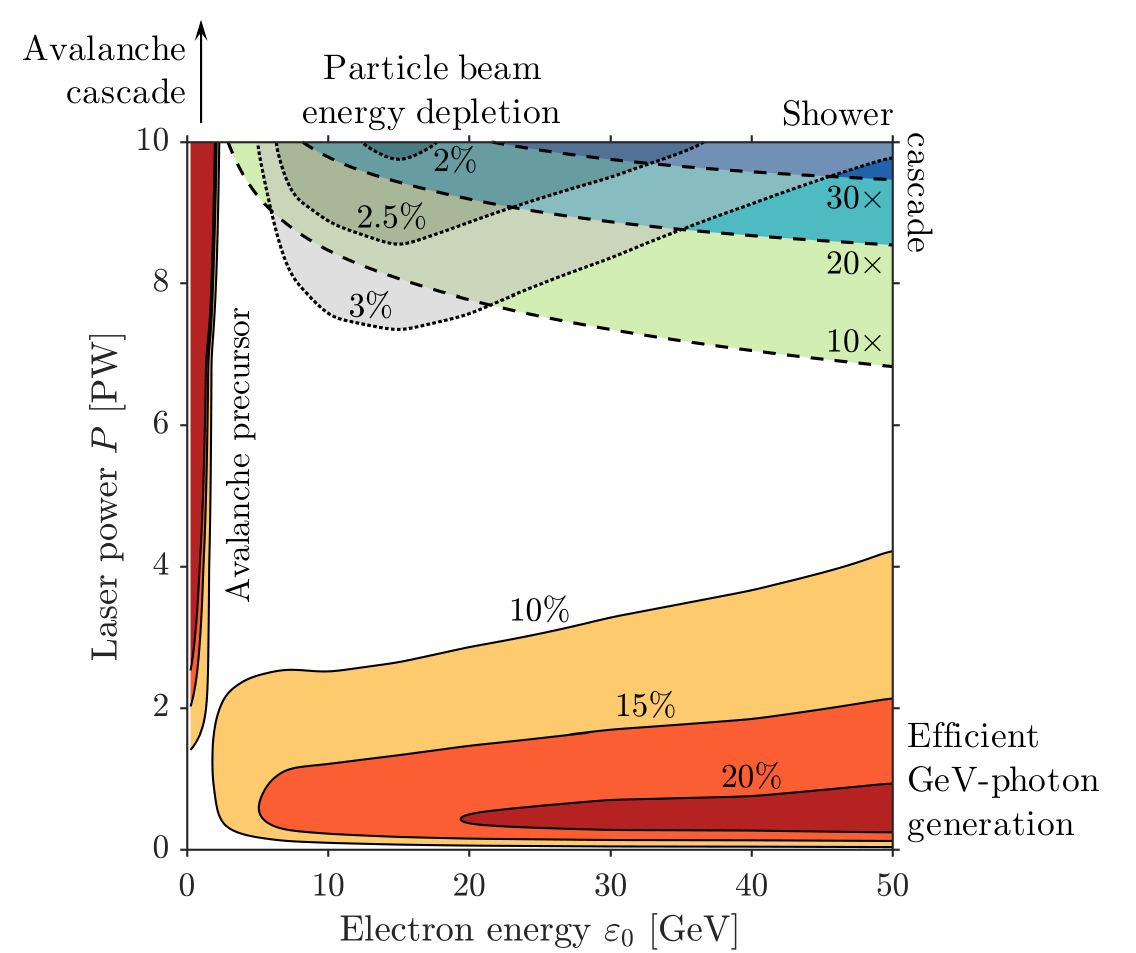}
    \caption{
        Interaction regimes in the collision of an electron beam of energy $\varepsilon_0$ with a dipole wave generated by MCLP with total power $P$:
        number of high-energy ($\hbar \omega > \varepsilon_0/2$) photons per incident electron (red-orange color scale, solid contours);
        number of electron-positron pairs per incident electron (blue-green color scale, dashed contours);
        final electron-beam energy, per cent of initial value (gray color scale, dotted contours).
        Reproduced from \citet{magnusson.pra.2019}.
    }
    \label{fig:regimes}
    \end{figure}

The dipole wave could also be used in an advanced laser-electron-beam collision geometry (compare to \cref{fig:LaserSchemes}):
the advantage is the sub-wavelength-scale confinement of the strong-field region, which means that electrons can penetrate to the center before their energy is exhausted by radiation emission.
This opens up a possibility to study SFQED phenomena, from single-emission to multi-staged processes, including extreme cases of electron-beam energy depletion, and shower- and avalanche-type cascades, where there is significant transformation of the laser or particle-beam energy into secondary particles ~\cite{magnusson.prl.2019, magnusson.pra.2019} and radiation (see \cref{fig:RadiationGeneration}). These different regimes are illustrated in \cref{fig:regimes}. This shows two distinct regions of relevance for high-energy photon generation: an avalanche precursor for lower electron energies, characterized by partial trapping and reacceleration of particles; and a shower-type process for higher electron energies and reduced power. Given the advantages that the MCLP concept provides for SFQED studies, one can argue that future laser facility designs should be build around this concept (see \cref{sec:FutureFacilities}).

\section{Applications}
\label{sec:Applications}

\subsection{Radiation generation}
\label{sec:RadiationGeneration}

In this section, we discuss the practical consequences of the charged particle dynamics discussed in \cref{sec:Dynamics}, with respect to laser-driven radiation sources.
The interest in using high-intensity lasers for this purpose is motivated by the fact that both the total power of the emitted radiation, and the characteristic frequency thereof, increase with the magnitude of the applied acceleration.
This may be characterized by the quantum parameter $\chi$, the transverse acceleration of an EM field in natural units (see \cref{sec:KeyParameters}), which increases with $\gamma$ and $a_0$, the laser normalized amplitude.
The expected scalings for the number of emitted photons, their characteristic frequency, and the total power are: $a_0^2$, $\gamma^2$ and $(\gamma a_0)^2$ in the linear regime $a_0 \ll 1$; and $a_0$, $\gamma^2 a_0$ and $(\gamma a_0)^2$ in the nonlinear regime $a_0 \gg 1$.
Beyond the straightforward increase of these qualities with laser intensity, the radiation properties depend on the interaction geometry.
In a laser-electron-beam collision, one of the two principal geometries discussed in \cref{sec:Geometries}, the radiation properties are largely inherited from those of the electron beam.
In a laser-driven plasma, where the laser fields are both accelerator and target, $\gamma$ depends on $a_0$ and the properties of the plasma (e.g. density and structure), which provides opportunities for optimization.
These topics are also reviewed in~\cite{vranic.pop.2019,gu.mre.2019}.

\subsubsection{Electron-beam driven radiation sources}

    \begin{figure}
    \includegraphics[width=\linewidth]{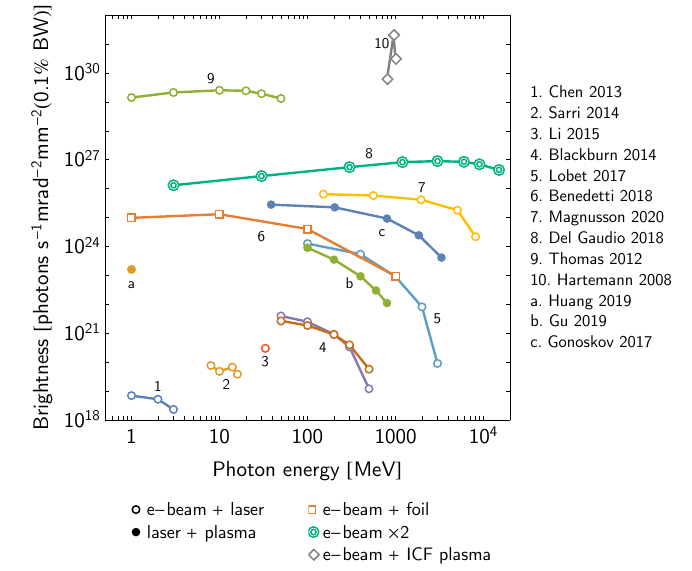}
    \caption{
        Brightness of radiation emitted in high-intensity laser interactions with electron beams (open circles) or plasma targets (filled circles), as well as in non-laser strong-field environments, as (1-2) measured in recent experiments and (3-) predicted by simulations.
    }
    \label{fig:RadiationGeneration}
    \end{figure}

Scattering of laser light by a beam of relativistic electrons is a well-known method to generate radiation with desirable properties~\cite{corde.rmp.2013,albert.ppcf.2016}.
In the linear regime $a_0 \ll 1$, this is generally referred to as inverse Thomson or Compton scattering~\cite{esarey.pre.1993}.
While the radiation is characterized by high monochromaticity, polarization purity and brightness~\cite{hartemann.prstab.2005}, the small size of the Compton cross section means that high electron densities, at the laser interaction point, are required for large yields.
Furthermore, the photon energy is kinematically limited to $\hbar\omega' \simeq 4 \gamma^2 \hbar \omega_0$, where $\omega_0$ is the laser frequency, or $\hbar\omega' [\text{MeV}] \simeq 20\, \Energy_0^2 [\text{GeV}] \lambda^{-1} [\micron]$.
The MeV range, which is of interest for nuclear physics applications, requires GeV-scale electron beams, as available from conventional RF accelerators located at dedicated Compton-photon source facilities, including HI$\gamma$S~\cite{weller.ppnp.2009}, NewSubaru~\cite{utsunomiya.npn.2015} and ELI-NP~\cite{gales.rpp.2018}.
However, the advent of compact laser-driven accelerators producing GeV-electron beams has meant Compton scattering has also been investigated all-optically~\cite{chen.prl.2013,sarri.prl.2014,yan.nphot.2017} (see also \citet{alejo.fp.2019}).

Increasing the laser intensity, i.e. the photon density, increases the yield of $\gamma$ rays but also induces nonlinear effects.
Redshifting (the `mass shift') and the onset of higher harmonics act to reduce and increase the characteristic photon energy respectively, transforming the emission spectrum from narrowband to broad and synchrotron-like.
This makes it possible to create high-flux $\gamma$-ray sources, with photon energies extending up to the initial energy of the electron beam.
A useful characteristic of the radiation is the \emph{brightness}, which is a measure of the photon phase-space density.
Conventionally measured in units of $\text{photons}\,\text{s}^{-1}\text{mrad}^{-2}\text{mm}^{-2}$ (at 0.1\% bandwidth), it gives the energy $\Energy$ radiated per unit photon energy $\hbar\omega$, solid angle $\Omega$, area $A$ and time $t$:
    \begin{align}
    \text{brightness} &=
        10^{-3} \frac{\partial^5 \Energy}{\partial \hbar\omega \, \partial \Omega \, \partial\! A \, \partial t}
    \label{eq:Brightness}
    \\
        &\simeq
        10^{-3} \hbar\omega \frac{1}{\pi \sigma_x \sigma_y} \frac{1}{\pi \theta_x \theta_y} \frac{1}{\tau}
        \frac{\rmd N_\gamma}{\rmd \hbar\omega}
    \label{eq:ApproxBrightness}
    \end{align}
In \cref{eq:ApproxBrightness}, $\frac{\rmd N_\gamma}{\rmd \hbar\omega}$ is the number of photons radiated per unit energy, $\sigma_{x,y}$ is the source size in the transverse directions $x$ and $y$, $\theta_{x,y}$ is the photon divergence in the same directions, and $\tau$ is the duration of the radiation burst.

In the beam-driven, nonlinear regime $\gamma \gg a_0 \gg 1$, the radiation emitted by a beam of electrons that collides with a linearly polarized laser pulse is directed into an ellipse with opening half-angles $\theta_x \simeq a_0/\gamma$ and $\theta_y \simeq 1/\gamma$, where the major axis $x$ is oriented along the laser polarization~\cite{harshemesh.ol.2012}.
The source size $\sigma_{x,y}$ may be estimated the smaller of the radius of the laser focal spot or the radius of the electron beam and is generally a few microns in magnitude.
Similarly, the duration $\tau$ is inherited from the duration of the electron bunch.
For LWFA electron beams, energies up to a few GeV, durations of tens of femtoseconds~\cite{lundh.nphys.2011}, beam radii at focus of a few microns~\cite{schnell.prl.2012} are now achievable.
In \cref{fig:RadiationGeneration} we show recent experimental results~\cite{chen.prl.2013,sarri.prl.2014} and theoretical predictions~\cite{li.prl.2015,blackburn.prl.2014,lobet.prab.2017,thomas.prx.2012} for the $\gamma$-ray brightness achievable in a collision of an electron beam and a single laser pulse.

At electron energies and laser intensities large enough for $\chi \gtrsim 1$, it is likely that the highest energy photons decay to electron-positron pairs before escaping the laser pulse.
This loss can be reduced by reducing the effective size of the interaction region.
Thus we also show predictions for an electron beam colliding with a dipole wave~\cite{magnusson.prl.2019} (approximate $4\pi$ focusing achieved by colliding multiple laser pulses, see \citet{gonoskov.pra.2012}), which maximises the field strength in a region that is sub-wavelength in size.
The same processes may be driven in non-laser strong-field geometries:
in \citet{benedetti.nphot.2018}, the magnetic fields generated by filamentation of an electron beam travelling through a solid foil are shown to induce bright, synchrotron-like emission;
\citet{delgaudio.prab.2019} study the collision of two dense electron bunches, each emitting photons in the boosted Coulomb fields of the other bunch (see \cref{sec:BeamBeam}).

For applications that demand monoenergetic $\gamma$ rays, but also high photon flux, alternative approaches must be explored.
Redshifting of the harmonics may be explained as the electron acquiring an intensity-dependent effective mass $m_\star = m \sqrt{1 + a_\text{rms}^2}$~\cite{kibble.prl.1966,harvey.prl.2012}, where $a_\text{rms} = a_0$ (circular polarization) or $a_0/\sqrt{2}$ (linear polarization).
The variation of this redshift over the pulse envelope $g(t)$ spoils the monochromaticity of the individual harmonics once $a_0 \gtrsim 1$.
In order to benefit from the increased photon yield at higher laser intensity, it has been proposed to chirp the pulse such that the `local frequency' $\omega_0(t) = \omega_0 [1 + a_\text{rms}^2 g^2(t)]$~\cite{gheb.prstab.2013,terzic.prl.2014,rykovanov.prab.2016,seipt.prl.2019}.
This compensates for the effective mass increase, but is yet to be experimentally demonstrated.

\subsubsection{Laser-driven radiation sources}

In the electron-beam driven case, the energy of the radiation is largely supplied by the pre-accelerated electrons.
The laser acts only as the undulator (wiggler) target, thereby separating the acceleration and radiation stages from one another.
In the laser-driven case, these stages are combined and the energy of the radiation is drawn ultimately from the laser itself.
The combination of simultaneous energy gain and radiation is already a feature of laser-plasma interactions in the classical regime: for example, oscillation of electron bunches accelerated in laser wakefields gives rise to femtosecond bursts of bright, broadband (multi-keV) X rays, named \emph{betatron} radiation~\cite{kiselev.prl.2004,rousse.prl.2004} (see review by \citet{corde.rmp.2013}).

In the low-intensity limit, the response of a plasma is determined by the parameter $n_e / \ncrit$, where $\ncrit = \epsilon_0 m \omega_0^2 / e^2$ is the classical critical density at frequency $\omega_0$. Plasmas with $n_e / \ncrit > 1$ are \emph{overdense}, i.e. opaque to the laser light. However, this threshold is altered at high intensity when the electron motion becomes relativistic. In the case of a thin, $\ell \ll \lambda$, solid, $n_e\gg \ncrit$ , density foil the value of $a_0$ should exceed $\pi(n_e/\ncrit)(\ell/\lambda)$, where $\ell$ is the thickness of the target, for the foil to become transparent \cite{vshivkov.pop.1998}. In the case of extended ($\ell > c \tau_L$), near-critical density ($n_e\sim \ncrit$) targets, $a_0$ should exceed $13.5 (n_e/\ncrit)(\ell/c\tau_L)$ \cite{bulanov.pop.2010}, where $\tau_L$ is the laser pulse duration. In both cases realistic parameters for the targets lead to $a_0$ being around several hundred.

Thus, the relevance for radiation generation is as follows:
at higher intensity, increased penetration of denser targets means that a larger number of electrons are driven by the laser fields.
As the emission rate of each individual electron also increases with intensity, dramatic growth of the photon flux and conversion efficiency $\eta$ are expected.
We outline the derivation of an approximate scaling law for $\eta$, adapted from~\citet{popruzhenko.njp.2019}.
Consider a circularly polarized laser with amplitude $a_0$ and frequency $\omega_0$, illuminating a plasma with electron areal density $\sigma_e = n_e c / \omega_0$ (i.e $n_e$ electrons per unit volume in a single skin depth).
The radiation intensity, $I_\text{rad}$, is given by the product of the areal density and the radiation power of a single electron: $\Power_e = (2\alpha m c^2 / 3 \tau_C) \chi^2$, where for circular motion the quantum parameter $\chi = \gamma^2 [\hbar \omega_0 / (m c^2)]$.
Comparing $I_\text{rad}$ to the laser intensity $I_0$, we have $\eta = I_\text{rad} / I_0 = (2\alpha/3) (\gamma^4 /a_0^2) (n_e/\ncrit) [\hbar \omega_0/ (m c^2)]$.
A self-consistent solution for $\gamma(a_0)$, which accounts for radiation losses classically, is given in \cref{eq:Zeldovich}~\cite{zeldovich.spu.1975}.
For $n_e = a_0 \ncrit$, this predicts $\eta \propto a_0^3$ for $\erad a_0 < 1$ and $\eta \to 1$ as $a_0 \to \infty$.
The conversion efficiency is substantial for $a_0 \gtrsim 439\, \lambda^{1/3} [\micron]$, or equivalently $\Power [\text{PW}] \gtrsim 13.0\, w_0^2 [\micron] \lambda^{2/3} [\micron]$, where $w_0$ is the focal spot size.

    \begin{figure}
    \includegraphics[width=\linewidth]{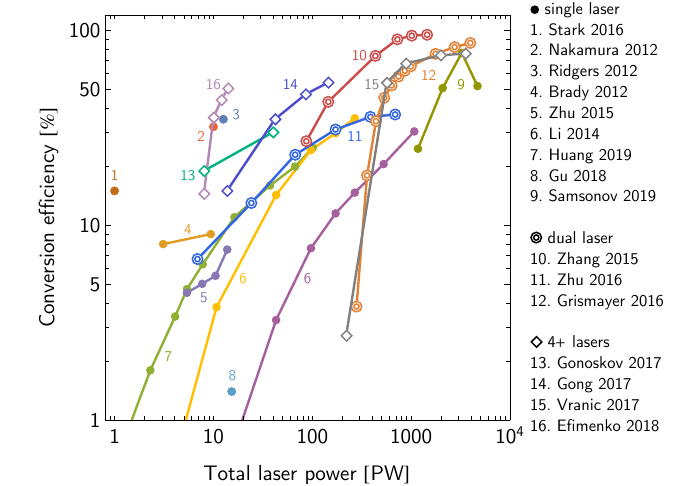}
    \caption{
        Efficiency of $\gamma$-ray generation in plasmas driven by single (filled circles), dual (double circles) and multiple (diamonds) lasers, as predicted by simulations.
    }
    \label{fig:ConversionEfficiency}
    \end{figure}

These estimates are borne out by a wide range of simulation studies in the recent literature.
In \cref{fig:ConversionEfficiency} we show the predicted $\gamma$-ray generation efficiency for a variety of laser-driven interaction geometries, as a function of the total laser power.
The majority of these results are obtained using PIC codes extended to include QED processes (see \cref{sec:Numerics}).
\Citet{ridgers.prl.2012} and \citet{nakamura.prl.2012} predict $\eta \simeq 0.3$ for a single, $10$-PW laser pulse interacting with an aluminium foil or an exponentially rising plasma density ramp, respectively: in both cases, the plasma density is somewhat larger than $a_0 \ncrit$ and the laser is linearly polarized, which maximises electron heating.
The interaction of circularly polarized lasers with thick plasma targets has been investigated by \citet{liseykina.njp.2016, popruzhenko.njp.2019}.

A disadvantage of single-laser irradiation is that the radiation pressure of a high-intensity laser causes longitudinal motion of the plasma electrons, reducing their $\chi$ parameter.
One way to overcome this is to irradiate a thin plasma target from both sides, with two equally intense laser pulses: the symmetric radiation pressure arrests longitudinal motion and by optimising the electron density, penetration of the laser fields throughout the plasma can be achieved, as well as high (50\%) conversion efficiency to gamma radiation~\cite{bashinov.pop.2013}.
The conversion of laser energy to $\gamma$ radiation can be so efficient that a target expected to be relativistically transparent can instead become optically opaque~\cite{zhang.njp.2015}.
The logic of multi-sided irradiation can be extended to four~\cite{gong.pre.2017,grismayer.pre.2017} or even twelve laser pulses~\cite{gonoskov.prx.2017}, creating a two or three-dimensional optical trap for energetic plasma.
Increased axial symmetry of the particle motion in such a field configuration can then enhance the directionality of the radiation emission.
The laser and plasma parameters can be chosen to optimise, e.g., electron-positron pair creation~\cite{vranic.ppcf.2017} or GeV photon production~\cite{gonoskov.prx.2017, efimenko.sr.2018}.

The use of structured targets provides an opportunity to achieve high conversion efficiency while reducing the required laser power.
For example, \citet{taphuoc.np.2012} report how a laser pulse propagating through a gas jet placed in front of a plasma mirror, accelerates, and then collides with, a trailing electron bunch, after the laser pulse is reflected from the mirror.
Compton scattering of the electrons then leads to the emission of bright, forward-directed x rays (see also \cite{dopp.ppcf.2016}).
This `retro-reflection' geometry has been studied via simulation, at much higher laser intensity, as a means of generating $\gamma$ rays and electron-positron pairs~\cite{gu.cp.2018,huang.njp.2019}, as counterpropagation of laser and electron beam maxmises the quantum parameter $\chi$.
Maintaining overlap between the laser pulse and the energetic electrons is a crucial part of achieving high conversion efficiency to synchrotron radiation.
One way to do so is to exploit RR-induced trapping (see \cref{sec:Trapping}) in a conical target, made of a high-density material such as gold and filled with a near-critical density hydrogen plasma, which is irradiated by laser pulse~\cite{zhu.njp.2015}.
Electrons are accelerated to GeV energies and confined close to the axis, where the cone has focused the laser pulse to higher amplitude~\cite{zhu.njp.2015}.
Two such cones, pointed toward each other and irradiated from both sides, could be used to generate $\gamma$ rays with energy densities of order $10^{18}~\text{J}\text{m}^{-3}$~\cite{zhu.ncomm.2016} or high orbital angular momentum~\cite{zhu.njp.2018}.
The relative importance of the laser and plasma fields in driving radiation emission is investigated by \citet{stark.prl.2016} in simulations of a plasma channel target.
It is shown that it is a self-generated, quasistatic magnetic field in the channel which allows for high conversion efficiency, as this prevents cancellation between the electric- and magnetic-field contributions to $\chi$ for a forward-propagating electron; tens of TW of $\gamma$ rays are predicted for a laser intensity of $5 \times 10^{22}~\Wcmsqd$~\cite{stark.prl.2016}.

Characteristic of the synchrotron radiation discussed in this section is that the emission is confined to the duration of the laser pulse. This is typically tens of femtoseconds, unlike the radiation from bremsstrahlung of hot electrons, which can persist for picoseconds after the initial interaction due to refluxing~\cite{compant.pop.2013}.
{Numerical solutions of high-intensity lasers impinging on plasma targets show that synchrotron emission dominates over bremsstrahlung particularly for thin targets, as well as at higher intensity~\cite{martinez.prr.2020}.}
Furthermore, the electron population necessary to support efficient radiation generation can be generated by the interaction itself: the development of a QED cascade, when the fields are strong enough that the radiated photons are likely to pair-create before leaving the interaction region, can lead to near-total absorption of the laser energy (see \cref{sec:Cascades}).

\subsection{Positron sources}
\label{sec:Positrons}

    \begin{figure}
    \includegraphics[width=\linewidth]{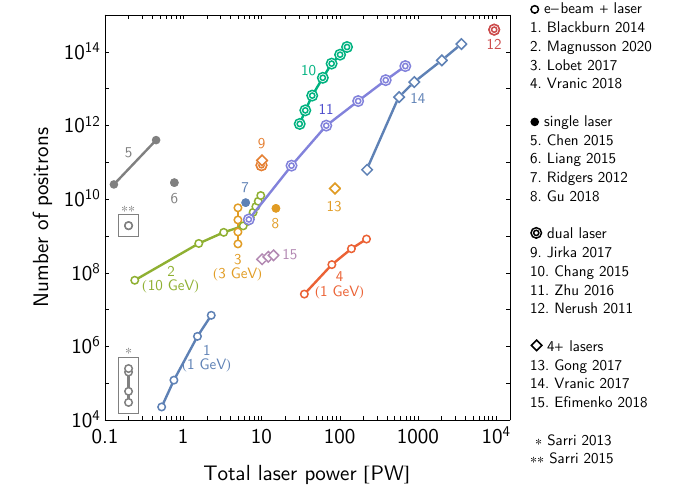}
    \caption{
        Number of positrons produced in high-intensity laser-plasma interactions.
        For laser-electron beam interactions (open circles), the energy of the electron beam is noted in brackets.
        {Points marked with asterisks indicate experimental results from LWFA electron-beam interactions with high-Z foils; in these cases the laser power is not indicated.}
    }
    \label{fig:PositronYield}
    \end{figure}

It is been long argued that the problem with the production of positrons in laser-plasma interactions might be one of the bottlenecks for a plasma-based lepton collider. 
There are several proposals how to create positrons in laser-plasma interactions for different applications. The most straightforward way was demonstrated in \citet{chen.prl.2009,chen.prl.2015,liang.sr.2015}, where a moderate intensity laser is used to irradiate a solid-density target several millimeters thick.
As a result, energetic electrons, created at the front surface of the target, travel through the target emitting bremsstrahlung photons. These photons are converted into electron-positron pairs by colliding with nuclei.
Another interesting effect connected with this positron source is that the positrons are accelerated by the sheath fields (typical for TNSA) at the back of the target.

The production of positrons from solid-density foils, but with an externally accelerated electron beam, is a typical procedure used in conventional accelerators.
It has recently been demonstrated using laser-wakefield accelerated electron beams~\cite{gahn.apl.2000,sarri.prl.2013,sarri.ncomm.2015}.
Such a setup might be advantageous for plasma-based positron acceleration since it allows for the efficient coupling of produced positrons to the acceleration stage, whether it is laser- or electron-beam-driven (see also \cite{roadmap.doe.2016,alegro.arxiv.2019,alegro.arxiv.2020}).

The use of high-intensity lasers and SFQED effects to produce dense bunches of electron-positron pairs is unavoidably coupled to the production of high-energy photons, which is discussed in \cref{sec:RadiationGeneration}.
A common point to these geometries is that high-energy photons are produced by Compton scattering inside a region of strong field: thus it is possible for the photon to `decay' via the nonlinear Breit-Wheeler process before escaping the interaction region.
The dependence on field strength, i.e. laser intensity, for pair production is much stronger than for photon emission. Note that the scale for the number of positrons in \cref{fig:PositronYield} covers ten orders of magnitude for only three orders of magnitude in laser power.
There are two reasons for this:
first, the probability of pair creation is exponentially suppressed at low energy and field strength (see \cref{sec:LCFA}), which means that only the highest energy photons are likely to decay;
second, the production of high-energy photons is itself sensitive to the highest field strength achieved in the interaction.

{As one electron is produced for every positron in a pair-creation event, sufficiently prolific pair production should lead to the formation of a `pair plasma'.
In experiments where LWFA electron beams interact with high-$Z$ materials, these manifest themselves as relativistic, neutral beams~\cite{sarri.ncomm.2015,warwick.prl.2017}.}
The possibility of dense positron bunch production via the EM avalanche, discussed in \cref{sec:Cascades}, was first pointed out in \citet{bell.prl.2008}.
Using a single laser and a solid target, thereby exploiting the reflection of the laser to form a standing wave, to generate positrons was proposed by \citet{ridgers.prl.2012}, in which a positron density of $10^{26}~\text{m}^{-3}$ is reported.
Since then, the combination of multiple laser pulses and structured plasma targets has become a rich field of study for numerical simulations: we present an overview of recent results in \cref{fig:PositronYield}, where the number of interacting laser pulses is one~\cite{ridgers.prl.2012,gu.cp.2018}, two~\cite{jirka.sr.2017,chang.pre.2015,zhu.ncomm.2016,nerush.prl.2011} or more than four~\cite{gong.pre.2017,vranic.ppcf.2017,efimenko.sr.2018}.
{These are compared with the results of experimental campaigns using LWFA electron-beam--driven~\cite{sarri.prl.2013,sarri.ncomm.2015} or laser-illuminated~\cite{chen.prl.2015,liang.sr.2015} high-$Z$ targets.}

Those results using electron beams and lasers correspond to the shower-type cascade described in \cref{sec:Cascades}, i.e. the energy to produce the positrons is supplied by the electron beam rather than the laser itself.
The advantages are that the positrons are produced as a collimated beam and the required laser intensity is much lower~\cite{blackburn.prl.2014,lobet.prab.2017,vranic.sr.2018,magnusson.prl.2019}.
Radiation reaction means that these positrons are expected to have lower energy than the initial electron beam.
We note, however, that a scheme to generate and \emph{accelerate} positrons is described in \citet{vranic.sr.2018}, in which simulations of an electron beam colliding at $90^\circ$ to an intense laser are presented.
It is worth noting that, if the goal of an experiment is to generate high-energy photons, pair creation instead represents a loss mechanism.
The use of highly confined field configurations, such as dipole waves, is proposed as a way to mitigate this loss~\cite{gonoskov.prx.2017,magnusson.prl.2019}.

\subsection{Polarized particle beams}
\label{sec:PolarizedBeams}

The elementary processes of photon emission and electron-positron pair creation depend on the spin of the participating particles.
While implementation of these processes in simulations using unpolarized rates, which are averaged (summed) over initial (final) spin states, remains standard, the desire to study polarization effects has prompted extension of these codes to include spin-dependent emission and spin precession, in particular.
The underlying theory is reviewed in \cref{sec:Spin}; here we discuss the results and implications of this work.

Electrons in a static magnetic field can have their spins aligned parallel or antiparallel to the magnetic field, taking this as the canonical quantization axis.
The asymmetric probability of photon emission associated with a change of the electron spin (a spin-flip transition) leads to radiative polarization of the electron population, in the so-called Sokolov-Ternov effect~\cite{SokolovTernov}.
In storage rings with GeV electron beams and magnetic fields of a few Tesla, the characteristic polarization time ranges from minutes to hours~\cite{mane.rpp.2005}.
An equivalent phenomenon is predicted to occur at the magnetic node of the standing wave formed by two counterpropagating, circularly polarized lasers~\cite{delsorbo.pra.2017}: the magnitude of the rotating electric field is such that the polarization time is only a few femtoseconds.
The orbit at this node is, however, unstable~\cite{kirk.ppcf.2016}, which restricts the degree of polarization attainable to approximately 30\%~\cite{delsorbo.ppcf.2018}.

Radiative polarization should also occur in the collision between a high-energy electron beam and a single laser pulse.
However, as the magnetic field in a linearly polarized laser oscillates, the polarization built up in one half-cycle is lost in the following half-cycle;
even for ultrashort laser pulses, where the asymmetry between field oscillations is greatest, the degree of polarization is limited to less than one per cent~\cite{seipt.pra.2018}.
To increase this, it is necessary to break that symmetry, which could be accomplished by using a two-colour laser pulse~\cite{song.pra.2019,seipt.pra.2019}.
Simulations indicate that an \emph{average} polarization degree of up to 17\% could be achieved in the collision of a 5-GeV electron beam with a 160-fs laser pulse that has $a_0 = 16$ and 30\% of its energy in the second harmonic~\cite{seipt.pra.2019};
or 11\% for a 2-GeV electron beam and a 30-fs laser pulse with $a_0 = 100$ and 25\% of its energy in the second harmonic~\cite{song.pra.2019}.
However, as the radiative energy loss and spin polarization are correlated, post-selection of electrons below a certain energy can be used to increase the polarization degree to 60\%, at the cost of reducing the yield~\cite{song.pra.2019,seipt.pra.2019}. {Recent studies also suggest that the spin-field coupling can be enhanced by the use of short, quasi-unipolar laser pulses \cite{aleksandrov.pra.2020, rosanov.jetl.2021}.}

Alternatively, spin-dependent radiation reaction in an elliptically polarized (EP) laser pulse can cause an initially unpolarized electron beam to split into two angularly separated beams with opposite polarization degrees:
provided the degree of ellipticity is a few per cent, and the laser pulse duration is approximately five cycles, post-selection of one of the beams yields a polarization degree of up to 65\% for an electron initial energy of 4 GeV and $a_0 = 100$~\cite{li.prl.2019}.
An asymmetry in the angular profile of the radiation, emitted when an initially polarized electron beam collides with such an EP laser pulse, has been proposed as a means of diagnosing the polarization degree of that electron beam~\cite{li.prap.2019}.
We note that in all the above cases, the spin polarization occurs in the transverse direction.

Polarized electron beams then can be used to generate $\gamma$-ray beams with certain polarization properties~\citet{li.prl.2020po}.
These photons could be used to produce polarized positron beams, because at high intensity they could `decay' to electron-positron pairs before escaping the laser pulse, and the pair creation process is itself spin-dependent (see \cref{sec:ComptonBreitWheeler}).
In a two-colour laser pulse, with $a_0 = 83$ and 25\% of its energy in the second harmonic, \citet{chen.prl.2019} show that, while an initially unpolarized 2-GeV electron beam acquires an average polarization degree of only 8\%, the positrons produced by the photons it emits have a much higher polarization degree of 60\%.
This is because the pair creation rate is more sensitive to spin than the photon emission rate~\cite{chen.prl.2019}.
The same cause underlies polarized positron production in a laser pulse with a small degree of ellipticity: \citet{wan.plb.2020} show that an initially unpolarized 10-GeV electron beam colliding with an EP laser pulse with $a_0 = 100$ drives the creation of two, angularly separated positron beams with oppositely oriented polarizations.
A yield of positrons equivalent to 1\% of the initial electron beam is estimated, with a polarization degree of 86\%~\cite{wan.plb.2020}.
If, however, a pre-polarized electron beam is available, photon emission and pair creation in a circularly polarized laser can transfer a high degree of \emph{longitudinal} polarization (40--65\%) to the created positrons~\cite{li.prl.2020pr}.

\subsection{Ion acceleration}
\label{sec:Ions}

Laser-based ion acceleration (see \citet{mourou.rmp.2006,macchi.rmp.2013,daido.rpp.2012,bulanov.pu.2014, bulanov.pop.2016} and references cited therein) has received considerable attention over the last two decades for the potential applications to diverse research areas: fundamental particle physics, inertial confinement fusion, warm dense matter, medical therapy, etc. It is expected that with the fast development of multi-PW laser facilities, laser ion acceleration will be able to generate ion beams with energies in excess of 100 MeV, as required by many applications. Until now, laser systems were only able to achieve the acceleration of ions with energies approaching 100 MeV~\cite{kim.pop.2016,wagner.prl.2016,higginson.ncomm.2018}. While most of the experimental results were obtained in the Target Normal Sheath Acceleration (TNSA) regime~\cite{snavely.prl.2000, maksimchuk.prl.2000, wilks.pop.2001, fuchs.nphys.2006}, higher ion energies are expected to be generated by employing advanced regimes of laser ion acceleration, as well as different targets ranging from nm-scale solid density foils to near-critical density (NCD) slabs, gas jets, and liquid jets. These regimes include, to name a few, \emph{radiation pressure acceleration} (RPA)~\cite{esirkepov.prl.2004,henig.prl.2009,kar.prl.2012,bin.prl.2015}, \emph{shockwave acceleration} (SWA)~\cite{haberberger.nphys.2012}, \emph{relativistic transparency} (RIT)~\cite{palaniyappan.nphys.2012}, \emph{magnetic vortex acceleration} (MVA)~\cite{kuznetsov.ppr.2001} and \emph{chirped standing-wave acceleration}~\cite{mackenroth.prl.2016}. Analytical and computer simulation estimates show that a PW or several PW laser system may be able to generate ions with energies ranging from several hundred MeV to GeV per nucleon in these regimes (see \citet{bulanov.pop.2016} and references cited therein). We note that NCD targets, as well as composite targets with NCD parts, have attracted a lot of attention recently not only for ion acceleration \cite{bulanov.prl.2015, bin.prl.2015}, but also for brilliant gamma-ray and electron-positron pair production: see \cref{sec:RadiationGeneration,sec:Positrons}.

All these results rely on the physics of intense laser pulse interactions with NCD plasma, the basics of which are best illustrated by MVA. At the intensities employed in the above mentioned papers, the effects of the radiation reaction should begin to manifest themselves. 
However, laser-driven ion acceleration usually does not attract a lot of attention, from the point of view of radiation reaction. Moreover, the analytical treatment of the laser ion acceleration is limited to phenomenological, one-dimensional models, which employ different simplifications to be solvable. The inclusion of the radiation reaction in these models is questionable, since it is not clear if the multi-dimensional effects that are neglected, for example, are more significant than those of radiation reaction. Nevertheless, computer simulation studies of different scenarios of laser ion acceleration, performed with PIC codes, make it significantly easier to include RR effects. Thus, most of the studies on RR effects in laser ion acceleration employ some kind of numerical description of RR, ranging from classical equations of motion (Landau-Lifshitz) coupled to a 1D PIC code, to fully 3D PIC codes with QED modules: see \cref{sec:Numerics}.

In most cases, two characteristic targets are considered: a thin, solid-density foil; and a near-critical density plasma. In the former, the most relevant mechanism is RPA; in the latter, it is either hole boring (HB) or MVA.

RPA comes into play when the laser is able to push the foil as a whole by its radiation pressure. The idea has a close analogy to the ``light sail'' scheme for spacecraft propulsion \cite{zander.tiz.1924,marx.nature.1966}. The RPA is the realization of the relativistic receding mirror concept~\cite{bulanov.ufn.2013}, where the role of the mirror is played by an ultra-thin solid-density foil.
A simple 1D model of RPA describes the foil motion by the following equation~\cite{bulanov.pop.2016}:
\begin{equation}
\frac{1}{\left(1-\beta^2\right)^{3/2}}\frac{d\beta}{dt}=\frac{(2|R|^2+|A|^2)}{2}\frac{m_e}{m_i}\frac{a^2(\psi)}{\varepsilon_p}\frac{1-\beta}{1+\beta}
\end{equation}
Here $\psi=x- c t$, $m_i$ is the ion mass, $\beta$ is the foil velocity, $R$ is the reflection coefficient, and $A$ is the absorption coefficient. These coefficients are connected through the energy conservation condition, $|R|^2 +|T|^2 +|A|^2=1$, where $T$ is the transmission coefficient. The parameter $\varepsilon_p$ is defined as $ \varepsilon_p= (2 \pi e^2 n_e \ell)/(m_e \omega c)$, where $\ell$ is the target thickness. It governs the transparency of the foil, i.e. $R= R(\varepsilon_p)$: if $a_0 \ll \varepsilon_p$, the foil is opaque to radiation; if $a_0\gg \varepsilon_p$, the foil is transparent to radiation.

It is plausible that the RR affects both the absorption and reflection of the incident laser pulse. The absorption will be modified through the multiphoton Compton process, i.e., the emission of high energy photons by the electrons in strong EM field, thus, providing the transformation of laser energy into $\gamma$'s. The reflection will be modified by the creation of electron-positron pairs, through the multiphoton Breit-Wheeler process.
Since the absorption is less effective in coupling laser energy to ions and increased opacity leads to a similar effect, it can be expected that the RR effects lead to a decrease in maximum attainable ion energy. This decrease of ion energies has been observed in 1D, 2D, and 3D PIC simulations with classical RR taken into account~\cite{tamburini.njp.2010,tamburini.nima.2011,chen.ppcf.2011,wu.pop.2015,delsorbo.njp.2018}.
These studies show the contraction of the particle phase space due to RR, which manifests itself as a cooling mechanism. However, simulations show that the effects of RR were significant only in the case of linear polarization of the laser pulse, in which case they lead to a significant reduction of the maximum ion energy. The circular polarization case does not demonstrate any strong influence of RR effects on the process of ion acceleration. 

Further studies employed QED-PIC codes to model the interaction of intense laser pulses with solid-density targets~\cite{ridgers.prl.2012, brady.pop.2014,zhang.njp.2015,delsorbo.njp.2018,duff.ppcf.2018,samsonov.sr.2019}, with the production of bright $\gamma$-rays, laser pulse depletion, and electron-positron cascade development being observed.
In \citet{zhang.njp.2015}, a model for the relativistic transparency evolution in the presence of SFQED processes is developed. It predicts that a foil, initially transparent for radiation, will turn opaque in strong fields. A significant reduction of the maximum ion energy in the case of linear polarization, as well as no effect in the case of circular polarization, during RPA are also observed in simulations. 

Modelling of the laser ion acceleration in the HB regime ~\cite{nerush.ppcf.2015,delsorbo.njp.2018,yano.pop.2019}, as well as in underdense plasma~\cite{ji.njp.2014,wang.pre.2017,wu.pop.2018}, shows effects similar to those in thin-foil target interactions, i.e. laser energy depletion due to cascade development, QED-induced opacity, and the decrease of charge separation fields in plasma.

\bib

%% file: conclusions.tex
\section{Outlook}

\subsection{Open questions}
\label{sec:OpenQuestions}

The state-of-the-art understanding of SFQED phenomena is based on a number of approximations.
While in most forseeable experiments, it should be sufficient to characterize the observed phenomena at least at the qualitative level, there are many indications that we are reaching the limits of validity of these approximations.
How to go beyond these {approximations are} among the most pressing questions in the field.

\subsubsection{Theoretical questions}

The plane-wave approximation is a backbone of almost all analytical results in SFQED theory. It is connected with the fact that the Dirac equation for an electron moving in strong EM plane wave has an exact solution, which allows derivation of analytical formulae for the probabilities of quantum process. The existence of an exact solution in this case is due to the fact that plane waves are null fields with a high degree of symmetry~\cite{heinzl.prl.2017}. However, the high-power laser facilities able to deliver high-intensity EM pulses needed for SFQED studies achieve extreme field strengths by focusing laser light to spots comparable in size to the laser wavelength.
Tight focusing leads to significant departures from the plane-wave idealisation.
Furthermore, there is considerable interest in the collisions of multiple laser beams, a field configuration which has many fewer symmetries. This is even more so for EM fields generated in plasma. Thus, looking for new (even approximate) analytic solutions to the Dirac equation in more realistic interaction configurations is necessary.
Initial progress in this direction includes focusing~\cite{dipiazza.prl.2014,dipiazza.prl.2016,dipiazza.pra.2017}, standing waves~\cite{heinzl.prd.2016,king.prd.2016,lv.prr.2021}, rotating electric fields~\cite{raicher.plb.2015,raicher.prr.2020} and plasma environments~\cite{raicher.pra.2013,mackenroth.pre.2019}.

The external field approximation is usually assumed when calculating almost all SFQED processes probabilities, i.e., the external field has infinite energy and does not change as a result of photon emission or pair production. However, at certain conditions these processes can lead to the depletion of the EM field energy, which invalidates the external field approximation~\cite{narozhny.pla.2004, bulanov.pre.2005, bulanov.prl.2010s, nerush.prl.2011, seipt.prl.2017, ilderton.prd.2018, heinzl.prd.2018}. Here one should distinguish between the quantum depletion, mentioned above, and the classical one, which is due to the particle acceleration by the EM field over longer timescales~\cite{meuren.prd.2016}.
A self-consistent analytical framework which accounts for the backreaction of quantum processes, and the particles they produce, on the EM fields needs to be created~\cite{ilderton.prd.2018,heinzl.prd.2018}.

In most of the interactions between particles and high-intensity EM fields under consideration, it is plausible that particles undergo multiple quantum processes.
Depending on the interaction setup, this could constitute either an avalanche- or shower-type cascade (see \cref{sec:Cascades}).
The cascades give rise to a number of interesting phenomena, such as radiative trapping, attractors and chaos in charged particle motion, the generation of positron and photon sources, and the extreme case of plasma structures. Analytical and computational studies of these processes usually rely on the LCFA. A full QED description is limited to two-stage processes such as trident and double Compton: see references in \cref{sec:HigherOrderProcesses}. 
A full QED treatment of multi-staged cascade process is still lacking, but recently progress has been made in considering beyond-two-vertex processes~\cite{dinu.prd.2020}.

The importance of higher order processes is intrinsically linked to the Ritus-Narozhny conjecture of pertubative SFQED breakdown at $\alpha \chi^{2/3} > 1$~\cite{ritus.ap.1972,narozhny.prd.1980,fedotov.jpcs.2017}.
If one considers the probabilities of Compton and Breit-Wheeler processes at such field strength, one finds that the mean free paths for electrons, positrons, and photons are of the order of the Compton wavelength.
It is questionable, therefore, whether a classical particle trajectory is a useful concept anymore.
However, recent studies show that there is no universal behavior for Compton and Breit-Wheeler processes as $\chi\rightarrow\infty$ \cite{ilderton.pra.2019,podszus.prd.2019}. Further studies are needed to determine the exact behavior of SFQED processes in extreme fields and their scalings, which should depend on various initial conditions.

\subsubsection{Simulation developments}

Simulations of the SFQED phenomena are generally based on the LCFA. As discussed in \cref{sec:LCFA}, this is due to the fact that the formation length/time of QED processes is much smaller than the typical inhomogeneities of the EM field in laser-plasma interactions, at high intensity.
Thus, these processes can be treated as occurring instantaneously and the local value of the EM field strength can be used to calculate the probabilities.
This is expected to be valid if $a_0\gg 1$ and $a_0^3/\chi \gg 1$~\cite{ritus.jslr.1985,dinu.prl.2016, dipiazza.pra.2019}.
Otherwise, there are contributions from different formation regions to the same final state~\cite{harvey.pra.2015}:
for instance, there always can be photons with the formation length of the order of the EM field inhomogeneities, even if $a_0 \gg 1$ (see \cref{sec:FormationLength}).
Note that the LCFA is usually accompanied by the assumption of collinear emission, i.e. that the photon is emitted along the instantaneous direction of electron/positron momentum~\cite{blackburn.pop.2018}; this is a good approximation for the highest energy photons.
Moreover, at the higher intensities are considered in laser-plasma interactions, stronger, ultrashort, and higher frequency EM fields can be generated~\cite{naumova.prl.2004,teubner.rmp.2009,gonoskov.pre.2011}, so that the simple model of a monochromatic carrer wave no longer suffices.

Validity checks of the LCFA have been performed by \citet{harvey.pra.2015,dinu.prl.2016,blackburn.pop.2018,dipiazza.pra.2018}.
Proposed corrections to the rates include:
modification of the spectrum at low energy~\cite{dipiazza.pra.2018,dipiazza.pra.2019};
accounting for the dependence on field gradients~\cite{ilderton.pra.2019};
and discarding of photons with excessively long formation lengths~\cite{blackburn.pra.2020}.
These proposals have yet to be implemented in the PIC codes used to study laser-matter interactions at high intensity.
{An alternative strategy is to treat the background field as a \emph{locally monochromatic plane wave}, rather than a locally constant, crossed field~\cite{heinzl.pra.2020}: this exchanges the flexibility of the LCFA for increased accuracy in modelling interactions with plane-wave-like fields~\cite{blackburn.njp.2021} and has already been used in connection with experiments~\cite{bamber.prd.1999}.}

The SFQED emission rates depend on the spin of electrons/positrons, as well as on the polarization of photons, as discussed in \cref{sec:Spin} and in \citet{ilderton.prd.2020, seipt.pra.2020}. This can result in the generation of polarized electron, positron, and photon beams (see, e.g., \citet{chen.prl.2019, seipt.pra.2019, king.pra.2020n, tang.plb.2020}, and the discussion in \cref{sec:PolarizedBeams}), which can be important for high-energy physics applications. Studies of spin and polarization effects in the framework of SFQED have started only recently and more effort is required to advance this topic. Of particular interest is the dependence of the higher order SFQED effects, such as cascades, on spin and polarization \cite{seipt.njp.2021}.

There are many other important theoretical problems connected with radiative corrections, vacuum polarization, 4-wave mixing, photon-photon scattering Cherenkov radiation in strong fields, and beam-beam interaction~\cite{macleod.prl.2019, bulanov.prd.2019, yakimenko.prl.2019}. We note that beyond the PIC-QED framework used to study SFQED effects, there is also the real-time lattice QED approach \cite{hebenstreit.prd.2013,shi.pre.2018}, which needs to be advanced.

\subsection{Experimental programs}
\label{sec:FutureFacilities}

    \begin{figure*}[tbp]
    \centering
	\includegraphics[width=1.0\textwidth]{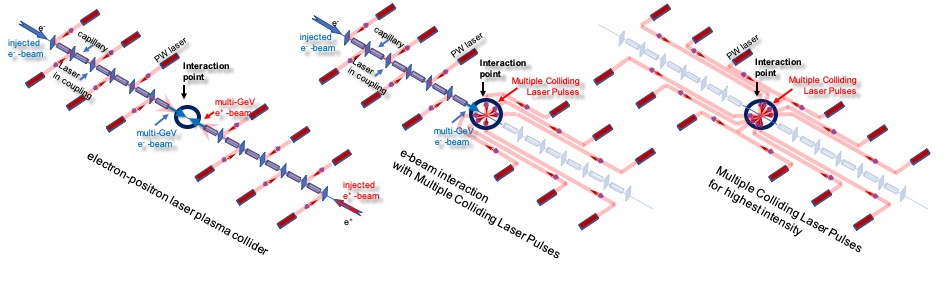}
	\caption{
	    An dedicated SFQED facility, housing multiple PW-class lasers.
	    The facility can operate in several modes, including:
	    (i) $e^+e^-$ collider, with all lasers used to drive the staged acceleration of electron and positron beams;
	    (ii) laser-electron-beam collider, where half of the lasers drive the staged acceleration of the electron beam and the remaining half provides the high-field region via the multiple colliding pulses configuration;
	    and (iii) all the laser pulses are brought to the interaction point to generate highest intensity possible.
	    Reproduced from \citet{zhang.pop.2020}.
    }
	\label{fig:facility}
    \end{figure*}

In \cref{sec:Geometries} we introduced two basic interaction geometries for the exploration of SFQED effects: a high-energy electron beam colliding with a high-intensity laser pulse and two colliding, high-intensity laser pulses.
Based on this review's discussion of SFQED interaction properties and the availability of lasers and electron beams, we can envisage a three-stage program, with associated facility development, for experimental research~\cite{zhang.pop.2020}.
The first two stages correspond to the two basic interaction configurations, whereas the third provides a setup flexible enough to accommodate both at much higher energies and intensities.
{This three-stage program for laser-based facilities is well aligned with the strategic plans for plasma-based advanced accelerators and novel radiation sources and follows a path from a 10-GeV demonstrator facility to the plasma based electron-positron collider  (for details on plasma based collider strategic plans see \cite{roadmap.doe.2016,alegro.arxiv.2019,alegro.arxiv.2020}).}

First, the high-power laser facility should be able to investigate the basic quantum processes of SFQED, described in \cref{sec:ComptonBreitWheeler}, by colliding electron beams and high-intensity laser pulses.
This could be realized with two laser beamlines, where one is for particle acceleration \cite{bulanov.nima.2011,roadmap.doe.2016,alegro.arxiv.2019,alegro.arxiv.2020}, or by colocating a PW-class laser and an existing conventional lepton accelerator~\cite{meuren.arxiv.2020}.
In both cases the laser should be capable of being focused to a spot size comparable to a single wavelength.
The parameters required for $\chi \sim 1$ are within reach of existing laser facilities, as 8-GeV electron beams from LWFA have been demonstrated~\cite{gonsalves.prl.2019}, as has an intensity of $10^{23}~\Wcmsqd$ from a PW of laser power~\cite{yoon.optica.2021}.
A 10-GeV LWFA module paired with a tightly focused PW-class laser, or a GeV-class electron beam paired with a multi-PW laser can satisfy the energy-power requirements, {which places the required total laser power into the 1-10 PW range.}
We note that this setup is similar to that of E144 experiment at SLAC~\cite{bula.prl.1996} and two recent ones at GEMINI~\cite{cole.prx.2018, poder.prx.2018}. 
However, it is of critical importance to study SFQED effects in such configuration at higher precision and energy.
As such, it is important that the facility deliver stable, high-repetition-rate collisions, with the ability to tune the electron energy and laser intensity across a few orders of magnitude. 
This would permits exploration of the classical-to-quantum, perturbative-to-nonperturbative transitions, as well as tests of the approximations underlying analytical and numerical predictions of SFQED (see \cref{sec:OpenQuestions}).
{There are a number of existing, being built, and planned laser facilities, which satisfy the above mentioned requirements of 1-10 PW power (see \cref{fig:LaserHR}): L4 at ELI Beamlines in Czech Republic, HPLS at ELI NP in Romania, Apollon (F1) at LULI in France, PEARL-X at IAP in Russia, SULF at SIOM in China, J-EPOCH at ILE or QST in Japan, EP-OPLA at LLE in USA, CoReLS at IBS in South Korea, ZEUS at CUOS in USA, and ATLAS-3000 at CALA in Germany.}

In the second stage, the high-power laser facility would reach the radiation-dominated regime and produce avalanche-type cascades.
The simplest configuration is two, counterpropagating laser pulses; however, dividing the same amount of power among \emph{multiple} colliding laser pulses that arrive at the same interaction point is more beneficial from the point of view of reaching extreme field strengths.
Reaching the radiation-dominated regime, assuming a focal spot size of a few wavelengths, demands a total laser power in the range 10--100 PW.
Scientific goals at this stage would be higher order SFQED processes and the rich dynamical phenomena that accompany them (see \cref{sec:Dynamics}).
The coupling between classical, relativistic plasma physics and quantum effects will put numerical models into the test, particularly as significant field depletion is expected.

In the third stage, the facility would be flexible enough to accommodate both accelerator and SFQED studies.
A sketch is shown in \cref{fig:facility}.
Three modes of operation are possible: (i) a plasma-based electron-positron collider; (ii) a plasma-based electron/positron and laser collider; (iii) multiple colliding laser pulses.
The first mode corresponds to a compact means to achieve electron-positron collisions relevant for high-energy physics.
Dense electron or positron beams can themselves induce SFQED phenomena due to the strong fields that accompany them, as discussed in \cref{sec:BeamBeam}.
In the second mode, the lasers from one arm are rerouted to the interaction point to form a standing EM wave, which is probed by the high-energy electron beam accelerated by the lasers in the other arm.
This type of interaction is discussed in \citet{magnusson.prl.2019, magnusson.pra.2019} as a means of studying both shower and avalanche-type cascades and as a high-brightness source of multi-GeV photons.
The third mode should provide the ultimate test for SFQED, by achieving highest possible intensity with all the lasers focused at the interaction point. 

\section{Conclusions}
\label{sec:Conclusions}

The study of charged particle motion, induced by electromagnetic fields of different origin, forms one of the pillars that support our understanding the natural world.
It extends from from basic electrostatics and magnetism to classical and quantum electrodynamics, and is unavoidable in extreme plasma environments as found in terrestrial laboratories and in space.

In this review, we have summarized the results of recent research on charged particle motion and radiation in strong electromagnetic fields.
While these strong fields can be encountered in highly magnetized astrophysical environments, during beam-beam interactions in particle colliders, and in heavy-ion collisions, we focus on the interactions involving high intensity lasers.
Due to the fast development of laser technology, lasers now produce some of the strongest electromagnetic fields available in the laboratory.
While present-day lasers can efficiently accelerate particles to relativistic energies and produce sources of high frequency radiation, their next-generation counterparts will provide EM field strengths so much higher that they will access the so-called radiation dominated regime, where the motion of particles will be determined not by just by external fields, but also by the self-generated, i.e. radiation, fields.

Though the problem of radiation reaction in classical electrodynamics dates back to Lorentz, Abraham, and Dirac (see \cref{sec:ClassicalRR}), and the first results in SFQED were obtained in the 1960s, the advance in laser technology (see \cref{fig:LaserHR}) has prompted significant recent progress, in theory (\cref{sec:StrongFieldQED}), simulations (\cref{sec:Numerics}), and experiments (\cref{sec:Experiments}).
The coupling between SFQED processes and classical, relativistic plasma physics has unveiled new complex dynamics and raised the possibility of developing new, high-flux particle and radiation sources.
All these effects are of interest for quantum field theory and will dominate the operation of next-generation laser facilities, whether they are designed specifically for the study of SFQED effects, for particle acceleration at multi-GeV and TeV level, or for secondary radiation generation. 

Despite the tremendous progress achieved, there are many open questions in SFQED (see \cref{sec:OpenQuestions}).
Solving these will require the coordinated effort of theoretical, simulation, and experimental groups and is likely to occupy the SFQED community for several years to come.
Undoubtedly many new puzzles will arise as the next generation of laser facilities comes into operation, when long-held theoretical expectations are challenged by experimental evidence.
We view this as the most exciting prospect of our field.

\bib

%% file: definitions.tex
\section*{List of commonly used symbols}

\begin{longtable}{p{0.1\linewidth}p{0.4\linewidth}p{0.4\linewidth}}
\hline
\hline
 & Value & Definition \\
\hline
$e$ & $1.602\times 10^{-19}$~C & elementary charge \\
$m$ & $0.511$ MeV/c$^2$ & electron mass \\
$c$ & $299~792~458$ m s$^{-1}$ & speed of light in vacuum \\
$\alpha$ & \makecell[lt]{$e^2/(4\pi\varepsilon_0 \hbar c)$ \\ $\quad \simeq 1/137$} & fine-structure constant \\
$\lc$ & \makecell[lt]{$\hbar/mc$ \\ $\quad \simeq 386$~fm} & Compton wavelength \\
$\Ecrit$ & \makecell[lt]{$m^2 c^3 / (e \hbar)$ \\ $\simeq 1.323 \times 10^{18}~\mathrm{V}\mathrm{m}^{-1}$} & critical field of QED \\
$\Bcrit$ & \makecell[lt]{$m^2 c^2 / (e \hbar)$ \\ $\simeq 4.4 \times 10^{9}$~T} & critical magnetic field of QED \\
$\Ecrit^\text{clas}$ & \makecell[lt]{$\Ecrit/\alpha$ \\ $\simeq 1.813 \times 10^{20}~\mathrm{V}\mathrm{m}^{-1}$} & classical critical field \\
$p^\mu$ & $\gamma m (c, \vec{v})$ & particle four-momentum \\
$\gamma$ & $(1 - v^2/c^2)^{-1/2}$ & Lorentz factor \\
$\hbar k^\mu$ & $\hbar \omega / c (1, \vec{k})$ & photon four-momentum \\
$\hbar \kappa^\mu$ & $\hbar \omega_0 / c (1, \vec{k})$ & laser four-momentum \\
$F_{\mu\nu}$ & & EM field tensor \\
$\tilde{F}_{\mu\nu}$ & & dual EM field tensor \\
$\InvariantF$ & $-F_{\mu\nu} F^{\mu\nu} / (2 \Ecrit^2)$ & 1st Poincare invariant \\
$\InvariantG$ & $-F_{\mu\nu} \tilde{F}^{\mu\nu} / (2 \Ecrit^2)$ & 2nd Poincare invariant \\
$\chi_e$ & $\abs{F_{\mu\nu}p^\nu} / (mc \Ecrit)$ & lepton quantum nonlinearity parameter \\
$\chi_\gamma$ & $\abs{F_{\mu\nu} \hbar k^\nu} / (mc \Ecrit)$ & photon quantum nonlinearity parameter \\
$a_0$ & \makecell[lt]{$e \abs{F_{\mu\nu} p^\nu} / (m c^2 \kappa_\mu p^\mu)$ \\ $\quad = e E_0 / (m c \omega_0)$} & dimensionless EM vector potential \\
$\gamma_K$ & $\omega_0 \sqrt{2 m V_0} / (e E_0)$ & Keldysh parameter \\
$\acrit$ & \makecell[lt]{$m c^2 / (\hbar \omega_0)$ \\ $\quad \simeq 4.1 \times 10^5 \lambda\,[\micron]$} & $a_0$ corresponding to a QED-critical field \\
$\Power_\text{rad}$ & & power radiated by an electron accelerated by an EM field \\
$L_i$ & $\gamma m c^3 / \Power_\text{rad}$ & depletion length, distance over which electron radiates away energy equal to its own energy \\
$L_q$ & & quantization length, typical distance between photon emissions \\
$L_f$ & \makecell[lt]{$\avg{\theta^2}^{1/2} \chi_e \lc / \gamma^2$ \\ $\simeq \frac{c \chi^{1/3}}{a_0\omega_0}\left(\frac{1-f_c}{f_c}\right)^{1/3}$} & formation length, typical distance over which photon is emitted \\
$P_b$ & & pair creation rate \\
$L_p$ & $c / P_b$ & pair creation length \\
$\erad$ & \makecell[lt]{$4 \pi r_e / (3 \lambda)$ \\ $\simeq 1.17\times 10^{-8}/\lambda\,[\micron]$} & dimensionless parameter which characterises the role of radiation reaction in charged particle motion \\
$\acrr$ & \makecell[lt]{$\erad^{-1/3}$ \\ $\quad \simeq 440\lambda^{1/3}\,[\micron]$; \\ $(4\pi\erad\gamma)^{1/2}$ \\ $\simeq 5.9 \lambda^{1/2}\,[\micron] / \Energy_0\,[\text{GeV}]$} & classical RR threshold (laser-laser and laser--electron-beam respectively) \\
$\aqrr$ & \makecell[lt]{$4\alpha^2/(9\erad)$ \\ $\quad \simeq 2000\lambda\,[\micron]$; \\ $\alpha/(3\erad\gamma)$ \\ $\simeq 105 \lambda\,[\micron] / \Energy_0\,[\text{GeV}]$} & quantum threshold (laser-laser and laser--electron-beam respectively) \\
$g^\mu_\text{LAD}$ & & RR term in Landau-Lifshitz equation \\
$g^\mu_\text{LL}$ & & RR term in Landau-Lifshitz equation \\
\makecell[lt]{$\frac{d\Energy}{d\hbar\omega}$} & & energy radiated per unit photon energy \\
\makecell[lt]{$\frac{d W}{d\hbar\omega}$} & & spectral emission rate \\
$f_c$ & $\hbar\omega / (\gamma m c^2)$ & energy transfer from electron to photon in Compton scattering \\
$\xi_q$ & \makecell[lt]{$\frac{2}{3\chi_e}\frac{f_c}{1-f_c}$} & Compton spectrum shape parameter\\
$\psi_{pr}$ & & Volkov solution of Dirac equation \\
$P^c$ & & The probability rate of the Compton process \\
$P^b$ & & The probability rate of the Breit-Wheeler process \\
$\varphi_e$ & & The electron distribution function  \\
$G(\chi)$ & \makecell[lt]{$\frac{3\sqrt{3}}{\pi}\int\limits_0^\infty K_{2/3}(u)\times$ \\ $\frac{8u+15\chi u^2+18\chi^2 u^3}{(2+3\chi u)^4} du$} & The ratio between the instantaneous radiation powers predicted by QED and classical theory. Also referred to as 'Gaunt factor'\\
$\varepsilon_p$ & $(2\pi e^2n_e l)/(m_e\omega_0 c)$ & The parameter governing the transparency of the thin solid density foil under the action of EM wave. \\

\hline
\hline
\label{tbl:Definitions}
\end{longtable}

\newpage
\onecolumngrid
\section*{Parameters of the laser facilities shown in Fig. 2}

\begin{longtable}{p{0.2\linewidth}p{0.15\linewidth}p{0.15\linewidth}p{0.15\linewidth}p{0.1\linewidth}p{0.1\linewidth}p{0.1\linewidth}p{0.1\linewidth}}
\hline
\hline
 Name & Facility & Country & wavelength, $\mu$m & energy, J & duration, fs & power, PW \\
\hline
Station of Extreme Light & SIOM & China	& 0.9 & 1500 & 15 & 100 \\
EP-OPAL & LLE & USA & 0.8 & 500 & 20 & 25 \\
SULF	& SIOM & China	&	0.8	& 220	& 21 &	10 \\
Apollon (F1) & LULI & France & 0.82 & 150 & 15 & 10 \\
PEARL-X	& IAP &	Russia	& 0.527 & 200 & 20 & 10 \\
L4 & ELI-Beamlines & Czech Republic & 1.057 & 1500 & 150 & 10 \\
HPLS & ELI-NP & Romania & 0.814 & 200 & 20 & 10 \\
J-EPOCH & ILE+QST & Japan & 0.8 & 200 & 20 & 10 \\
SULF (current) & SIOM & China & 0.8 & 130 & 24 & 5.4 \\
SG-II 5 PW & SIOM  & China & 0.808 & 150 & 30 & 5 \\
CAEP-PW & CAEP & China & 0.8 & 91 & 19 & 4.8 \\
CoReLS & IBS & South Korea & 0.8 & 83 & 20 & 4.15 \\
ZEUS & CUOS & USA & 0.8 & 60 & 20 & 3 \\
ATLAS-3000 & CALA & Germany & 0.8 & 60 & 25 & 2.4 \\
Qiangguang & SIOM  & China & 0.8 & 52 & 26 & 2 \\
SG-II 5 PW & SIOM & China & 0.808 & 37 & 21 & 1.8 \\
Apollon & LULI & France & 0.82 & 38 & 22 & 1.7 \\
Nova Petawatt & LLNL & USA & 1.054 & 660 & 440 & 1.5 \\
BELLA & LBNL & USA & 0.8 & 40 & 30 & 1.3 \\
LFEX  & GEKKO XII & Japan & 1.054 & 2000 & 1500 & 1.3 \\
J-KAREN-P & KPSI & Japan & 0.82 & 40 & 30 & 1.3 \\
PETAL & CEA & France & 1.053 & 850 & 700 & 1.2 \\
Xtreme Light III & NLCM & China & 0.8 & 32 & 28 & 1.1 \\
CoReLS & IBS & South Korea & 0.8 & 33 & 30 & 1.1 \\
Z-Petawatt & Sandia & USA & 1.054 & 500 & 500 & 1 \\
Vulcan Petawatt & CLF, RAL & UK & 1.054 & 500 & 500 & 1 \\
Orion & AWE & UK & 1.053 & 500 & 500 & 1 \\
Apollon (F2) & LULI & France & 0.82 & 15 & 15 & 1 \\
PEnELOPE & HZDR & Germany & 1.03 & 150 & 150 & 1 \\
VEGA-3 & CLPU & Spain & 0.8 & 30 & 30 & 1 \\
CETAL & INFLPR & Romania & 0.8 & 25 & 25 & 1 \\
L2 & ELI-Beamlines & Czech Republic	 & 0.82 & 15 & 15 & 1 \\
L3 (HAPLS) & ELI-Beamlines & Czech Republic & 0.82 & 30 & 30 & 1 \\
SG-II UP (Shenguang II) & SIOM  & 	China & 1.054 & 1000 & 1000 & 1 \\
PWM (Petawatt Module) & GEKKO XII & Japan & 1.054 & 420 & 470 & 0.89 \\
Texas Petawatt & UoT at Austin & USA & 1.057 & 120 & 140 & 0.86 \\
J-KAREN & KPSI & Japan & 0.8 & 28 & 33 & 0.85 \\
Xingguang-III (fs beam) & LFRC & China & 0.8 & 20 & 27 & 0.74 \\
OMEGA-EP (short pulse) & LLE & USA & 1.053 & 500 & 700 & 0.71 \\
Diocles & ELL & USA	 & 0.8 & 20 & 30 & 0.67 \\
NIF & LLNL & USA & 0.3516 & $1.80 \times 10^6$ & $3.00 \times 10^6$ & 0.6 \\
PEARL & IAP & Russia & 0.91 & 24 & 45 & 0.53 \\
HERCULES & CUOS & USA & 0.8 & 15 & 30 & 0.5 \\
Gemini & CLF & UK & 0.8 & 15 & 30 & 0.5 \\
Laser Megajoule & CEA & France & 0.351 & $1.50 \times 10^6$ & $3.00 \times 10^6$ & 0.5 \\
PHELIX & HI GSI & Germany & 1.054 & 200 & 400 & 0.5 \\
Titan (west) & LLNL & USA & 1.053 & 300 & 700 & 0.43 \\
SCAPA & Univ. Strathclyde & UK & 0.8 & 8.75 & 25 & 0.35 \\
Scarlet & Ohio State Univ. & USA & 0.8 & 10 & 30 & 0.33 \\
Jeti200 & HI Jena & Germany & 0.8 & 4 & 17 & 0.24 \\
ALEPH & Colorado State Univ. & USA & 0.4 & 10 & 45 & 0.2 \\
POLARIS & HI Jena & Germany & 1.03 & 17 & 98 & 0.17 \\
Arcturus & Heinrich-Heine Univ. & Germany & 0.8 & 3.5 & 30 & 0.12 \\
DRACO & HZDR & Germany & 0.8 & 3.5 & 30 & 0.12 \\

\hline
\hline
\label{tbl:LaserData}
\end{longtable}
\twocolumngrid